\documentclass[universe,article,accept,pdftex,moreauthors]{Definitions/mdpi} 
\newcommand{\re}{$R_e$}
\newcommand{\Lsig}{$L-\sigma$}

\newcommand{\MRa}{$R_e$-$M^*$}
\newcommand{\MVBmV}{$M_V$-$(B-V)$}
\newcommand{\nL}{$n$-$L$}

\newcommand{\Ie}{$I_e$}

\newcommand{\IeRe}{$I_e - R_e$}

\newcommand{\LnV}{$\log(L)-\log(n)$}

\newcommand{\BmV}{$\rm{B-V}$}

\newcommand{\bfilt}{$\rm{B}$}
\newcommand{\vfilt}{$\rm{V}$}

\newcommand{\kms}{km\, sec$^{-1}$}

\firstpage{1} 
\pubvolume{1}
\issuenum{1}
\articlenumber{0}
\pubyear{2025}
\copyrightyear{2025}
\datereceived{ } 
\daterevised{ } 
\dateaccepted{ } 
\datepublished{ } 
\hreflink{https://doi.org/} 
\pdfoutput=1 

\Title{The scaling relations of galaxies with different morphology: comparison among WINGS, MANGA and Illustris data samples.}

\TitleCitation{Scaling relations and morphologies}

\Author{Mauro D'Onofrio $^{1}$\orcidA{}, Francesco Brevi $^{2}$\orcidB{}, Cesare Chiosi $^{2}$\orcidC{} and Paola Marziani$^{3}$\orcidD{}}

\AuthorNames{Mauro D'Onofrio, Francesco Brevi, Cesare Chiosi and Paola Marziani}

\isAPAStyle{%
       \AuthorCitation{D'Onofrio, M., Brevi, F., Chiosi, C., \& Marziani, P.}
         }{%
        \isChicagoStyle{%
        \AuthorCitation{Mauro, D'Onofrio, Francesco, Brevi, Cesare Chiosi and Paola Marziani.}
        }{
        \AuthorCitation{D'Onofrio, M.; Brevi, F.; Chiosi, C.; Marziani, P.}
        }
}

\address{%
$^{1}$ \quad Department of Physics and Astronomy G. Galilei, University of Padua, Italy; mauro.donofrio@unipd.it\\
$^{2}$ \quad Department of Physics and Astronomy G. Galilei, University of Padua, Italy; cesare.chiosi@unipd.it and francesco.brevi@studenti.unipd.it\\
$^{3}$ \quad Istituto Nazionale di Astrofisica, Osservatorio Astronomico di Padova, Italy; paola.marziani@inaf.it}

\corres{Correspondence: mauro.donofrio@unipd.it.com; Tel.: +39-049-8278233 (M.D.)}

\abstract{We present a panoramic view of several scaling relations (ScRs) of galaxies of different morphology. The ScRs are obtained from the data of two large surveys (WINGS and MANGA). We analyze the distribution (parameterized by the percent over the total) of galaxies in each region of the diagnostic planes that are set up by means of suitable physical quantities. In addition to this, we discuss the origin of the differences observed in the ScRs between the two samples. Finally, we compare the observational data with the theoretical ones taken from two subsets of the Illustris large scale simulations (TNG50 and TNG100) and  we discuss how the comparison should be performed for a correct statistical answer.}

\keyword{galaxies: structures; galaxies: scaling relations; galaxies: morphology.} 

\begin{document}


\section{Introduction}
{Over the past decades, astronomers have measured important  properties of galaxies, such as their masses, luminosities, radii, velocities, shapes, metallicities and morphology. All these measurements characterize a galaxy by means of an array of structural parameters. Of course across time, several definitions of the structural parameters and methods to obtain them have been proposed.}

The observational data reveals that many structural parameters are mutually correlated, e.g. luminosity and velocity dispersion (the Faber-Jackson relation, \cite{FaberJackson1976}), mass and radius (the Mass-Radius \MRa\ relation \cite{MoMaoWhite98}), surface brightness and radius (the \IeRe\ relation \cite{Kormendy1977}). {Sometimes the correlations involve up to three parameters} (e.g. radius, surface brightness and velocity dispersion, i.e. the Fundamental Plane \cite{Dressleretal1987,DjorgovskiDavis1987}). These correlations take today the name of scaling relations (ScRs).

ScRs among galaxy properties are key diagnostic tools for understanding the physical mechanisms driving galaxy formation and evolution. These correlations have been used to probe the mass distribution, stellar populations, and assembly histories of galaxies. The tightness of these correlations suggests that galaxies evolve along well-defined physical paths governed by a limited number of parameters, such as stellar mass, velocity dispersion, and specific angular momentum.

Most of the studies devoted to the analysis of the ScRs have addressed various aspects of the 2D correlations, such as the degree of correlation, the slope and the intercept of the fitted relation and the scatter around it. The ScRs are commonly used in different astrophysical contexts. One can, for instance, use them to measure  the distance of galaxies, or to derive peculiar velocities, or more in general, to compare theory with observations. Customarily, a theory is accepted when it reproduces the main observed ScRs of galaxies.

A less explored method for approaching the ScRs concerns the analysis of the global distribution of galaxies in the 2D parameter space defined by the chosen variables. Indeed, galaxies do not distribute uniformly in the ScRs, but their density varies considerably in different areas of the parameter space and in general it depends on the mass and luminosity of the galaxy.

The 2D analysis can reveals: 
i) Population bimodality and morphological segregation, such as: early vs late types, compact vs extended galaxies, quenched vs star-forming systems, pseudo-bulges vs classical bulges. Different regions of the 2D plane correspond to different formation channels (e.g., dissipative collapse vs. disk growth). 
ii) Environmental effects. By comparing densities in the 2D planes given by the two surveys one can note shifts in the average size, increased compactness in cluster cores, presence of environmental compaction or stripping, and a morphology–density relation in structural space. 
iii) Intrinsic scatter caused by the underlying physics. The scatter of the 2D distribution carries major physical significance. Large scatter often implies diversity in merger histories, feedback effects, angular momentum, stellar population age, and metallicity dependence of the  dissipation process at the stage of galaxy  formation. 
iv) Identification of outliers and rare objects. The 2D planes naturally show the occurrence of  outliers such as ultra-compact dwarfs, super-compact massive galaxies, cored vs cuspy systems, highly extended diffuse galaxies, rejuvenated systems or compaction remnants. 
v) Evolutionary tracks and structural transformations. The density distribution reflects typical paths followed by galaxies across the  structural space.
For example: the transition from the blue cloud to the green valley and the red sequence, the location of gas-rich mergers, compact starbursts, quiescent compact objects, size growth via dry mergers, and finally the disk building-up via secular bar and pseudo-bulge formation. 
vi) Constraints on the mass-assembly and quenching. The density of objects and its variation in diagnostic  planes reveals at what mass compaction occurs, at which structural stage quenching happens, whether quenching precedes or follows size growth, and the typical structural state of galaxies at quenching time;  
vii) Consistency (or inconsistency) between surveys. If the 2D distributions differ, the samples have different structural demographics. This tells us that either they experienced different formation histories or they are affected by different selection effects.

This information is extremely important, in particular for the validation of galaxy simulations, that need to reproduce in a great detail both the galaxy properties   and the number counts of objects  with different morphology in each region of the parameter space. These data are indeed tightly  related to the statistics of the dark matter halos and the effects of the baryon physics  during the evolutionary history  of galaxies of different morphological type.
The shape of the 2D distribution is a fundamental diagnostic tool for simulations, by far more informative than the slope of any scaling relation. It may highlight if the simulations reproduce compactness at given mass, if the number of extended galaxies agrees with the observations, if the shapes and sizes are realistic, and finally if the effects of  quenching occurs in the correct structural regime. The 2D density maps may reveal transitional regimes such as: mass thresholds where structural properties steeply change, the emergence of a second sequence, and bending or curvature in the parameter space.

{Quite often the artificial galaxies are compared with the observed ones showing the superposition of models and real data in the most common ScRs. By looking qualitatively at such superpositions, many studies conclude for a satisfactory match between theory and observations, avoiding a much complex statistical analysis. The analysis is generally devoted to the correlations between the parameters and the comparison with that obtained by observations. The 2D analysis of the distribution is always not addressed.}

The aim of this paper is twofold. First, we analyze the 2D distribution of galaxies of different morphologies in various ScRs by exploiting two large datasets, WINGS (WIde-field Nearby Galaxy clusters Survey, \cite{Fasanoetal2006}) and MANGA (Mapping Nearby Galaxies at Apache Point Observatory \cite{Bundyetal2015}). Here, we tentatively address the problem of understanding the differences observed in the 2D distribution of galaxies of different morphologies in several ScRs. Second, we compare the observational data with artificial galaxies, by exploiting the data of the Illustris-TNG simulations.

As we will see, both goals are difficult to achieve. Indeed many effects alter the 2D distribution of galaxies in the 2D parameter space, in such a way that the same 2D plane obtained with different data-sets is often quite different from a statistical point of view.
At the origin of such differences there are various causes: i) Random and systematic errors. These affect all types of data. The use of different detectors, filters, calibrations, software, methodology, etc., generally introduce errors that affect all the measured parameters.  The typical error bars on the scaling parameters are of the order of 20\%. ii) The adopted procedures of data analysis are often quite different, each of which with advantages and disadvantages. A popular  example is the subtraction of the sky background. The different assumptions adopted for this parameter introduces systematic differences in the estimated radii.  iii) The definition of the structural parameters is far from univocal, and therefore the parameters themselves are not defined in a standard way. For example, the radius of a galaxy, one of the most important structural parameter, is defined in several different ways. One can use a  certain isophote level (e.g the radius where the surface brightness is 25 mag arcsec$^{-2}$) or a certain degree of concentration of light (e.g. 50\%, 90\%). In addition to it,  one can adopt different techniques to get those limits. Some additional remarks are worth here about the effective radius \re, one of the most used radii. It has indeed multiple definitions. Quite often it is defined as the radius of the circle that contains half the total luminosity of a galaxy, but many authors use the radius of the ellipse major axis that contains half the total luminosity or the radius derived from fitting of the light profiles with the aid of empirical formulae that contains \re\ as a parameter. Furthermore, the technique adopted to measure the radius can take into account or not the convolution with the seeing point spread function (PSF) or can use a deconvolved image or simply exclude the central regions. Sometimes alternative definitions are adopted to define the radius of a galaxy. For instance, one sets the radius at the edge limit where the surface brightness has a visible cut-off.  All this will affect the ScRs. iv) The galaxy samples under examination are not statistically similar. The different degree of completeness and homogeneity of the samples affect the statistical analysis. The galaxy samples are often very different in luminosity, morphology, redshift, volume density, environment and so on. Comparing apples with pears is not a good thing in general. v) The statistical analysis of the correlation changes
{when different fitting methods are used. The fitted parameters,
in particular slopes and intercept, change when orthogonal or bivariate or linear least square methods are applied. All these largely unsettled uncertainties engender ambiguity in the final study of the ScRs.}

The current situation is that as today we miss a reference distribution of  galaxies of different morphologies in each 2D parameter space. In other words, we cannot answer some simple questions, such as "How many galaxies of a given morphological type 
in the local Universe have a radius and surface brightness or mass or central velocity dispersion within a given range?". Such information is very important when we try to model our Universe. A precise answer to this question would imply that we are able to reproduce the correct number of galaxies of a given morphology in any  volume of the Universe and are also able to follow their evolution in luminosity, mass and radius. The task is difficult but it is not impossible. The first step toward this goal is to introduce some standard ways of working in order to minimize the differences observed in the galaxy distributions in the ScRs.

{\bf All the mentioned effects taken together are a severe obstacle for the scientific analysis. Efforts are necessary to create standard parameters and standard procedures of data analysis in such a way that scientific comparison is meaningful.
This study aims to highlight the importance of working with standard procedures, statistically well defined samples, and clean methods of analysis.} 

{To better emphasize how many factors influence the observed distributions of galaxies in the ScRs we analyze  two very different samples of galaxies (WINGS and MANGA), that cover different environment, volumes, luminosity range and are based on different technique of analysis. This choice allow us  to investigate the ScRs in a statistical perspective, and better understand what problems cause the discrepancies noted in the two 2D distributions.}

The comparison of real galaxies with theoretical simulations is also very important. Up to now most studies devoted to numerical simulations of galaxies, when dealing with ScRs focused their analysis on simple statistical correlations (slopes, intercepts and scatter). Little attention has been paid  to the detailed 2D distribution of galaxies on the planes in which the various ScRs are derived. This means that one cannot firmly assess that simulations correctly reproduce the observational ScRs. 

{To this aim, we adopted two different samples of artificial galaxies, extracted from the Illustris database, that is  TNG50 and TNG100. The data have different resolution, redshift and luminosity range. At the end of the work the ScRs of these data-sets, will be compared with WINGS and MANGA
and the origin of the observed difference will be addressed. We will
learn that it is not sufficient to work with similar volumes of the Universe to get similar ScRs. The consistency between real and observational data requires a careful analysis of the luminosity functions for all morphological types, the adoption of similar technique of data analysis and similar definition of the adopted scaling parameters.}

The paper is structured as follows. In Sec. \ref{sec:2}  we  present  the two databases we have used, namely WINGS and MANGA. In Sec. \ref{sec:3} we compare the range spanned by the measured structural parameters in galaxies of the two surveys. We then address the analysis of the different ScRs for the galaxies of different morphology according to the main groups: early-types (ETGs) and late types (LTGs). We begin with the \IeRe\ plane (Sec. \ref{sec:4}), and continue with the \MRa\ plane (Sec. \ref{sec:5}), the \MVBmV\ plane (Sec. \ref{sec:6}), the \LnV\ plane (Sec. \ref{sec:7}), and the \Lsig\ plane (Sec. \ref{sec:8}). 
We present a comparison of the main observed ScRs for ETGs and LTGs with those coming from the Illustris numerical simulations (Sec. \ref{sec:9}). The structural parameters extracted from the data of TNG50 and TNG100 obtained respectively by \cite{Rodriguez-Gomezetal2019} and \cite{Ferreiraetal2025} for ETGs and LTGs are compared with  our datasets. Some 
conclusions are drawn in Sec. \ref{sec:10}.

In the paper we adopted the standard $\Lambda$CDM model of the Universe, with $H_0 = 70 km\, s^{-1}\, Mpc^{-1}$, 
$\Omega_\Lambda = 0.7274$, $\Omega_m = 0.2726$,   and $\Omega_b =0.0456$ \cite{Vogelsberger_2014a, Vogelsberger_2014b}.

\section{The galaxy samples}\label{sec:2}

The observational data used in this study are extracted from the WINGS \citep{Fasanoetal2006,Varela2009,Cava2009,Valentinuzzi2009,Moretti2014,Donofrio2014,Gullieuszik2015,Morettietal2017} and MANGA databases
\citep{Bundyetal2015,Sanchezetal2022}.

WINGS (WIde-field Nearby Galaxy-cluster Survey) is the largest data sample for galaxies in nearby clusters ($0.04 \le z \le 0.07$). The database includes magnitudes, morphologies, effective radii and surface brightness, stellar velocity dispersions, star formation rates, and many other useful measurements.
The optical photometric catalog in the \bfilt\ and \vfilt\ bands is 90\% complete at \vfilt\ $\sim 21.7$ mag. The spectroscopic sample, with $\sim12,000$ redshifts is $\sim80\%$ complete down to $V=20$. The photometric database includes more than 390,000 galaxies detected by Sextractor. Those with an area larger than 100 pixels ($\sim33,000$) were analyzed with the software GASPHOT \citep{Pignatelli} that provided the effective parameters \re, \Ie, the total luminosity $L$ in the \bfilt\ and \vfilt\ bands, and the S\'ersic index $n$. The final $B-V$ color is referred to the whole galaxy and was corrected for galactic absorption and k-correction.

The radius \re\ is that of the circle that encloses half the total V-band luminosity. Using a single S\'ersic law convolved with a space-varying PSF given by a multi-gaussian curve, GASPHOT performed a simultaneous $\chi^2$ best-fit of the major- (a) and minor-axis (b) luminosity growth curves of the galaxies (\cite{Pignatelli}). The semi-major axis radius $a$ was then circularized (\re=$a*\sqrt{b/a}$) and converted in kpc units adopting the distance modulus of each galaxy on the basis of the redshift of the cluster (\cite{Donofrio2014}). 

The average surface brightness \Ie\ comes from the relation linking $L$ and \re\ ($L=2\pi I_eR^2_e$). This means that there are no systematic biases entering the \IeRe\ plane due to the different morphologies, flattening, and shapes of the light profiles of the galaxies. The errors on \re\ and \Ie\ are $\simeq20\%$. The size of these errors does not affect the distribution of galaxies in the \IeRe\ plane that span several order of magnitudes. 

{The total luminosity of the WINGS galaxies is based on the distance of the cluster, but for the galaxies of the spectroscopic sample, which are based on the measured redshifts (\cite{Cava2009} and \cite{Morettietal2017}). The error on the spectroscopic measurements on z is also close to 20\%.}

The aperture corrected velocity dispersions $\sigma$ for 1,729 ETGs are those measured by the Sloan Digital Sky Survey (SDSS) and by the National Optical Astronomical Observatory (NOAO) survey.  New measured velocity dispersions were derived by \cite{Bettoni2016} for a few dwarf galaxies. These were obtained within the same aperture used by SDSS. The measured velocity dispersion spans a wide luminosity range, so that dwarf and giant objects are present in the sample, but the data sample is by no means complete. LTGs are almost absent. The huge difference in richness of the sub-sample mirrors the difficulty in measuring the velocity dispersion compared to surface brightness and half-light radius.

The stellar masses $M^*$ of the WINGS catalog were derived {by means of } the SED fitting of the WINGS spectra by \cite{Fritzetal2007}. For both these variables the error is expected around 20\%. In this paper however, in order to avoid systematic differences with MANGA, we preferred to use the same definition of stellar mass adopted by the authors of the MANGA catalog (see below).

The morphology of the WINGS galaxies was obtained with the software MORPHOT \cite{Fasanoetal2012}. MORPHOT exploits 21 morphological diagnostic parameters, directly and easily computable from the galaxy image, to provide two independent classifications, one based on a maximum likelihood (ML) semi-analytical technique and the other one on a neural network (NN) machine. A suitably selected sample of $\sim1000$ visually classified WINGS galaxies was used to calibrate the diagnostics for the ML estimator and as a training set for the NN machine. The final estimator of the morphological type $T$ combines the two techniques. 

{In summary, the WINGS samples used here are two. The first one includes all photometric measurements ($\sim 33,000$ objects) while
the second one includes the spectroscopic measurements ($\sim 1,800$ objects). The first sample is used to build the \IeRe, the \MRa, the \BmV\ and the \nL\ planes, the second one is used for the \Lsig\ plane. The distances of the galaxies of the first sample are assumed to be the same of the cluster center, while for the galaxies of the second sample the distances are those derived from the measured redshift. The photometric sample might includes $\sim 1\%$ of background galaxies (generally eliminated from the catalog being much smaller in angular size and redder in color).}

MANGA (Mapping Nearby Galaxies at APO) is one of the projects included in the fourth version of SDSS (SDSS-IV; \cite{Blantonetal2017}). The sample of galaxies in the web catalog contains $\sim 10,800$ and was selected from the NASA-Sloan Atlas (NSA) and were observed with an IFU spectrograph {(see, http://vizier.cds.unistra.fr/viz-bin/VizieR-2)}. This catalog contains several structural parameters of local galaxies (with $z\le0.1$), derived from the combination of UV (GALEX), optical (SDSS), and near-infrared (2MASS) images (see for details \cite{Sanchezetal2022}). The selected sample contains a representative population of galaxies of the nearby Universe. The sample includes $\sim60$\% of the objects selected so that the fields of view (FOVs) of the adopted integral field units (IFUs) cover at least a 1.5 galaxy effective radius (\re), a second sample, with $\sim30$\% of the objects, such that the FOVs of the IFUs cover at least 2.5 \re; and a third sample, comprising $\sim10$\% of the { galaxies and aiming} to observe galaxies in the so-called the green valley.

MANGA employs dithered observations with 17 fiber-bundle integral field units that vary in diameter from 12" (19 fibers) to 32" (127 fibers). Two dual-channel spectrographs provided simultaneous wavelength coverage over 3600-10300 \AA\ at $R\sim2000$. With a typical integration time of 3 hr, MANGA reached a target r-band S/N ratio of 4-8 at 23 AB mag arcsec$^{-2}$. The selected targets have masses $M^* \ge 10^9 M_\odot$ using SDSS-I redshifts and i-band luminosity to achieve uniform radial coverage in terms of the effective radius, an approximately flat distribution in stellar mass, and a sample spanning a wide range of environments.

The large MANGA sample aims to study rare populations such as mergers, AGNs, post-starbursts, galaxies with strong outflows, etc., especially new classes of phenomena that can only be detected with resolved spectroscopy. 

From the MANGA catalog \citep{Sanchezetal2022} we extracted the absolute V-band luminosity $M_V$, the effective radius \re, the central velocity dispersion $\sigma$, the stellar masses $M^*$, the S\'ersic index $n$, the morphological types $T$ and the $B-V$ colors.

The MANGA morphological types are not highly accurate. The galaxy classification is good only in terms of statistical properties. The classification started from the accurate visual morphology determinations of SDSS galaxies by \cite{Vazquez-Mataetal2022}, which includes $\sim6000$ galaxies. This sample was used to train and test a machine-learning algorithm and to classify the rest of the galaxies. The parameters used to test the morphology are the S\'ersic index, the stellar mass, the line-of-sight velocity-to-velocity dispersion ratio at the effective radius, the ellipticity, the concentration index, and the colors. A Gradient Tree Boosting algorithm was used to assign the final morphology. To each morphological type a probability index P(MORPH) was assigned. This gives the probability that {a galaxy belongs to one of the  13 MORPH groups} (cD, E, S0, Sa, Sab, Sb, Sbc, Sc, Scd, Sd, Sdm, Sm, and Irr). A final parameter $T$ running from $-2$ for cD to 10 for Irr galaxies gives the final classification.

The galaxy’s V-band surface brightness at its effective radii R50 and \re, encircling half of the light within the FOV and half of the total integrated light of the galaxy, respectively, was obtained by averaging the flux values in elliptical rings (using the known position angle and ellipticity of the object) of width 0.15 (in units of \re\ or R50). The final effective radius \re\ used here takes into account the $b/a$ ratio to obtain a circularized value. The effects of the PSF are probably taken into account in the fit of the major and minor axis made with a S\'ersic law, but the catalog and the associated paper do not explain this clearly. 

MANGA gives the broadband photometry in the Gunn u, g, r, and i and the Johnson B, V, and R filters, adopting the AB photometric system redshifted to the rest frame of each object. The k-correction was not considered and the $B-V$ color was not referred to the whole extension of the galaxies. The absolute magnitudes used the redshift to estimate the cosmological distance. The catalog provides several data, most of them of interest for the spectroscopic analysis. 

WINGS and MANGA are two somewhat complementary samples. WINGS contains galaxies in clusters, while MANGA is dominated by galaxies in the field. {WINGS contains ETG preferentially while  MANGA is dominated by LTGs.}
Here we choose to compare galaxies of different morphological types in the same range of redshift ($0.04\le z \le 0.07$). From the above description it is clear that the two samples are not statistically complete (for example they are not volume limited and the range of luminosity considered is not the same). In addition to it,  the adopted techniques for the data analysis used to derive the galaxy parameters are quite different. With these premises, on the basis of the above discussion in the introduction, we expect to observe quite large differences in the galaxy ScRs. {In this way one can better appreciate which effects play the strongest role in altering the 2D distribution of galaxies in the ScRs and understand which preliminary analyses are required for a correct statistical comparison of two data samples.}

Since our aim here is to highlight the effects of the non univocal definition and determination of the structural parameters, we do not attempt any correction to the data. The observed differences will therefore reflect the effects mentioned above and will stress the need to work with standard procedures of data analysis. This is necessary if we want to make a good statistical work. We just made sure that both surveys used the same cosmological framework to obtain the distance dependent parameters.

\section{The range of galaxy parameters}\label{sec:3}
\begin{figure}[]
\includegraphics[width=12.0 cm]{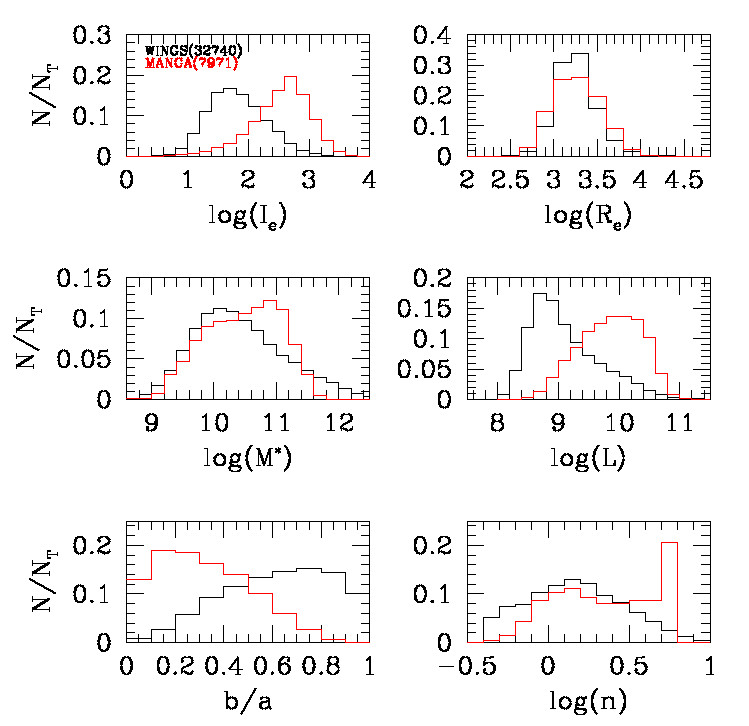}
\caption{Histograms of the measured parameters for the WINSG and MANGA samples. Black histogram: WINGS. Red histogram: MANGA.\label{fig1}}
\end{figure}

In order to quantify the differences between the properties of the two samples we start showing Figure \ref{fig1}. The figure presents the direct comparison of the parameter space for six variables: \Ie, \re, $M^*$, $L_V$, $b/a$ and $n$. In both samples the stellar mass $M^*$ was derived through the equation: $log(M^*/M_\odot) = -0.95 + 1.58(B - V) + 0.43(4.82 - M_V)$ adopted by the MANGA authors. {Some things are noteworthy}: i) The circularized radii \re\ {seem} to cover the same range in pc units; ii) MANGA contains more objects of high luminosity $L_V$ and mass $M^*$ with respect to WINGS ($L_{peak}\sim10^{10}$ and $M^*_{peak}\sim10^{11}$). It follows that the effective surface brightness \Ie, calculated from $L$ and \re, is systematically higher in MANGA than WINGS. Note, in addition, that the lower MANGA luminosity coincides with the peak of the WINGS distribution, so that the different distribution of galaxies in each range of luminosity will be the primary reason for the very different distributions of galaxies in many ScRs; iii) The peak of the $b/a$ ratio is very different for the two samples: MANGA contains more flattened objects ($b/a_{peak}\sim0.2$); iv) The S\'ersic index $n$ saturates for the MANGA sample at $\log(n)\sim0.8$.

\begin{figure}[]
\includegraphics[width=12.0 cm]{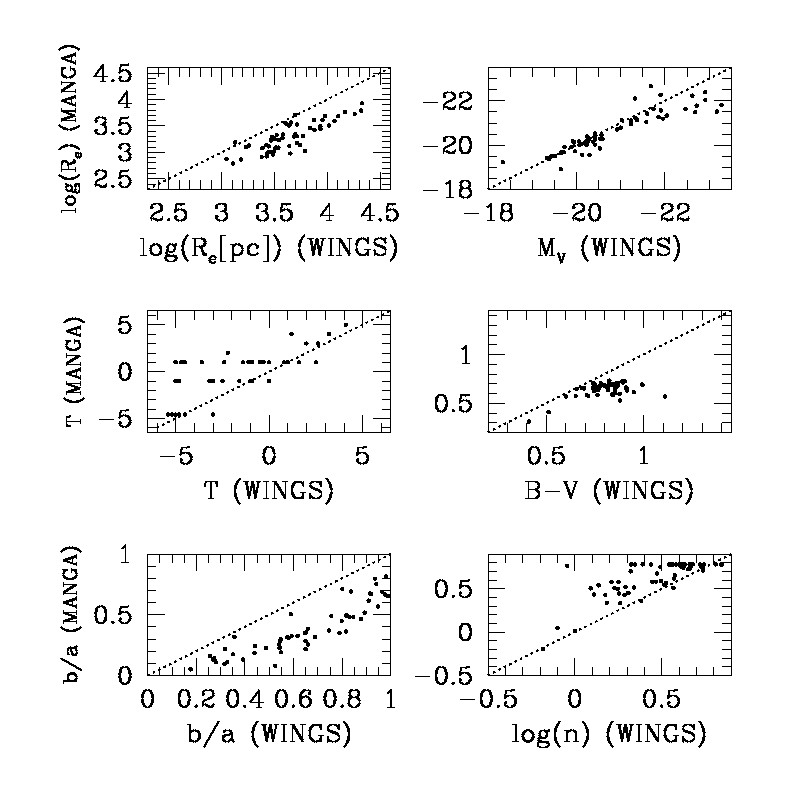}
\caption{Comparison of the measured parameters for the 58 galaxies in common between WINGS and MANGA.\label{fig2}}
\end{figure}

Unfortunately, these two  datasets have only 58 galaxies in common. Figure \ref{fig2} shows the direct comparison of \re, $M_V$, the morphological type $T$, the $B-V$ color, $b/a$ and $\log(n)$.
The comparison indicates that some discrepancies are present in all parameters. i) The WINGS galaxies {have  systematically larger  \re\ with respect to MANGA; ii) A small systematic difference in color exists and the WINGS objects are a bit redder; iii) The morphology of some objects is different. Some  ETGs in WINGS are classified as LTGs in MANGA. iv). The scatter in $M_V$ becomes larger  as the luminosity increases. v) The WINGS objects are systematically rounder than those in MANGA. 6. The S\'ersic index is systematically larger in MANGA and saturates toward $n\sim6$. }                                       

The origin of such different measurements for the same galaxies is difficult to establish. The larger \re\ of WINGS is not surprising. The accuracy with which the sky background was removed in these data likely determines systematically {larger} radii. The different morphology is probably due to the different diagnostic parameters used for the classification or the training algorithm of the automatic method. The different $B-V$ color might {depend on the area of the galaxy in which the color is measured and on the  adopted corrections for } galactic extinction and k-correction.

It is important to keep in mind these differences when discussing the ScRs. The intrinsic differences in the measured parameters and the different statistical properties of the two data samples are responsible of the different ScRs we observe.

\begin{figure}[]
\includegraphics[width=12.0 cm]{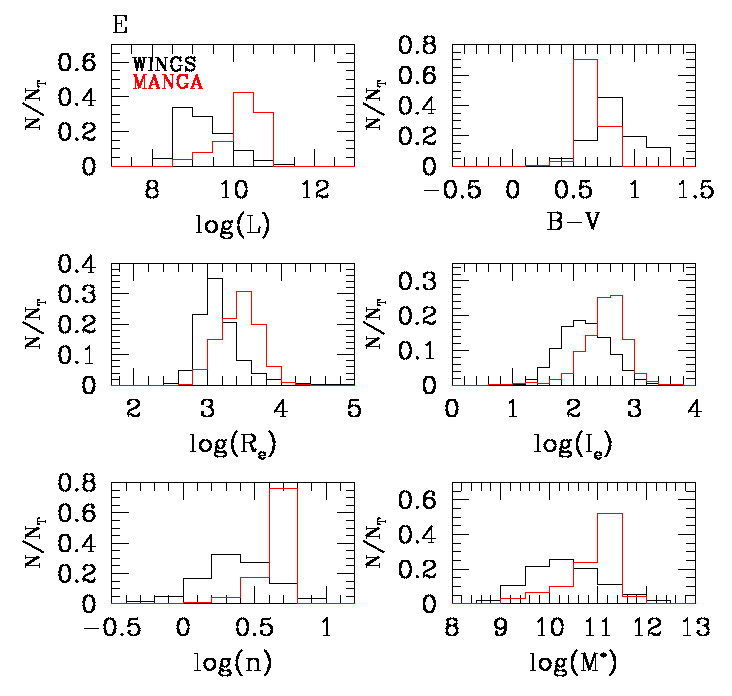}
\caption{Histograms of the measured parameters for E galaxies in the WINSG and MANGA samples. Black histogram: WINGS. Red histogram: MANGA.\label{fig3}}
\end{figure}

\begin{figure}[]
\includegraphics[width=12.0 cm]{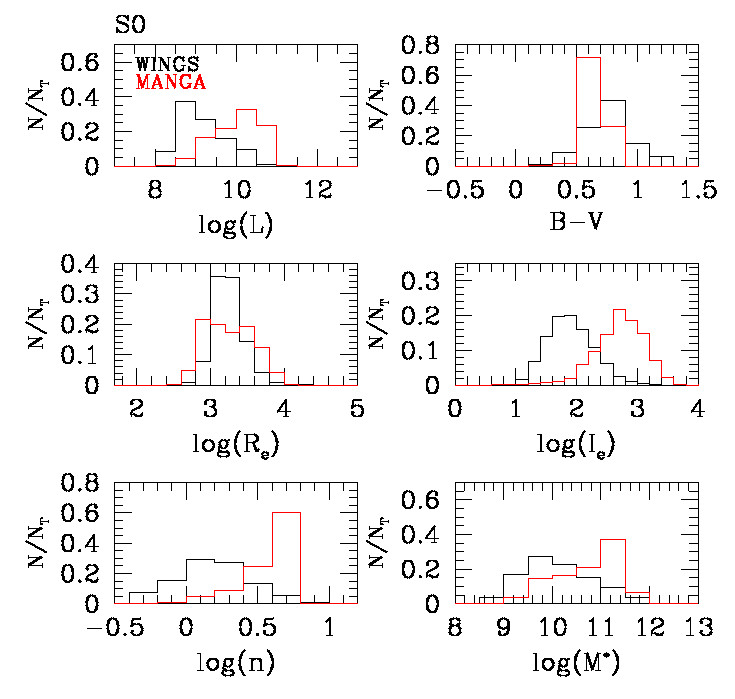}
\caption{Histograms of the measured parameters for S0 galaxies in the WINSG and MANGA samples. Black histogram: WINGS. Red histogram: MANGA.\label{fig4}}
\end{figure}

\begin{figure}[]
\includegraphics[width=12.0 cm]{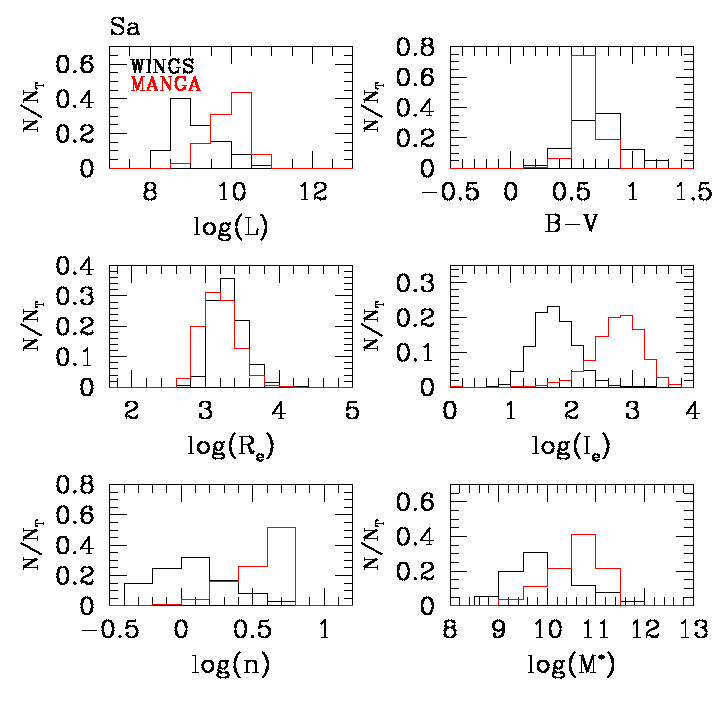}
\caption{Histograms of the measured parameters for Sa galaxies in the WINSG and MANGA samples. Black histogram: WINGS. Red histogram: MANGA.\label{fig5}}
\end{figure}

\begin{figure}[]
\includegraphics[width=12.0 cm]{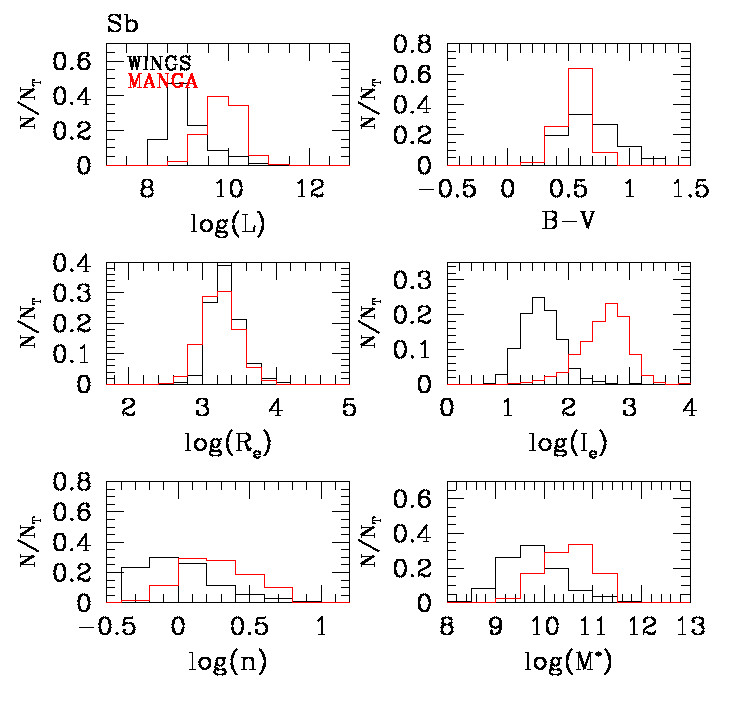}
\caption{Histograms of the measured parameters for Sb galaxies in the WINSG and MANGA samples. Black histogram: WINGS. Red histogram: MANGA.\label{fig6}}
\end{figure}

\begin{figure}[]
\includegraphics[width=12.0 cm]{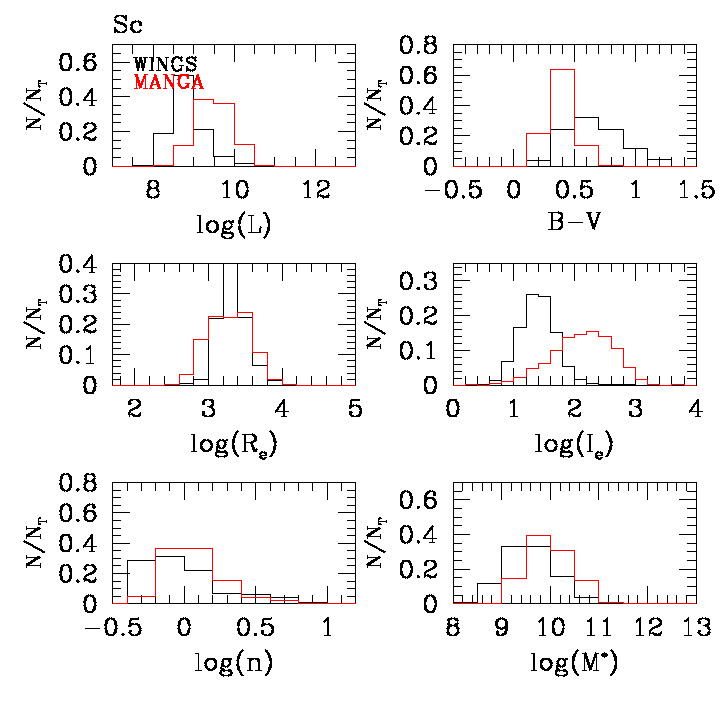}
\caption{Histograms of the measured parameters for Sc galaxies in the WINSG and MANGA samples. Black histogram: WINGS. Red histogram: MANGA.\label{fig7}}
\end{figure}

\begin{figure}[]
\includegraphics[width=12.0 cm]{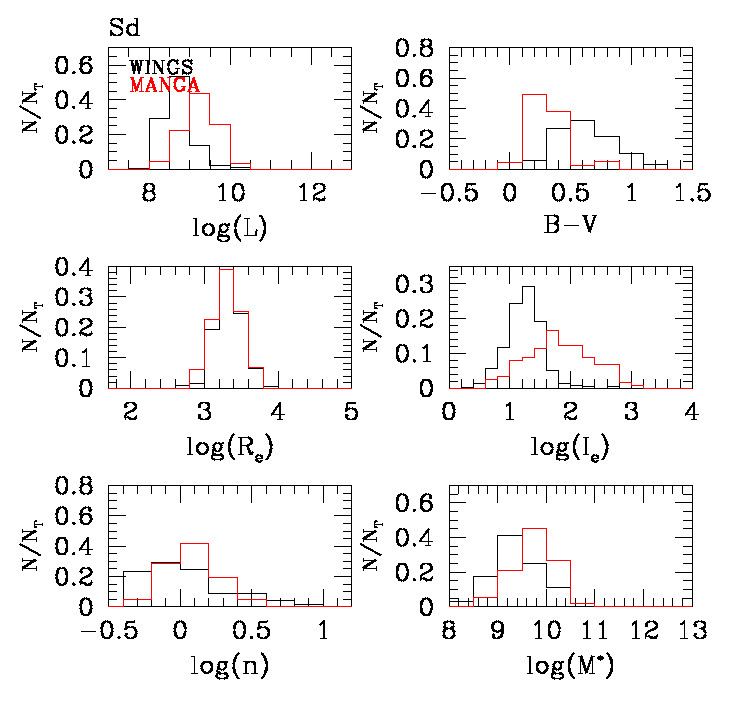}
\caption{Histograms of the measured parameters for Sd galaxies in the WINSG and MANGA samples. Black histogram: WINGS. Red histogram: MANGA.\label{fig8}}
\end{figure}

{In order to complete the analysis, the series of figures, \ref{fig3}, \ref{fig4}, \ref{fig5}, \ref{fig6}, \ref{fig7} and \ref{fig8}, shows the differences in the basic parameters of galaxies with the same morphological type. The whole WINGS and MANGA catalogs are examined with this criterion.  }

{Fig. \ref{fig3} presents the histograms of the distributions of $L$, $B-V$, \re, \Ie, $n$ and $M^*$ for the E galaxies. It is apparent that the MANGA sample contains only big ellipticals of high luminosity. Their color is systematically bluer than that of WINGS. The MANGA radii are now a bit larger,  the S\'ersic index peaks at $n\sim4$. and the stellar mass peaks around $10^{11}$ solar masses. }

Fig. \ref{fig4} compares the properties of galaxies classified as S0s. Again the MANGA galaxies are brighter. The effective radius is more or less the same. The effective surface brightness is systematically higher as well as the S\'ersic index. The masses peak at a larger value.

For Sa galaxies (Fig. \ref{fig5}) the situation is similar. The MANGA objects are systematically brighter and have larger S\'ersic index. The effective radius appears similar, so the final effective surface brightness appears systematically different. The $B-V$ color is more peaked in MANGA than in WINGS.

{For Sb galaxies,  the same considerations apply}. The MANGA luminosity is higher, the color is more peaked, the radius is similar, and the S\'ersic index is a bit shifted toward larger values (Fig. \ref{fig6}). \Ie\ is systematically different.
In the case of Sc and Sd galaxies we see more or less similar S\'ersic index, but the color is systematically bluer for MANGA objects (Fig. \ref{fig7} and \ref{fig8}). The MANGA galaxies are systematically brighter, so \Ie\ is quite different.

{To summarize, the histograms of the structural parameters for the various morphological types suggest  that the main difference } between the two samples is in the total number of galaxies in each interval of luminosity and mass. The radii are quite similar, but the different luminosity determines systematically different brightness. The color and the S\'ersic index present small differences. Probably the absence of k-correction and the different method used to derive colors and S\'ersic index $n$ are part of the reasons for the observed differences.

{A preliminary conclusion is evident. We are not dealing with observational ScRs that can be considered strictly equivalent from a statistical point of view. What is available are two sources of data that lead to similar ScRs which qualitatively look  identical but actually are not. The question soon arises: can we remove those differences by improving the data acquisition methods or are they telling that two populations of galaxies with different properties and histories are present? 
To cast light on this issue, we first examine one by one a few key ScRs. }

\section{The \IeRe\ plane}\label{sec:4}
The Kormendy relation (KR; \citep{Kormendy1977}) links the effective radius of ETGs to their mean surface brightness within \re. The relation is typically written as:
$<\mu>_e=\alpha+\beta\log(R_e)$ (when $<\mu>_e$ in mag arcsec$^{-2}$ is used instead of \Ie\ given in $L_\odot pc^{-2}$) and shows that larger elliptical galaxies have lower average surface brightness, reflecting systematic structural differences among spheroids. The KR is essentially a 2D projection of the Fundamental Plane of ETGs.

{The  KR has been the subject of many studies, among which we recall: i) To trace the structural evolution of galaxies. By comparing slopes and intercepts at different redshifts, one can tests whether ETGs grow in physical size or simply fade due to passive stellar evolution; ii) To examine galaxy assembly mechanisms. Different galaxy formation scenarios (dissipative collapse vs. dry mergers) predict different KR behaviors. The goal was to constrain formation pathways and merger history of spheroids; iii) To differentiate bulge types. Classical bulges follow the KR of elliptical galaxies, while pseudo-bulges deviate from it; iv) To study  the environmental effects. By comparing the KR in clusters, groups, and fields environments one can test whether environment drives structural evolution. The idea was to determine whether dense environments produce more compact or different ETGs;  v) To connect the KR  to the Fundamental Plane. The KR can be seen as a simpler projection of the FP (the relations between \re, \Ie\ and $\sigma$). By studying deviations from the KR helps identify non-homology (changes in S\'ersic index, structural variance). The aim was to understand deviations from virial expectations and stellar population effects. }

The analysis of the KR requires: { i) to consider the Cosmological surface-brightness dimming {and the k-corrections} (when used at high redshift $(1+z)^4$); ii) to apply a good sky subtraction and PSF correction (especially at high redshift); iii) to understand the selection effects (e.g., magnitude-limited samples); iv) to use the profile fitting method (S\'ersic vs. de Vaucouleurs) when fitting the light profiles and in general an univocal definition of \re. }

Over the years, the  KR has been the subject of many studies. To mention a few, we recall: \cite{HamabeKormendy1987} extended the KR, comparing elliptical galaxies and bulges; \cite{Capacciolietal1992} discovered that the \IeRe\ plane contains two different families of galaxies: the ordinary and the bright galaxies, that have quite different distribution in this parameter space and therefore quite different evolution; \cite{Ziegleretal1999} studied the KR at intermediate redshift, an early study of the evolutionary effects; \cite{Waddingtonetal2002} addressed the KR at high redshift, trying to determine passive evolution vs size growth; \cite{LaBarberaetal2003} analyzed environmental effects (cluster vs field ETGs); \cite{Nigoche-Netro2008} showed that the KR slope depends on the sample selection and magnitude range.

\begin{figure}[]
\begin{adjustwidth}{-\extralength}{-3cm}
\centering
\includegraphics[width=8.0cm]{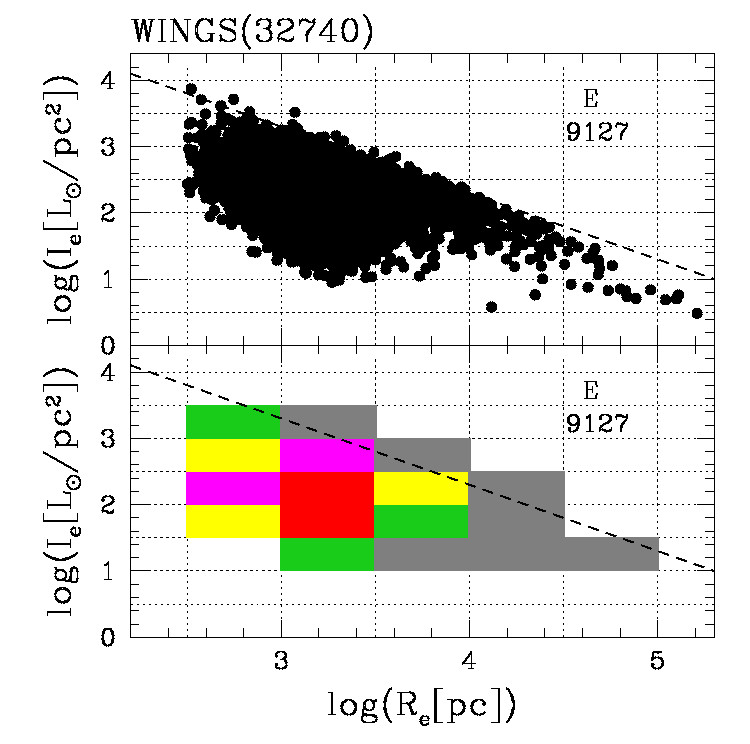}
\includegraphics[width=8.0cm]{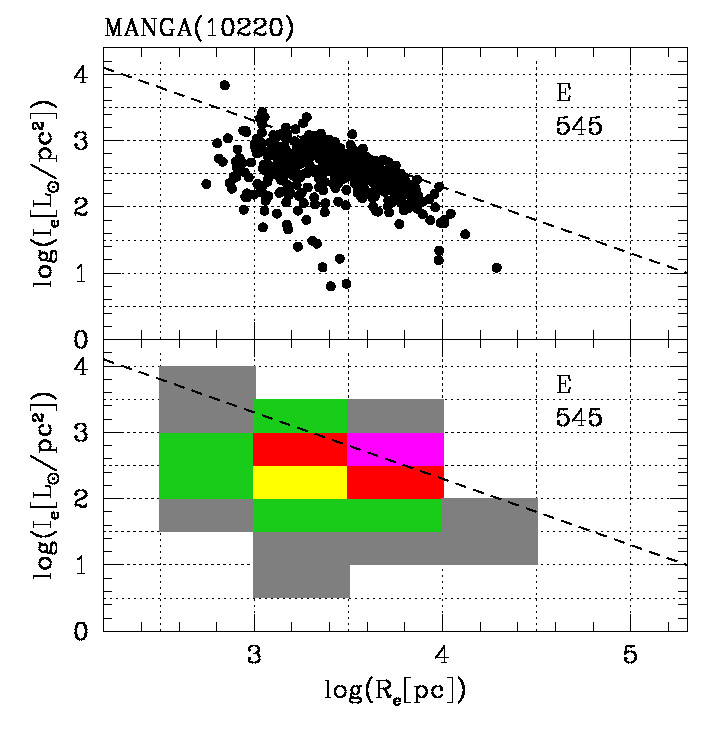}\\
\includegraphics[width=8.0cm]{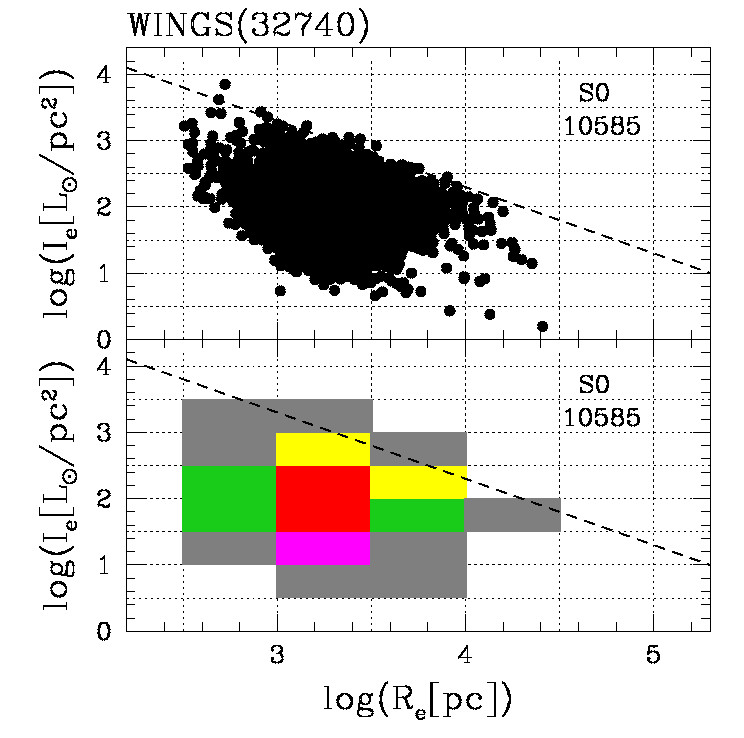}
\includegraphics[width=8.0cm]{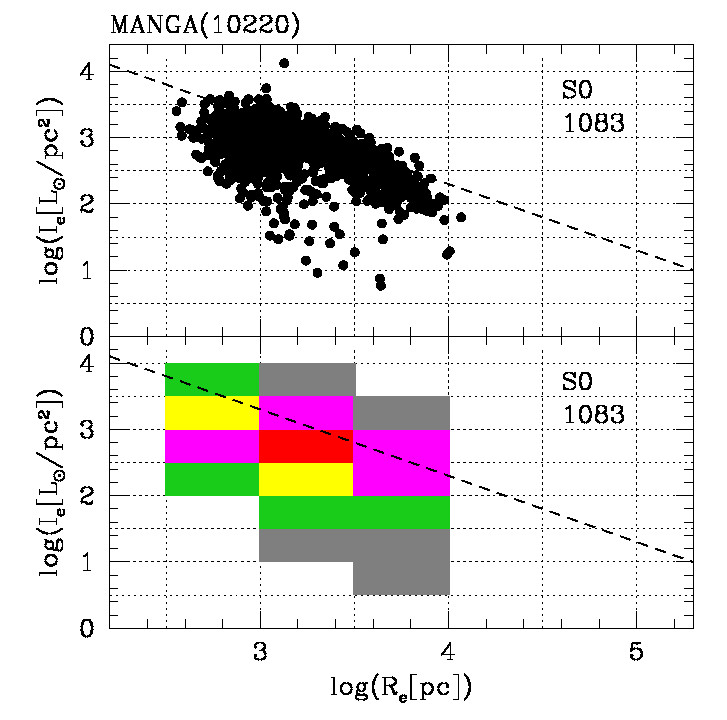}
\end{adjustwidth}
\caption{The \IeRe\ plane for E (top panels) and S0 (bottom panels) galaxies: (\textbf{a}) The top panel shows with black dots the whole distribution of galaxies of the WINGS survey classified as E. The bottom panel shows with different colors the areas with the large number density of objects given in percent. Red corresponds to regions where more than 20\% of the galaxies are found. Magenta gives the interval 10\% - 20\%, yellow 5\% - 10\%, green 1\% - 5\%, and gray 0.1\% - 1\%. (\textbf{b}) The same as in panel a) for the E galaxies of the MANGA survey. (\textbf{c}) The same as in panel a) for S0 galaxies of WINGS. (\textbf{d}) The same as in panel a) for S0 in MANGA. On top of the figure, we indicate the size of the whole sample for both data-sets. Within the box, we provide the total number of galaxies of that morphological type. The dashed line marks the ZoE.\label{fig9}}
\end{figure}

{Here we try to analyze the number density, defined as the percentage of galaxies over the total number in the sample that is observed in each boxy region of the \IeRe\ plane\footnote{The plane is divided in a number of sub-areas $\Delta X$ vs $\Delta Y$ of suitable size}. We begin with the WINGS and MANGA data-sets. Fig. \ref{fig9} shows the distribution of E and S0 galaxies in this diagram. The WINGS sample contains 9127 E galaxies of this type, while MANGA only 545. The S0s galaxies are 10585 for WINGS and 1083 for MANGA. Keep in mind that we consider here only the galaxies of the MANGA sample with a quality index Q=0\footnote{The authors of the MANGA catalog provide the quality index Q to identify the galaxies that have the best measured parameters}, i.e. those with the best determination of the parameters. }

{Qualitatively one can note  the similarity between the two \IeRe\ distributions in both morphological types. However, the analysis of the number density of objects in each region of the diagram (see the lower color panels) indicate a quite different story.}

The peak of the E distribution (shown in the colored panels) in WINGS is between 3 and 3.4 in $\log(R_e)$ (1 and 2.5 kpc), with only less than 1\% very big Es with large radii. MANGA, on the other hand, suggests two peaks of galaxy concentration: one between 3.5 and 4 (3 and 10 kpc) and one at the same WINGS interval, but with higher surface brightness.
The tail of Es with very large radii (\re\ >10 kpc) is present only in the WINGS database. This is due to the special data analysis made by the WINGS team for the subtraction of the sky background (see \cite{Fasanoetal2006}). {The detection of very faint halos around the central big Es, provided the very large values for \re. MANGA could not apply the same approach. The main goal of this survey was the spectral analysis of nearby galaxies. Necessarily they choose luminous objects and considered only the inner bright parts of the galaxies. So from the very beginning, the two data-sets stand on a different selection criterion. This is an important point to keep in mind: only the surveys that are based on similar data and similar data analysis should be compared. }

For S0s WINGS indicates a distribution quite similar to that of Es, but with no objects of very large size. MANGA provides approximately the same trend, but their galaxies are a bit brighter in  surface brightness (see the lower colored panels). Now the two distributions stop approximately at the same effective radius.

In both surveys the idea of the two families proposed by \cite{Capacciolietal1992} seems confirmed. The bright galaxies follow the KR, while the fainter galaxies are distributed in a cloud with radii smaller than $\sim2$ kpc in a wide interval of \Ie. {The $\Lambda$-shape of the \IeRe\ plane should be seriously considered  when the KR is analyzed. The choice of the sample,} in particular when high redshift galaxies are used, is fundamental for a correct comparison with local galaxies. As soon as we go up in redshift we loose progressively the fainter galaxies that do not follow the KR.

The Kolmogorov-Smirnov test made with the two vectors that contain the percentage of galaxies in each boxy-area of the \IeRe\ plane (those marked in the figures by small dotted lines), gives a very low probability for the hypothesis that the two distributions {(both that of Es and that of S0s)} are drawn from the same population. In other words, despite the similarity, we cannot say from a statistical point of view that the two distributions are similar. We cannot therefore answer the question posed {in the introduction}: how many Es or S0s (in percent over the total in the local Universe) exist in each area of the \IeRe\ plane?

\begin{figure}[]
\begin{adjustwidth}{-\extralength}{-3cm}
\centering
\includegraphics[width=8.0cm]{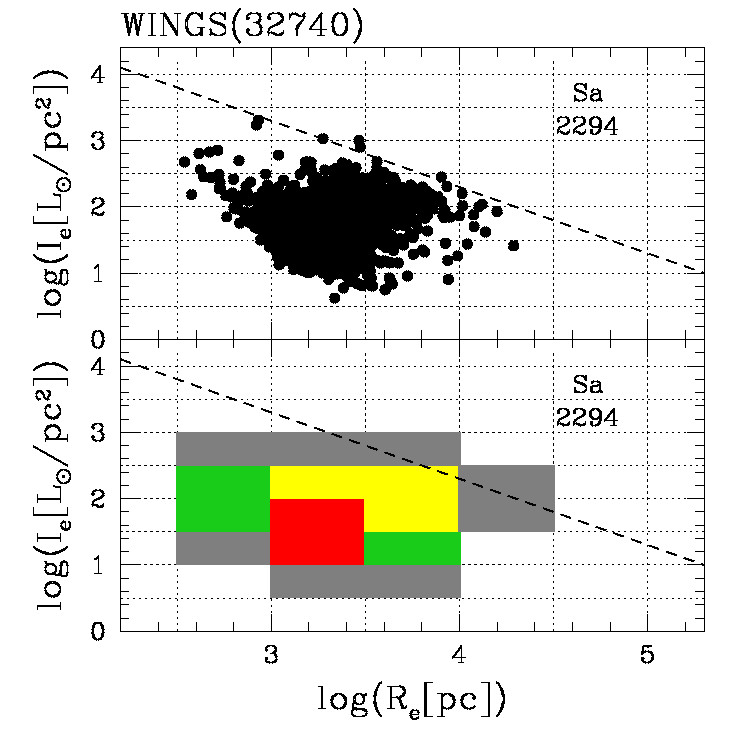}
\includegraphics[width=8.0cm]{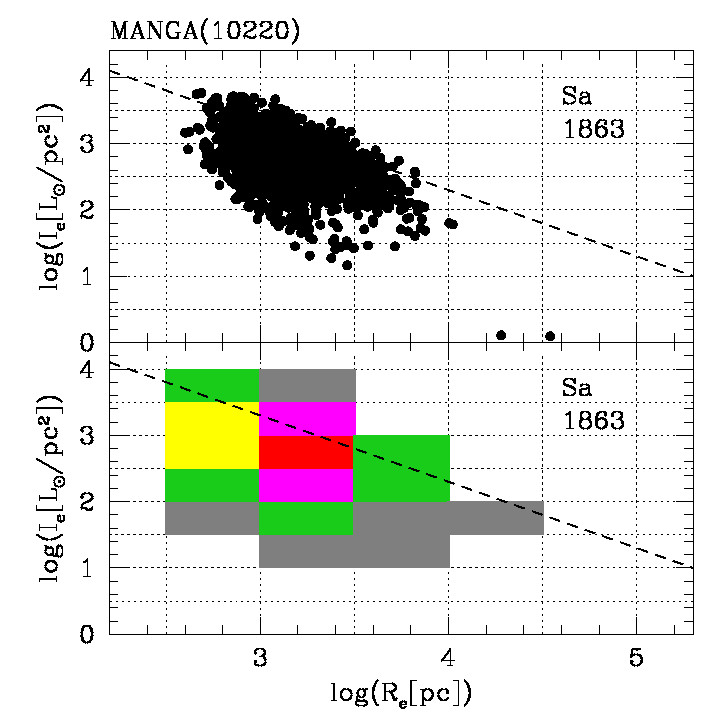}\\
\includegraphics[width=8.0cm]{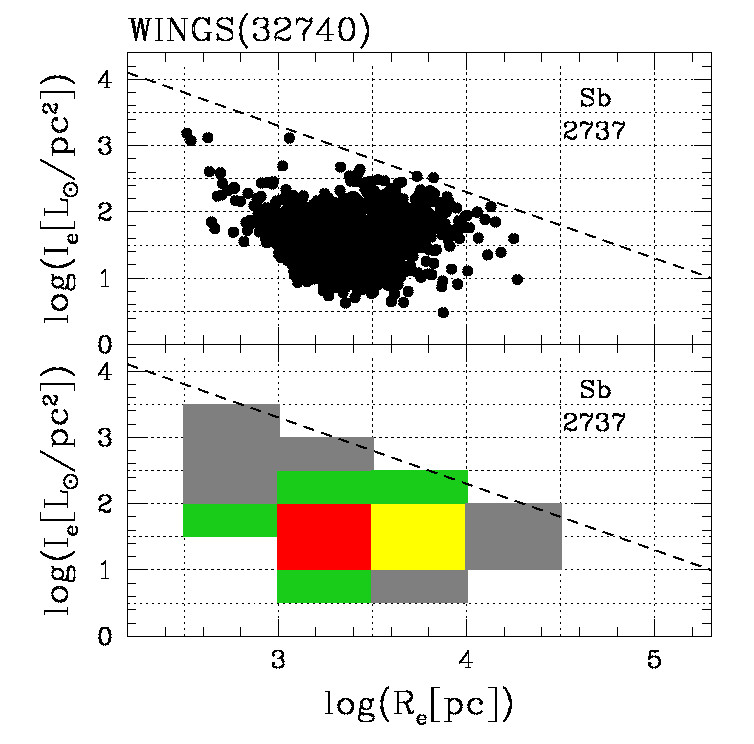}
\includegraphics[width=8.0cm]{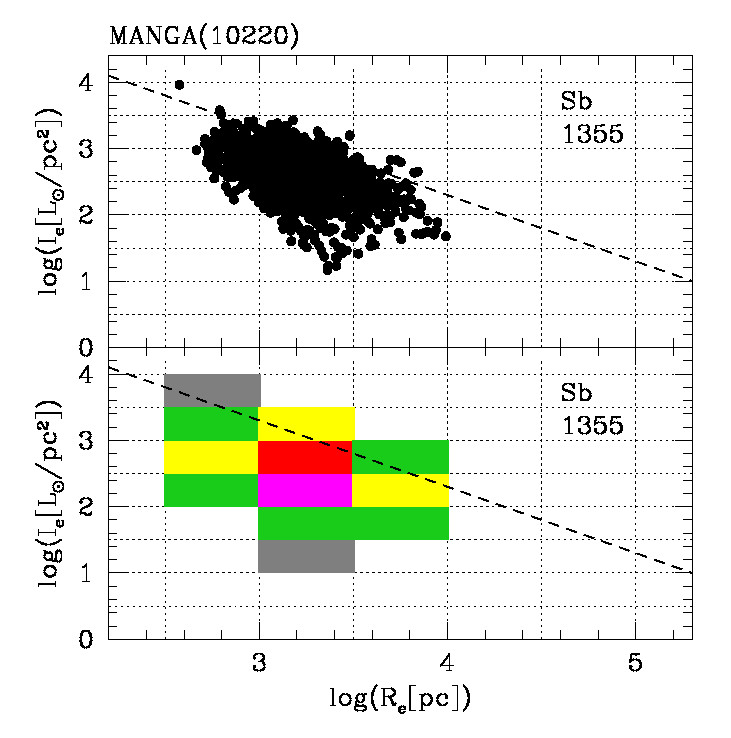}
\end{adjustwidth}
\caption{The \IeRe\ plane for Sa (top panels) and Sb (bottom panels) galaxies: (\textbf{a}) The top panel shows with black dots the whole distribution of galaxies of the WINGS survey classified as Sa. The bottom panel shows with different colors the areas with the large number density of objects given in percent. Red corresponds to regions where more than 20\% of the galaxies are found. Magenta gives the interval 10\% - 20\%, yellow 5\% - 10\%, green 1\% - 5\%, and gray 0.1\% - 1\%. (\textbf{b}) The same as in panel a) for the Sa galaxies of the MANGA survey. (\textbf{c}) The same as in panel a) for Sb galaxies of WINGS. (\textbf{d}) The same as in panel a) for Sb in MANGA. On top of the figure, we indicate the size of the whole sample for both data-sets. Within the box, we provide the total number of galaxies of that morphological type. The dashed line marks the ZoE.\label{fig10}}
\end{figure}

Figure \ref{fig10} shows the \IeRe\ for Sa and Sb galaxies. WINGS is on the left {panels and MANGA on the right panels}. Here we note that in WINGS spirals with small radii and high surface brightness are rare with respect to the MANGA sample. We believe that this is a real effect since most of WINGS spirals are members of clusters. {The environment stops the star formation activity by sweeping the gas content out}. It follows that in clusters spirals with high surface brightness are more difficult to observe. As before the location of the higher density peaks of the 2D distributions is quite different. MANGA has always a higher surface brightness with respect to WINGS. 
This is another effect to keep in mind when comparing the ScRs. { Environment can play a role. }

\begin{figure}[]
\begin{adjustwidth}{-\extralength}{-3cm}
\centering
\includegraphics[width=8.0cm]{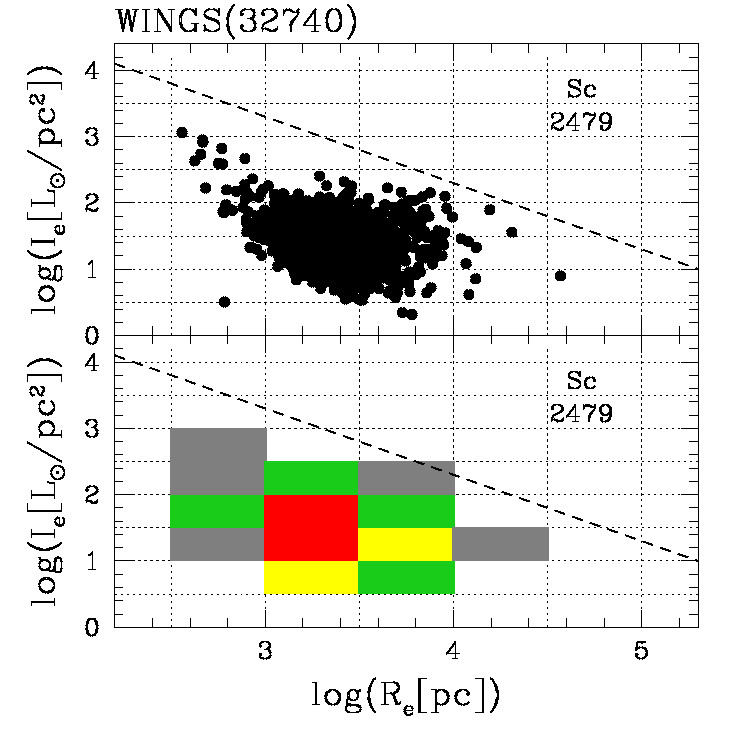}
\includegraphics[width=8.0cm]{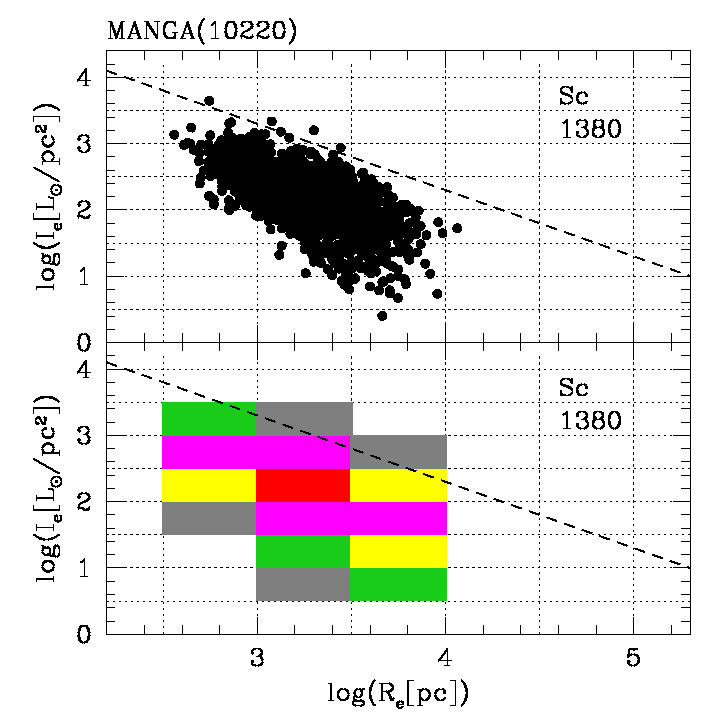}\\
\includegraphics[width=8.0cm]{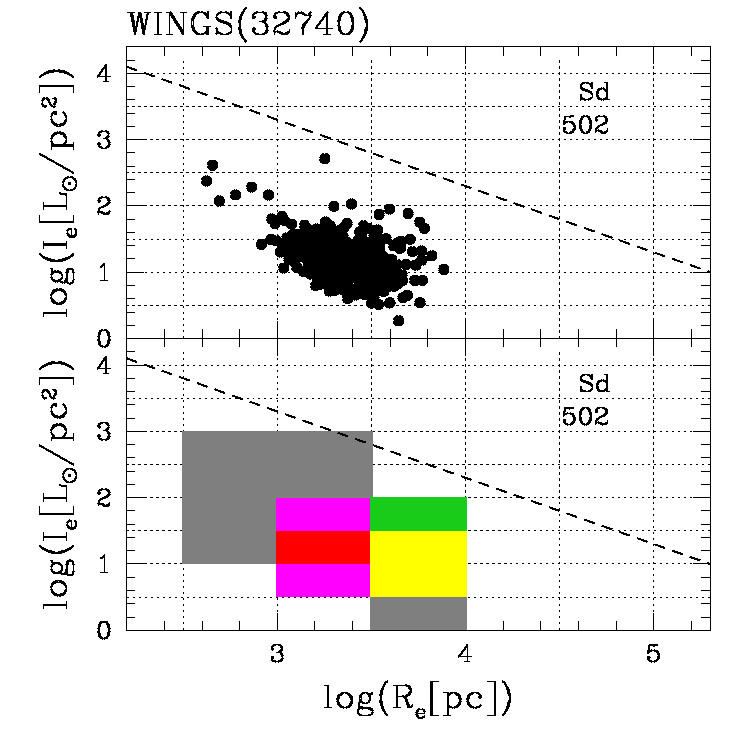}
\includegraphics[width=8.0cm]{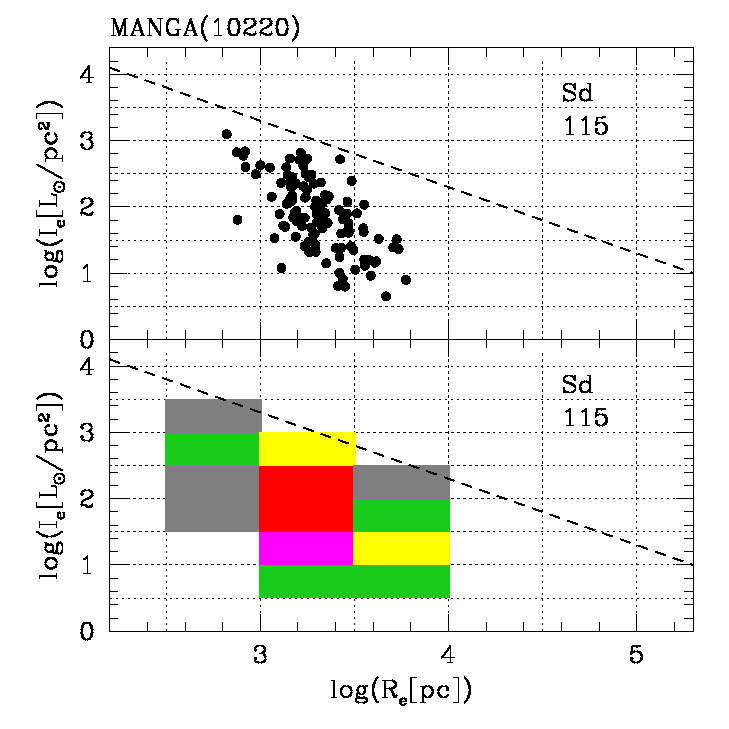}
\end{adjustwidth}
\caption{The \IeRe\ plane for Sc (top panels) and Sd (bottom panels) galaxies: (\textbf{a}) The top panel shows with black dots the whole distribution of galaxies of the WINGS survey classified as Sc. The bottom panel shows with different colors the areas with the large number density of objects given in percent. Red corresponds to regions where more than 20\% of the galaxies are found. Magenta gives the interval 10\% - 20\%, yellow 5\% - 10\%, green 1\% - 5\%,  and gray 0.1\% - 1\%. (\textbf{b}) The same as in panel a) for the Sc galaxies of the MANGA survey. (\textbf{c}) The same as in panel a) for Sd galaxies of WINGS. (\textbf{d}) The same as in panel a) for Sd in MANGA. On top of the figure,  we indicate the size of the whole sample for both data-sets. Within the box, we provide the total number of galaxies of that morphological type. The dashed line marks the ZoE.\label{fig11}}
\end{figure}

Figure \ref{fig11} {shows} the distribution of Sc and Sd galaxies in the \IeRe\ plane. {The two samples appear now to distribute in a different way} even from a qualitative point of view. WINGS does not show the typical trend between \Ie\ and \re\ with slope around $-1$. {The galaxies are distributed in a cloud and no tail  along the ZoE is visible. The \re\ of the WINGS and MANGA  galaxies peak  at nearly equal values, but MANGA contains  objects with higher values of \Ie. }

\begin{figure}[]
\begin{adjustwidth}{-\extralength}{-3cm}
\centering
\includegraphics[width=8.0cm]{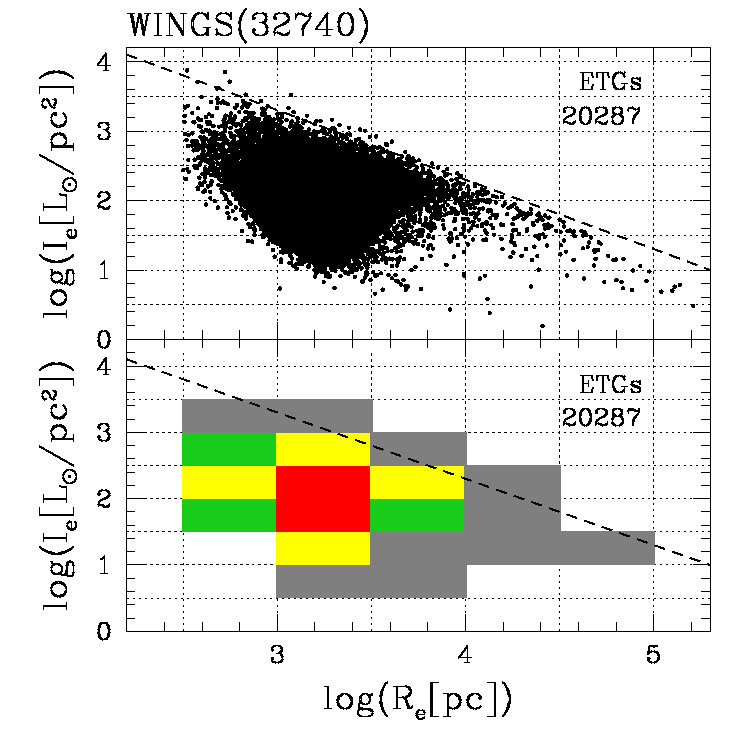}
\includegraphics[width=8.0cm]{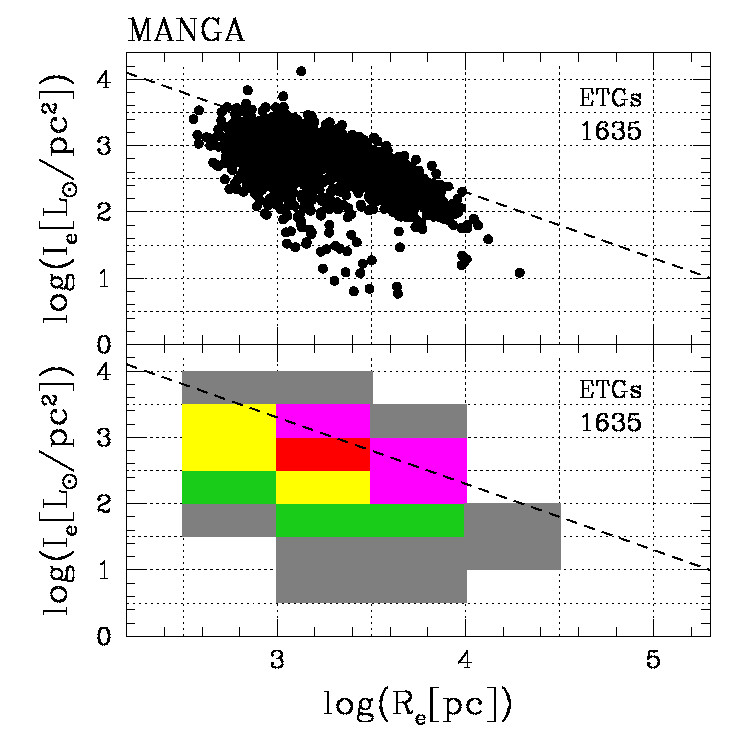}\\
\includegraphics[width=8.0cm]{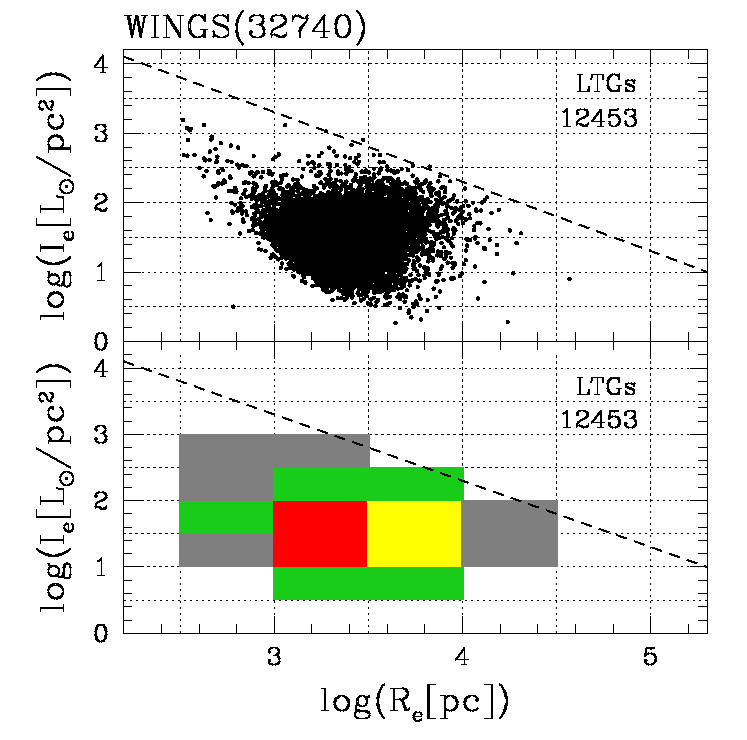}
\includegraphics[width=8.0cm]{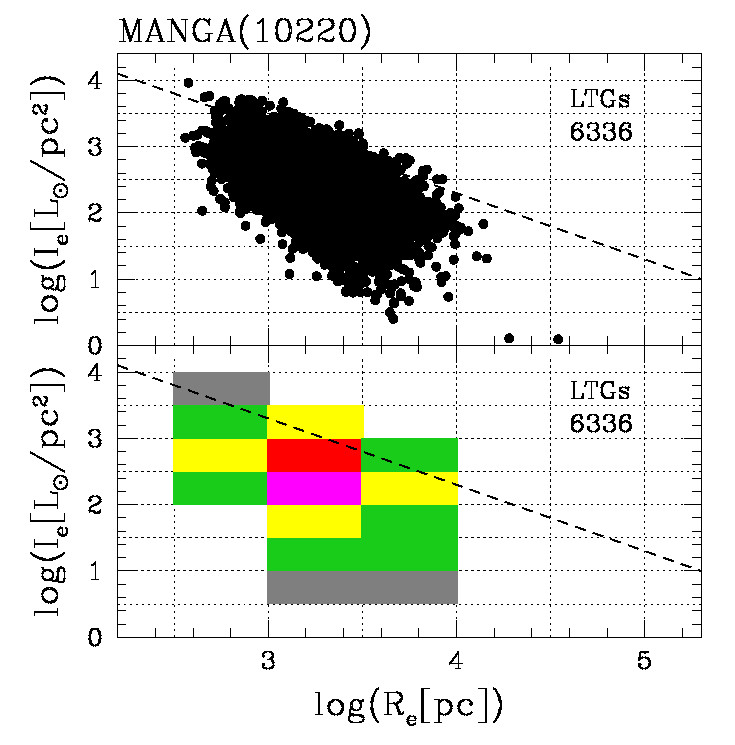}
\end{adjustwidth}
\caption{The \IeRe\ plane for ETGs (top panels) and LTGs (bottom panels) galaxies: (\textbf{a}) The top panel shows with black dots the whole distribution of galaxies of the WINGS survey classified as early-types. The bottom panel shows with different colors the areas with the large number density of objects given in percent. Red corresponds to regions where more than 20\% of the galaxies are found. Magenta gives the interval 10\% - 20\%, yellow 5\% - 10\%, green 1\% - 5\%,  and gray 0.1\% - 1\%. (\textbf{b}) The same as panel a) for the ETGs galaxies of the MANGA survey. (\textbf{c}) The same as in panel a) for LTGs galaxies of WINGS. (\textbf{d}) The same as in panel a) for LTGs in MANGA. On top of the figure, we indicate the size of the whole sample for both data-sets. Within the box, we provide the total number of galaxies of that morphological type. The dashed line marks the ZoE.\label{fig12}}
\end{figure}

Figure \ref{fig12} presents the global distribution in the \IeRe\ plane for ETGs and LTGs. WINGS contains very large galaxies with extended halos, high \re\ and low \Ie. These objects are absent in the MANGA sample. The LTGs in WINGS form a sort of cloud, and objects with high surface brightness are absent. MANGA, on the contrary, shows a correlation between the two variables extended to very large \Ie. { Even in this case, the statistical Kolmogorov-Smirnov test indicates a very low probability for the hypothesis that the galaxies have been extracted from the same population. }

\begin{table}[]
    \centering
    \begin{tabular}{|c|c|c|c|}
    \hline
    \multicolumn{4}{|c|}{E galaxies} \\
    \hline
    Lum. interval &  Weight. Energ. Dist. &  Nr. of Gal. &   Permutation p-val. \\
    \hline\hline
     7-13  & 0.327  & 2832-545 & 0.27 \\
     8-12  & 0.327  & 2846-545 & 0.27 \\
     9-11  & 0.190  & 1752-523 & 0.54 \\
10.0-10.5  & $2.4\times10^{-2}$  & 258-232 & 0.91 \\
    \hline
    
    \hline
    \multicolumn{4}{|c|}{S0 galaxies} \\
    \hline
    Lum. interval &  Weight. Energ. Dist. &  Nr. of Gal. &   Permutation p-val. \\
    \hline\hline
     7-13  & 0.795  & 3304-1083 & $9.0\times10^{-2}$ \\
     8-12  & 0.804  & 2800-1083 & $9.0\times10^{-2}$ \\
     9-11  & 0.568  & 1720-1027 & $9.0\times10^{-2}$ \\
10.0-10.5  & 0.327  & 309-354 & 0.54 \\
    \hline   

    \hline
    \multicolumn{4}{|c|}{Sa galaxies} \\
    \hline
    Lum. interval &  Weight. Energ. Dist. &  Nr. of Gal. &   Permutation p-val. \\
    \hline\hline
     7-13  & 1.269  & 721-1863 & $9.0\times10^{-2}$ \\
     8-12  & 1.255  & 713-1863 & $9.0\times10^{-2}$ \\
     9-11  & 1.065  & 349-1805 & $9.0\times10^{-2}$ \\
10.0-10.5  & 0.998  & 66-812 & 0.272 \\
    \hline   

    \hline
    \multicolumn{4}{|c|}{Sb galaxies} \\
    \hline
    Lum. interval &  Weight. Energ. Dist. &  Nr. of Gal. &   Permutation p-val. \\
    \hline\hline
     7-13  & 1.300  & 836-1355 & $9.0\times10^{-2}$ \\
     8-12  & 1.244  & 840-1355 & $9.0\times10^{-2}$ \\
     9-11  & 0.998  & 316-1322 & $9.0\times10^{-2}$ \\
10.0-10.5  & 0.846  & 37-465 & 0.363 \\
    \hline   

    \hline
    \multicolumn{4}{|c|}{Sc galaxies} \\
    \hline
    Lum. interval &  Weight. Energ. Dist. &  Nr. of Gal. &   Permutation p-val. \\
    \hline\hline
     7-13  & 0.700  & 741-1380 & $9.0\times10^{-2}$ \\
     8-12  & 0.687  & 730-1380 & $9.0\times10^{-2}$ \\
     9-11  & 0.505  & 222-1208 & $9.0\times10^{-2}$ \\
10.0-10.5  & 0.442  & 11-171 & 0.454 \\
    \hline   

    \hline
    \multicolumn{4}{|c|}{Sd galaxies} \\
    \hline
    Lum. interval &  Weight. Energ. Dist. &  Nr. of Gal. &   Permutation p-val. \\
    \hline\hline
     7-13  & 0.532  & 154-115 & $9.0\times10^{-2}$ \\
     8-12  & 0.471  & 151-115 & $9.0\times10^{-2}$ \\
     9-11  & 0.484  & 21-84 & 0.182 \\
10.0-10.5  & 1.843  & 1-4  & $9.0\times10^{-2}$ \\
    \hline   
    \end{tabular}
    \caption{Results of the Energy test. For each morphological type the table provides: the luminosity interval considered in the comparison, the value of the weighted energy distance, the number of WINGS and MANGA objects available for the calculation of the percentage of galaxies in each box of the \IeRe\ plane and the values {of the similarity probability for the two} distributions when 10 random permutations are applied to the data.}
    \label{tab:1}
\end{table}

{In order to better analyze the similarity of the 2D distributions of galaxies of different morphology in the \IeRe\ plane we decided to apply the Energy Distance statistical test \citep{SzekelyRizzo}.}
The test works with two {data sets (arrays), each containing the same number of points. The two arrays are named $X$ and $Y$. Each point of the array  corresponds to an object in the \IeRe\-plane and is fixed by the values of the x and y coordinates of the boxes drawn in the \IeRe\ planes and a z coordinate giving the percentage of galaxies calculated in each box (the number of galaxies in the box divided by the total number of galaxies and multiplied per 100). Each point in the array is indicated by the same symbol of the array but in lower-case, that is $x_i$ and $y_j$ where $i=1,.....n$, and $j=1,.....m$ with $n$ and $m$ the number of points in each array. Actually, $n$ and $m$ are equal being given by the total number of boxes. Finally, let us call indicate the coordinates of the generic points in the array as $ x_i^{(1)}, x_i^{(2)}, x_i^{(3)} $. With aid of the above notation we may write:}
\[
X = \{x_1, \ldots, x_n\}, \qquad 
Y = \{y_1, \ldots, y_m\} \subset \mathbb{R}^3 ,
\]
where each point is given by its coordinates
\[
x_i = (x_i^{(1)}, x_i^{(2)}, x_i^{(3)}), 
\qquad
y_j = (y_j^{(1)}, y_j^{(2)}, y_j^{(3)}).
\]
The distances between generic points $u$ and $v$ are measured using the Euclidean norm
\[
\|u - v\| 
= \sqrt{(u_1 - v_1)^2 + (u_2 - v_2)^2 + (u_3 - v_3)^2 } .
\]

The empirical energy distance between the two samples is given by:
\[
\mathcal{E}_{n,m}
=
\frac{2}{nm}
\sum_{i=1}^{n}\sum_{j=1}^{m} \|x_i - y_j\|
-
\frac{1}{n^{2}}
\sum_{i=1}^{n}\sum_{i'=1}^{n} \|x_i - x_{i'}\|
-
\frac{1}{m^{2}}
\sum_{j=1}^{m}\sum_{j'=1}^{m} \|y_j - y_{j'}\|.
\]
This quantity is non-negative and equals zero if and only if the distributions
of $X$ and $Y$ are identical (under mild regularity conditions).

For the two-sample energy test, the statistic is
\[
T_{n,m} = \frac{nm}{n+m} \, \mathcal{E}_{n,m},
\]
and its significance is assessed by comparing $T_{n,m}$ to its distribution under permutations of the pooled sample. The final p-value gives the probability of similarity.

Table \ref{tab:1} provides the results of the test for the galaxies of different morphology. Before running the test we applied a bootstrap technique to randomly select 10220 galaxies from the WINGS data-set, that contains 32740 objects, in order to match exactly the number of galaxies of the MANGA data-set. {In Table \ref{tab:1}, column (1) is the interval in luminosity considered for the comparison, column (2)  yields the value of the weighted energy distance, column (3) gives the number of galaxies of the two data-sets, and column (4) is the value of the probability of similarity when 10 permutation are allowed.} High values (close to 1) of the p-value indicates that the two samples are identical, low values (below 0.5) that the two distributions are different.
Values around to 0.5 indicate that the two distributions are marginally statistically consistent.

{Looking at the entries of Table \ref{tab:1} we note the following}: {i)} When the luminosity interval is {large}, the similarity between the two samples is very low, in particular for late type galaxies. The two data-sets appear similar only when the luminosity interval considered is very small; {ii)} the number of objects of a given morphology, extracted from the new WINGS sample with 10220 objects, is always very different from the MANGA objects of the same morphology. This means that one cannot reproduce the MANGA sample by simply extracting in a random way a similar number of galaxies from the WINGS data-base. There is an intrinsic difference between the two data-sets and the only possible explanation is that the two environments are very different. {To conclude, one cannot compare the ScRS of cluster and field environment in this simple way}. The comparative analysis of the ScRs from different environments requires a careful statistical pre-analysis.

\section{The \MRa\ plane}\label{sec:5}

The {  mass–radius} relation (\MRa) links the stellar mass $M^*$ of a galaxy to its effective radius \re.
It is often written as: $\log(R_e)=\alpha+\beta\log(M^*)$, with $M^*$ given in solar units and \re\ in kpc (or pc). 
The \MRa\ expresses how galaxies become physically larger as their stellar mass grows. In general the slope of the relation is different for ETGs and LTGs.

{The \MRa\ is central to galaxy formation and evolution research. It provides information about: i) The structural evolution of a galaxy. It tells how galaxies grow in size with time, and it provides clues on whether they grow inside-out (stars added to outskirts) or undergo size inflation due to mergers or experience stellar redistribution via feedback or disc instabilities; ii) It highlights the  formation channels for early vs late types. In LTGs disks  gas accretion and star formation dominate, while in ETGs and spheroids mergers and dynamical heating prevail.
The different slope and scatter of the relation encode that history; iii) In the $\Lambda$CDM large scale simulations and also in semi-analytic models of galaxy formation and evolution, the \MRa\ is a workbench for the adequacy of the underlying physical processes.
The relation can be used to constrain the feedback efficiency, verify merger rates and test the angular momentum evolution; iv) It provides a test of the environmental influence, that is by comparing  results for clusters, groups, and fields it can cast light whether dense environments favor compactification, quenching and tidal stripping; v) It helps clarify the connection with other scaling laws, such as the KR, the Fundamental Plane, the Tully–Fisher \& Faber–Jackson relations, and finally the mass–concentration and halo–stellar mass relations. All together, these relations map the baryonic and dark matter growth. }

{  Various studies} have addressed thw \MRa\ relation. Among them we remind \cite{Shenetal2003} who used the SDSS to establish a benchmark local \MRa\ distinguishing disk vs spheroid slopes. \cite{Trujilloetal2006,Trujilloetal2007} who studied the size evolution of massive galaxies to $z\sim2$. \cite{vanderweletal2014} that worked with CANDELS to address the mass-, redshift-, and type-dependent \MRa. \cite{Mosleh2013} that studied the growth of disk sizes over cosmic time,
\cite{Newmanetal2012} who addressed the merger-driven size growth of quiescent galaxies and \cite{HuertasCompanyetal2013} that analyzed the morphology-dependent size evolution. \cite{Langetal2015} using the GAMA survey tried to get a precise local \MRa. Finally, \cite{Chiosietal2020} attempted an explanation of the \MRa\ shape in a cosmological context.

\begin{figure}[]
\begin{adjustwidth}{-\extralength}{-3cm}
\centering
\includegraphics[width=8.0cm]{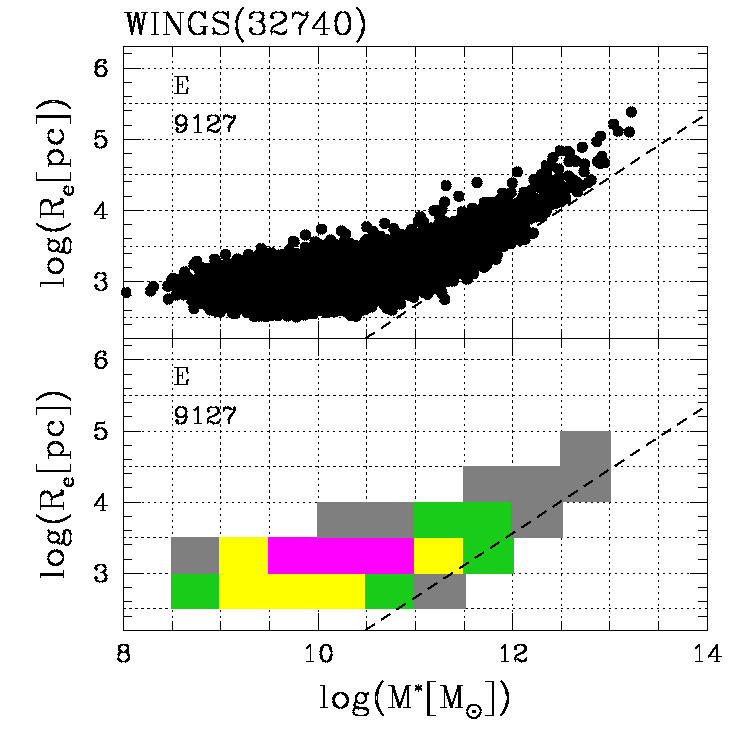}
\includegraphics[width=8.0cm]{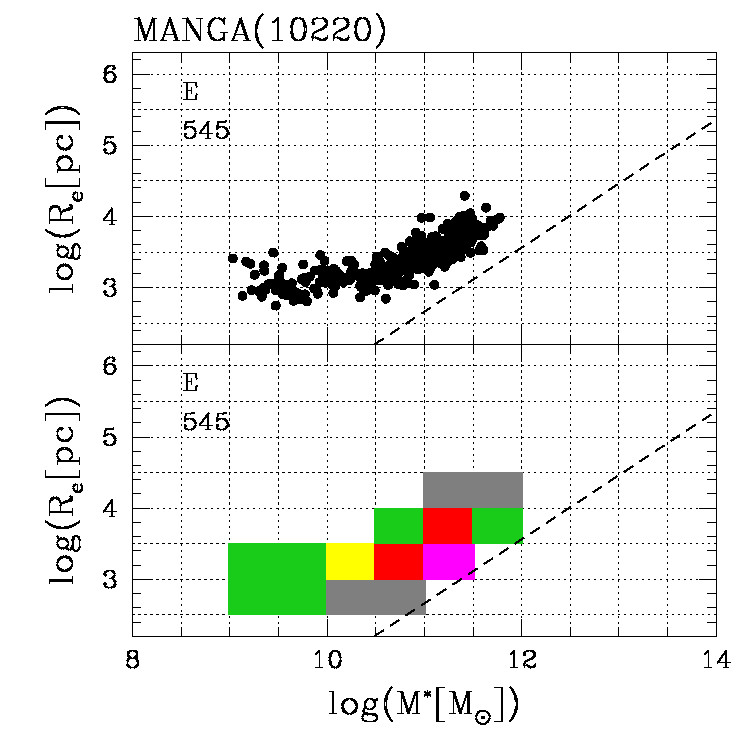}\\
\includegraphics[width=8.0cm]{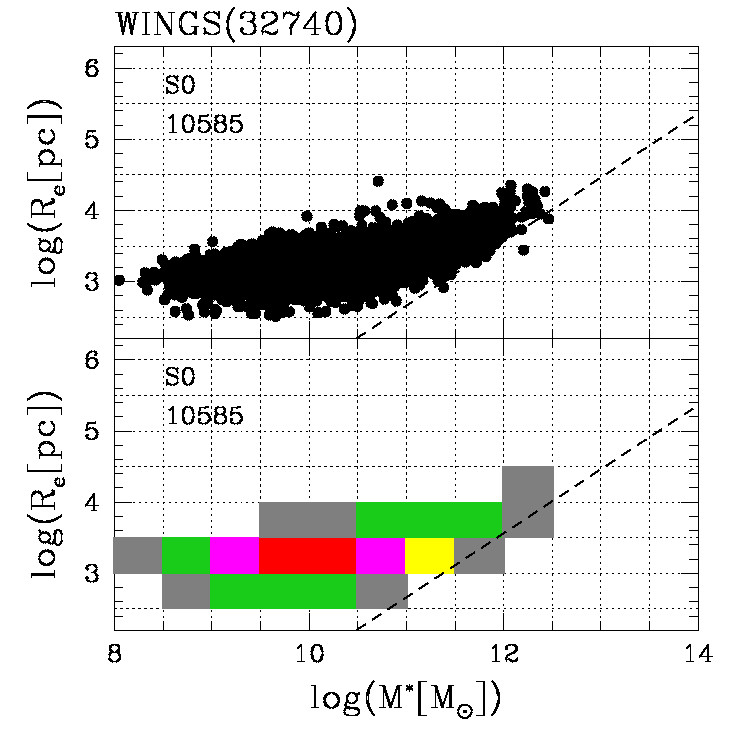}
\includegraphics[width=8.0cm]{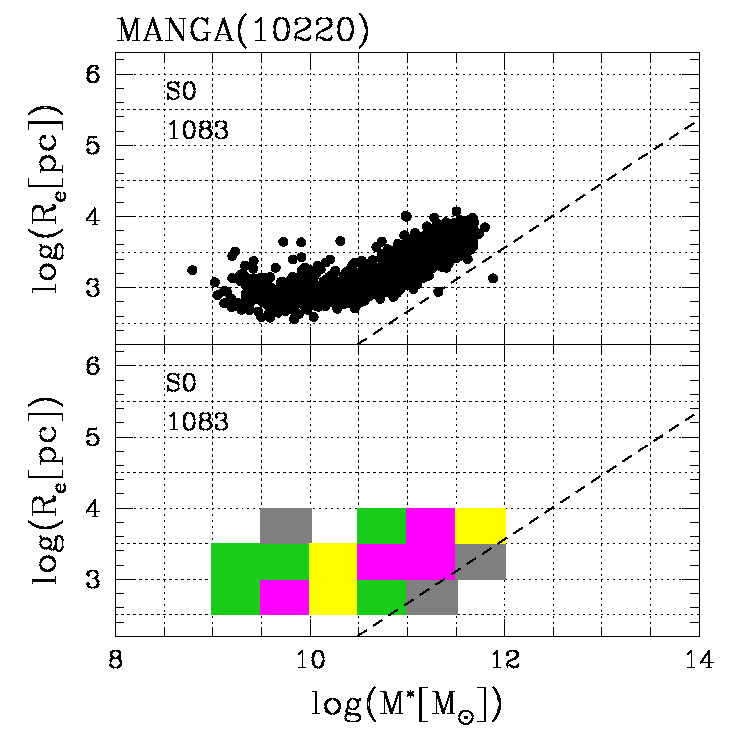}
\end{adjustwidth}
\caption{The \MRa\ plane for E (top panels) and S0 (bottom panels) galaxies: (\textbf{a}) The top panel shows with black dots the whole distribution of galaxies of the WINGS survey classified as E. The bottom panel shows with different colors the areas with the large number density of objects given in percent. Red corresponds to regions where more than 20\% of the galaxies are found. Magenta gives the interval 10\% - 20\%, yellow 5\% - 10\%, green 1\% - 5\%, and gray 0.1\% - 1\%. (\textbf{b}) The same as in panel a) for the E galaxies of the MANGA survey. (\textbf{c}) The ame as in panel a) for S0 galaxies of WINGS. (\textbf{d}) The same as in panel a) for S0 in MANGA. On top of the figure, we indicate the size of the whole sample for both data-sets. Within the box, we provide the total number of galaxies of that morphological type. The dashed line marks the ZoE.\label{fig13}}
\end{figure}

{In the following  we analyze} the \MRa\ plane with the same perspective used before for the \IeRe\ plane.  The different size of the two samples of E and S0 galaxies is clearly visible in Fig. \ref{fig13}. WINGS contains very massive and large Es. Both samples show a curved distribution in this plane. {The nearly horizontal trend at low mass hand starts to steep at masses around $3\times10^{10} M_\odot$.} The 2D distributions visible in the bottom color panels are not similar. The peaks of the distributions are different (higher $M^*$ in MANGA). In terms of radius the distributions are quite similar, but the range of mass is different.
The curvature of the distribution is stronger for Es than for S0s.

\begin{figure}[]
\begin{adjustwidth}{-\extralength}{-3cm}
\centering
\includegraphics[width=8.0cm]{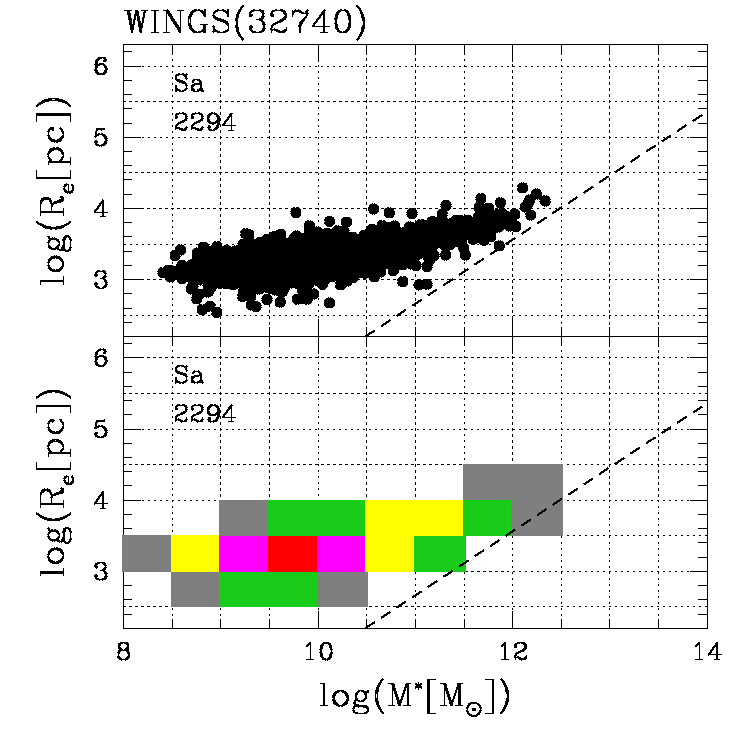}
\includegraphics[width=8.0cm]{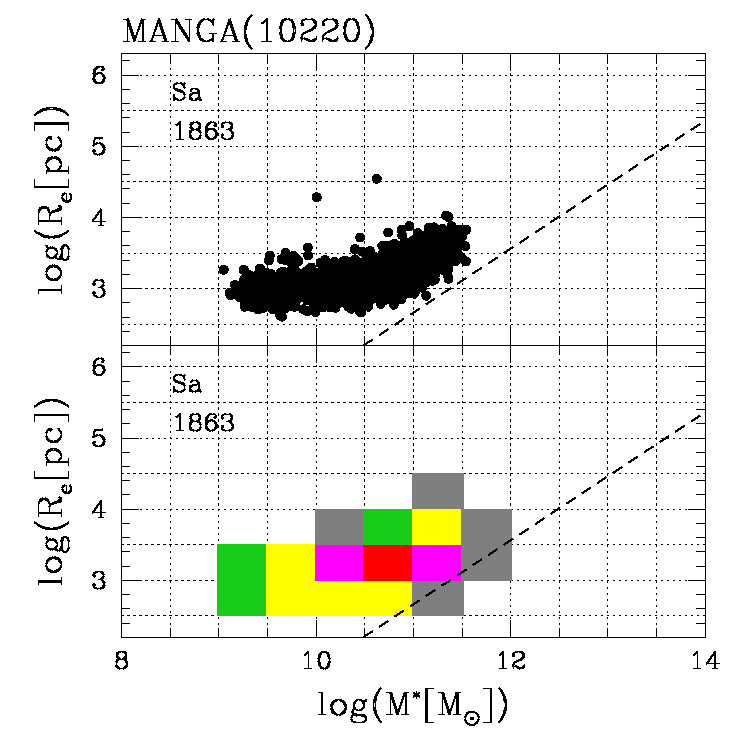}\\
\includegraphics[width=8.0cm]{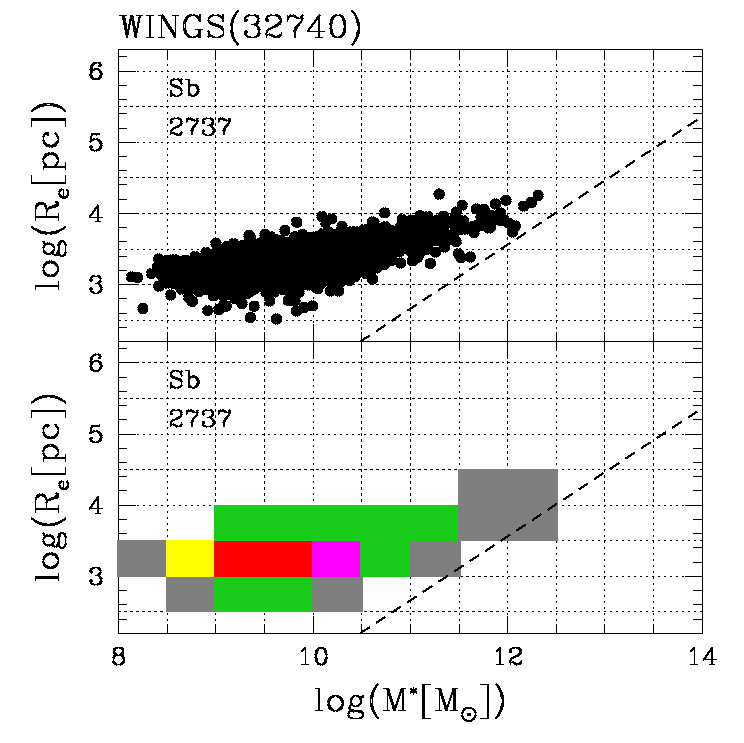}
\includegraphics[width=8.0cm]{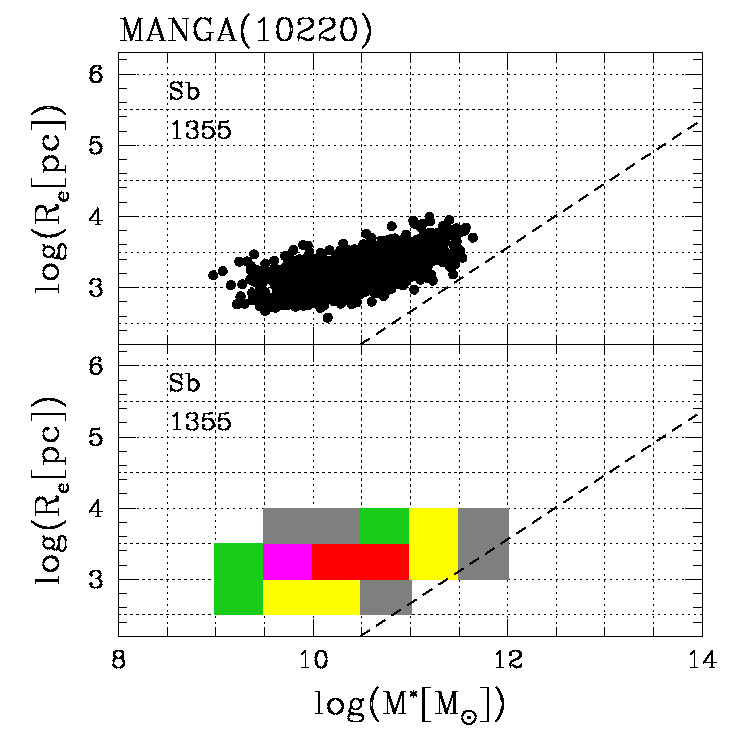}
\end{adjustwidth}
\caption{The \MRa\ plane for Sa (top panels) and Sb (bottom panels) galaxies: (\textbf{a}) The top panel shows with black dots the whole distribution of galaxies of the WINGS survey classified as Sa. The bottom panel shows with different colors the areas with the large number density of objects given in percent. Red corresponds to regions where more than 20\% of the galaxies are found. Magenta gives the interval 10\% - 20\%, yellow 5\% - 10\%, green 1\% - 5\%, and gray 0.1\% - 1\%. (\textbf{b}) The same as panel a) for the Sa galaxies of the MANGA survey. (\textbf{c}) The same as in panel a) for Sb galaxies of WINGS. (\textbf{d}) The same as in panel a) for Sb in MANGA. On top of the figure, we indicate the size of the whole sample for both datas-ets. Within the box, we provide the total number of galaxies of that morphological type. The dashed line marks the ZoE.\label{fig14}}
\end{figure}

Figure \ref{fig14} shows the \MRa\ plane for Sa and Sb galaxies. Both data-sets indicate a flattening of the observed distributions with a more linear trend of \re\ with $M^*$. Again the peaks of the distributions are a bit different in mass and more or less the same in radius. WINGS contains objects with bigger masses, not observed in MANGA.

\begin{figure}[]
\begin{adjustwidth}{-\extralength}{-3cm}
\centering
\includegraphics[width=8.0cm]{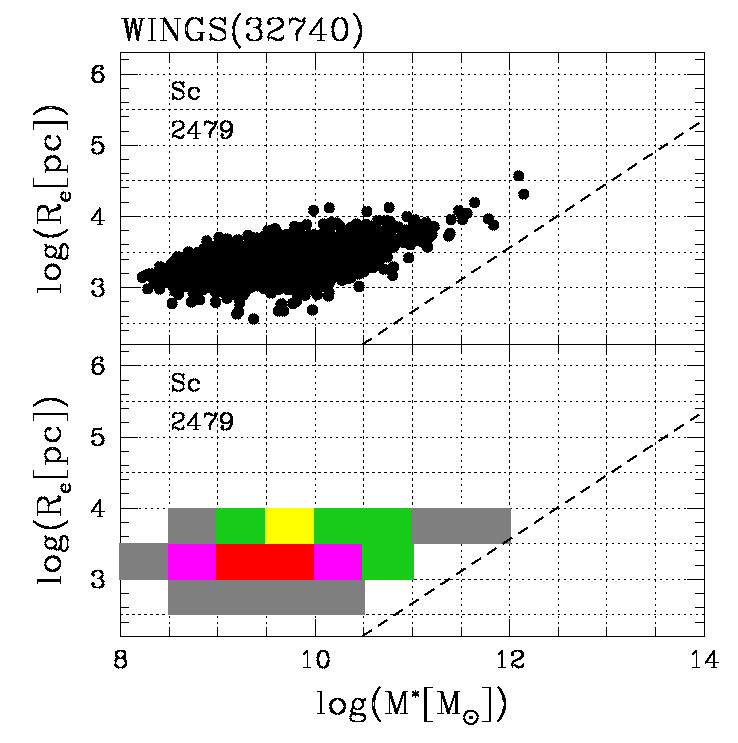}
\includegraphics[width=8.0cm]{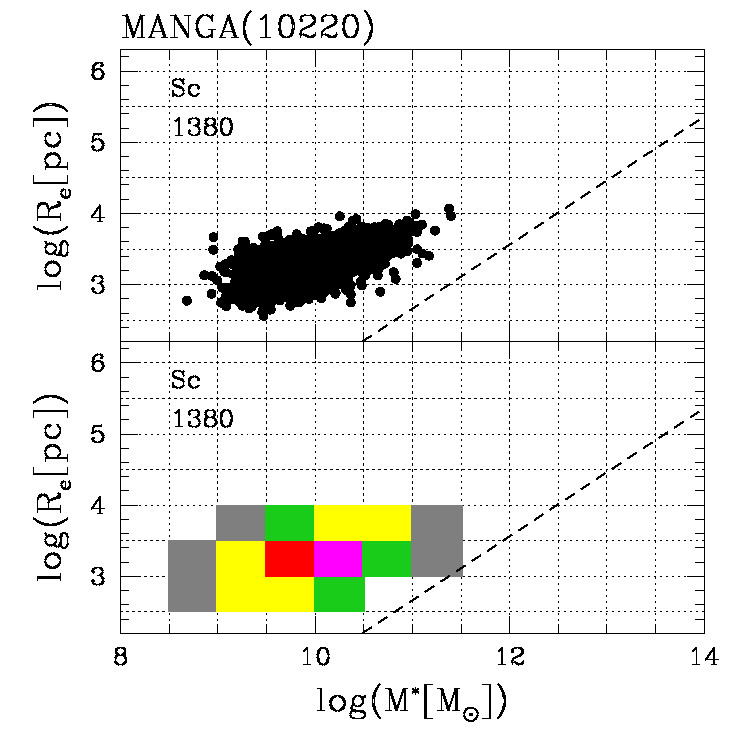}\\
\includegraphics[width=8.0cm]{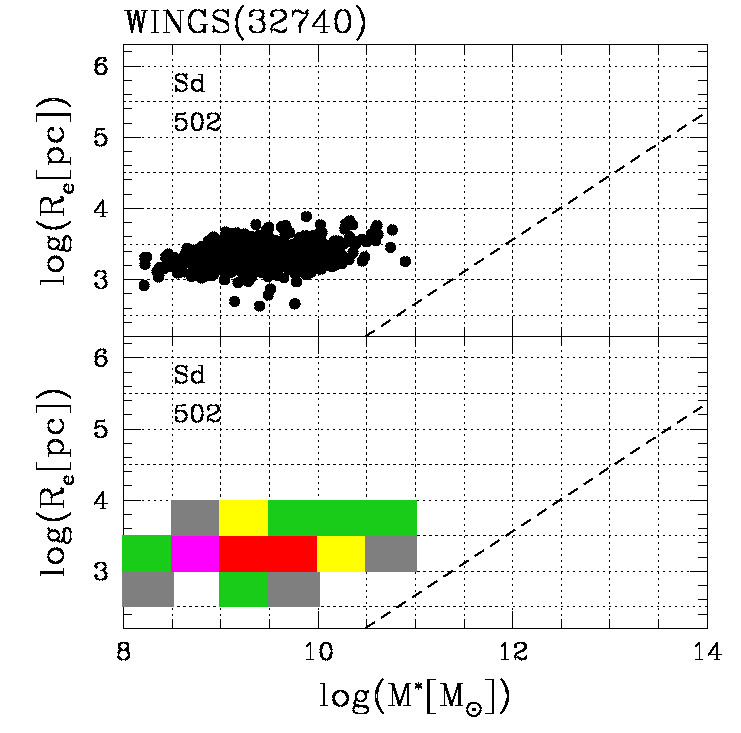}
\includegraphics[width=8.0cm]{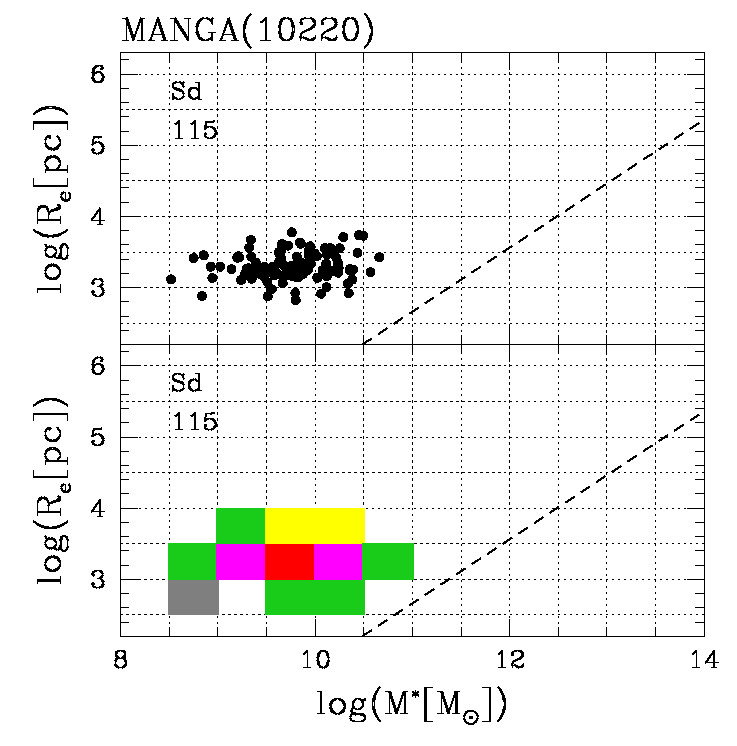}
\end{adjustwidth}
\caption{The \MRa\ plane for Sc (top panels) and Sd (bottom panels) galaxies: (\textbf{a}) The top panel shows with black dots the whole distribution of galaxies of the WINGS survey classified as Sc. The bottom panel shows with different colors the areas with the large number density of objects given in percent. Red corresponds to regions where more than 20\% of the galaxies are found. Magenta gives the interval 10\% - 20\%, yellow 5\% - 10\%, green 1\% - 5\%, and  gray 0.1\% - 1\%. (\textbf{b}) The same as in panel a) for the Sc galaxies of the MANGA survey. (\textbf{c}) The same as in  panel a) for Sd galaxies of WINGS. (\textbf{d}) The same as in panel a) for Sd in MANGA. On top of the figure, we indicate the size of the whole sample for both data-sets. Within the box, we provide the total number of galaxies of that morphological type. The dashed line marks the ZoE.\label{fig15}}
\end{figure}

For Sc and Sd galaxies (Fig. \ref{fig15}) the peaks of the distribution in the MANGA and WINGS samples occur approximately at the same \re\ and $M^*$. The correlation between the two variables is shallower, almost absent for Sd. This morphological type is very rare.

Figure \ref{fig16} shows the \MRa\ plane for the whole distribution of ETGs and LTGs. Here the difference between ETGs and LTGs is well visible. The curvature exists only for ETGs. For the MANGA sample the ETGs peak of the distribution is at higher $M^*$ and the very massive ETGs are not present. The distribution of LTGs is more {similar in the data-sets}, but again the peaks are different (at higher $M^*$ in MANGA). MANGA seems to suggest a slight curvature also for LTGs, but this can be originated by errors in the morphological classification.

\section{The \MVBmV\ plane}\label{sec:6}

The Color–Magnitude Relation \MVBmV\ describes how the color of a galaxy (typically the rest-frame optical color $B-V$) correlates with its absolute magnitude $M_V$ (a proxy for luminosity or stellar mass):
$(B-V)=a+b M_V$ (or equivalently $Color=a+b\log(L))$.

The general trend is that brighter (more massive) galaxies are redder {(i.e., they have higher $B-V$). This linear trend defines what is called the “red sequence” in the color–magnitude diagrams, } while star-forming galaxies populate the “blue cloud”, and intermediate systems occupy the “green valley.”

This relation reflects the connection between stellar population properties (age, metallicity, and star formation history) and galaxy mass or luminosity. The key physical drivers are: i) Metallicity effect: more massive galaxies retain metals more efficiently, making their stellar populations redder; ii) Age effect: older stellar populations are redder due to stellar evolution and the lack of young, blue stars; iii) Star formation quenching: galaxies move from the blue cloud to the red sequence when star formation shuts down.
Hence, the \MVBmV\ relation is a diagnostic of both stellar population evolution and mass assembly.

Researchers have used the color-magnitude diagram to trace galaxy evolution across cosmic time. The slope and scatter of the red sequence reveal when and how the stellar populations formed. The narrower the scatter the older and coeval is the stellar population. Broad scatter implies ongoing or recent star formation. Another use was that of constraining the metallicity–mass relation. The slope of the relation primarily arises from the mass–metallicity relation.
Comparing clusters at different redshifts shows chemical enrichment history. Studies of the galaxy environments can also be addressed.

The \MVBmV\ relation differs between clusters, groups, and field environments — cluster red sequences are tighter and redder.
The diagram can indicates environmental quenching efficiency. The identification of the galaxy populations (blue cloud, red sequence, green valley) can be used to separate passive and star-forming galaxies in big surveys.
Clearly, semi-analytic and hydrodynamic models must reproduce the observed \MVBmV\ slope, zero-point, and scatter to be realistic.

Many works have addressed the \MVBmV\ relation. We can cite
\cite{SandageVisvanathan1978a,SandageVisvanathan1978b}, that in a study of the Virgo and Coma clusters, first provided the interpretation of the color-magnitude diagram in terms of a metallicity-mass relation. \cite{BowerLuceyEllis1992} who measured the tightness of the relation concluding that ETGs formed their stars at $z>2$ and evolved passively since then. \cite{KodamaArimoto1997} that first reproduced the slope of the relation via metallicity–luminosity correlation in monolithic collapse models. \cite{Blantonetal2003} revealed the bimodal color distribution (red sequence and blue cloud) in large SDSS samples and \cite{Baldryetal2004} characterized the bimodality quantitatively and \cite{Fritzetal2014} studied the environmental dependence of the red sequence in nearby clusters. {  Finally, \cite{Sciarratta_etal_2019} addressed the issue of the cosmological Interpretation of the \MVBmV\ plane of cluster galaxies.} 

Figure \ref{fig17} shows the \MVBmV\ plane for E and S0 galaxies of the WINGS and MANGA surveys. {Here the difference in the number density distribution of the two morphological types between the two data-sets is large. MANGA provides a very thin distribution for Es and S0s not observed by WINGS.} The color of the WINGS galaxies appears spread on a large interval. This scatter could be in part due to the presence of some objects with non accurate redshift (we used here the photometric sample instead of the spectro-photometric one, attributing to all galaxies the same redshift of the cluster). The cut in the WINGS color is introduced by the redshift range of the clusters adopted to calculate the luminosity. Even in this plot the density distribution is different for the two samples. MANGA gives a peak at much higher luminosity.

A similar behavior is observed for Sa and Sb galaxies (Fig. \ref{fig18}). Again the scatter in color is {larger} for WINGS. Part of this scatter may be also due to the {k-correction} (not used by MANGA) and the size extension of the $B-V$ measurements. Anyway, the density distribution is again different.
For Sc and Sd galaxies (Fig. \ref{fig19}) the two samples give a quite different distribution: while WINGS {shows} a cloud of points, MANGA provides an horizontal distribution with a small scatter.

\begin{figure}[]
\begin{adjustwidth}{-\extralength}{-3cm}
\centering
\includegraphics[width=8.0cm]{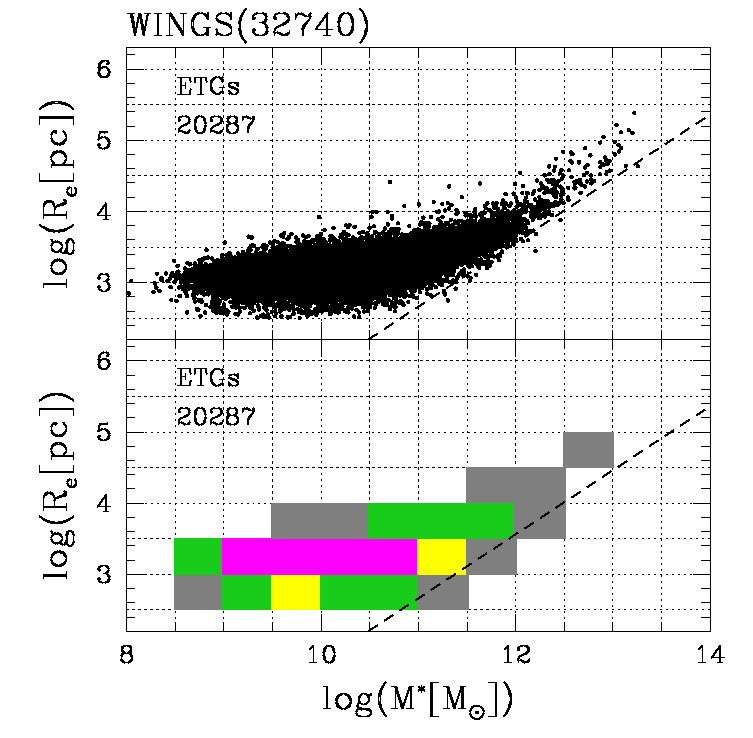}
\includegraphics[width=8.0cm]{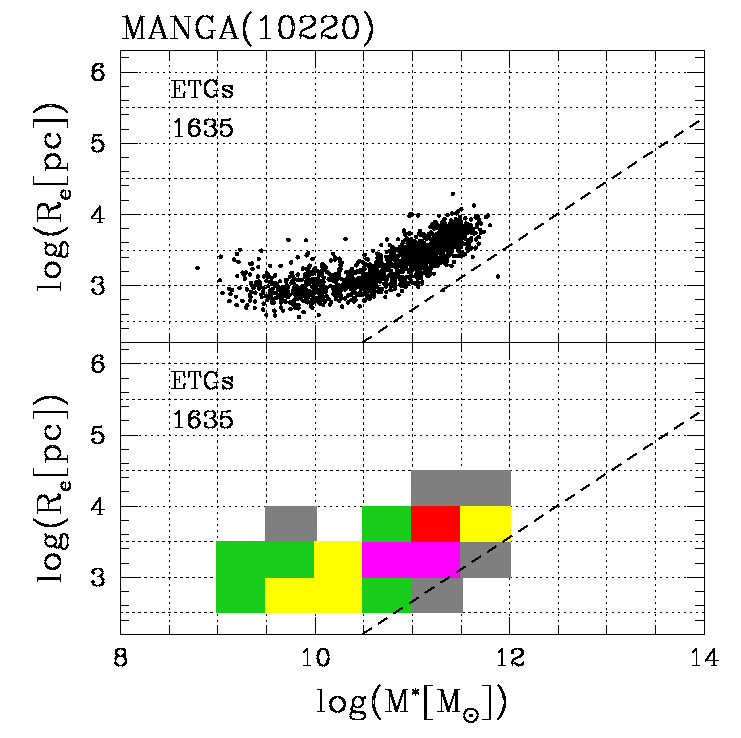}\\
\includegraphics[width=8.0cm]{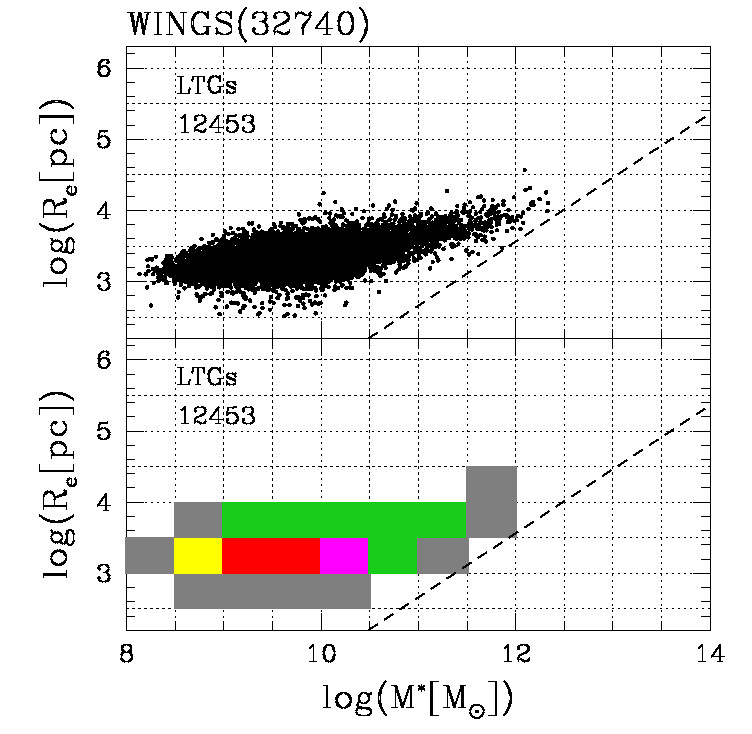}
\includegraphics[width=8.0cm]{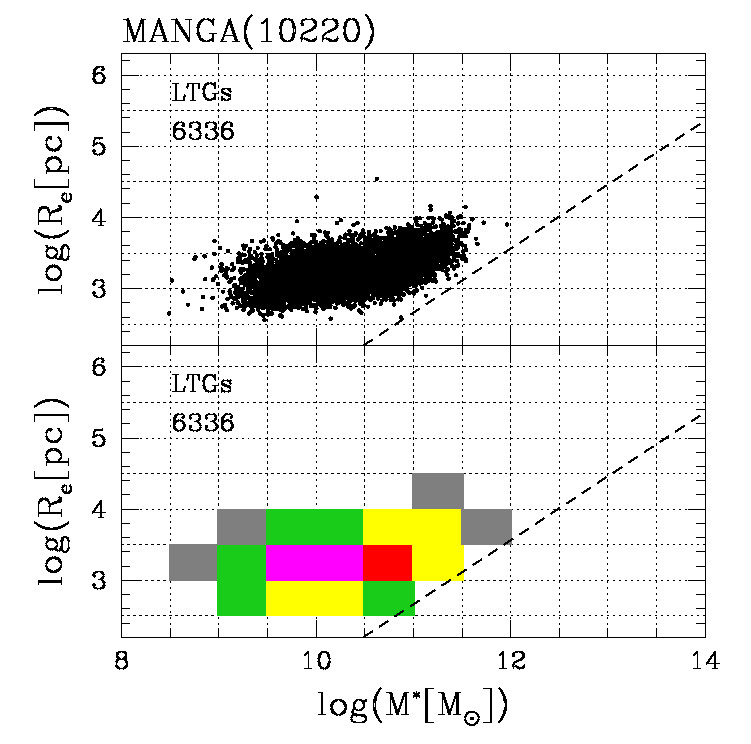}
\end{adjustwidth}
\caption{The \MRa\ plane for ETGs (top panels) and LTGs (bottom panels) galaxies: (\textbf{a}) The top panel shows with black dots the whole distribution of galaxies of the WINGS survey classified as early-types. The bottom panel shows with different colors the areas with the large number density of objects given in percent. Red corresponds to regions where more than 20\% of the galaxies are found. Magenta gives the interval 10\% - 20\%, yellow 5\% - 10\%, green 1\% - 5\%, and gray 0.1\% - 1\%. (\textbf{b}) Same as panel a) for the ETGs galaxies of the MANGA survey. (\textbf{c}) The same as in panel a) for LTGs galaxies of WINGS. (\textbf{d}) The same as in panel a) for LTGs in MANGA. On top of the figure, we indicate the size of the whole sample for both data-sets. Within the box, we provide the total number of galaxies of that morphological type. The dashed line marks the ZoE.\label{fig20}}
\end{figure}

\begin{figure}[]
\begin{adjustwidth}{-\extralength}{-3cm}
\centering
\includegraphics[width=8.0cm]{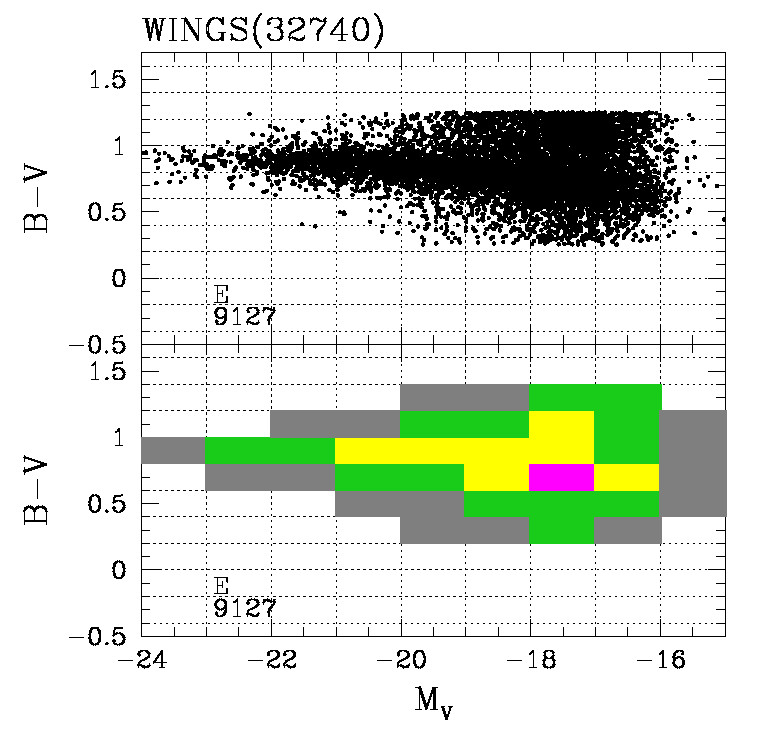}
\includegraphics[width=8.0cm]{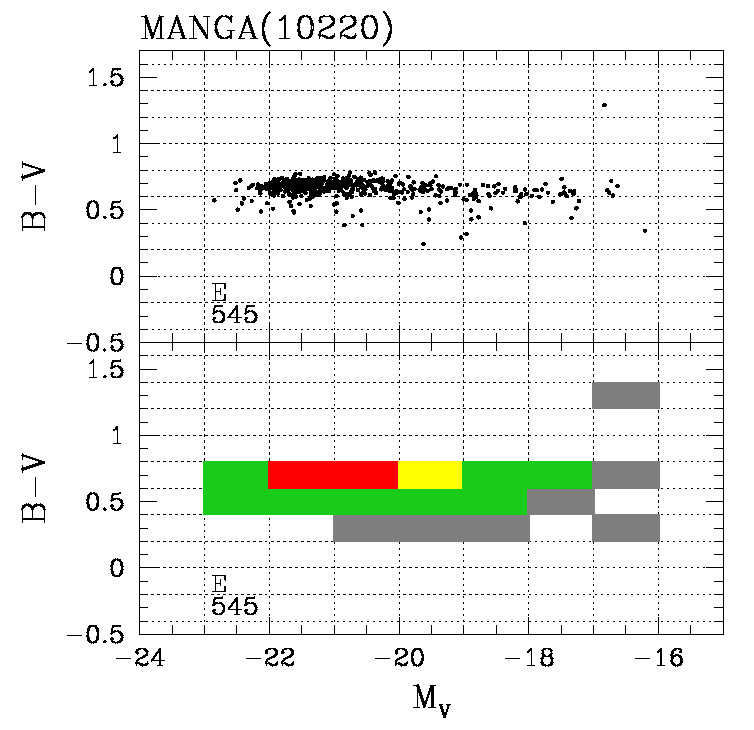}\\
\includegraphics[width=8.0cm]{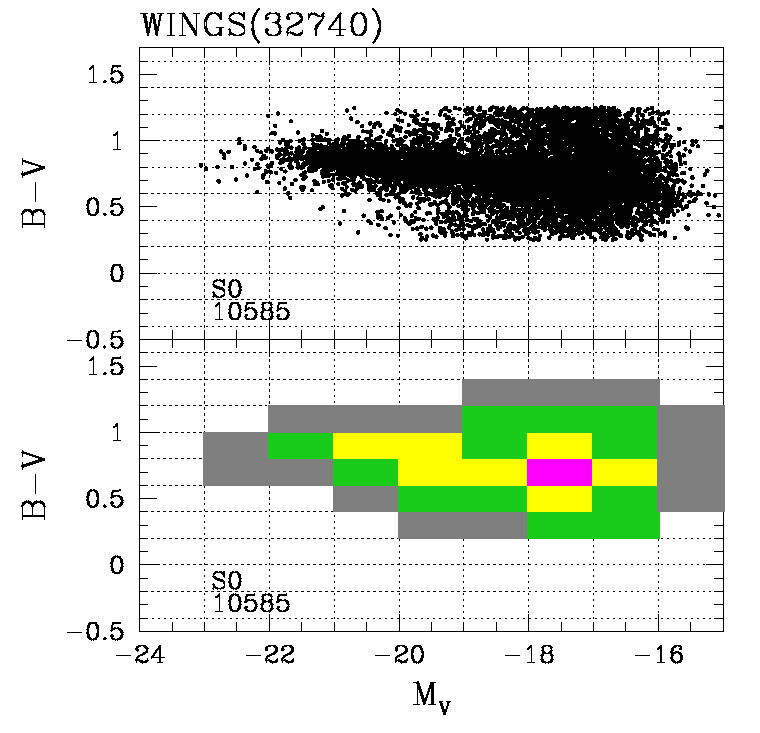}
\includegraphics[width=8.0cm]{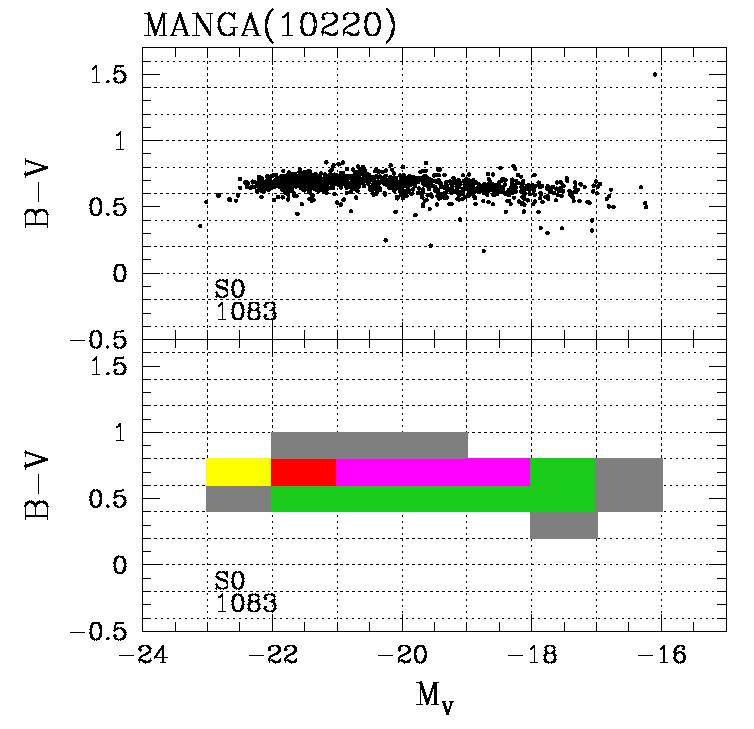}
\end{adjustwidth}
\caption{The \MVBmV\ plane for E (top panels) and S0 (bottom panels) galaxies: (\textbf{a}) The top panel shows with black dots the whole distribution of galaxies of the WINGS survey classified as E. The bottom panel shows with different colors the areas with the large number density of objects given in percent. Red corresponds to regions where more than 20\% of the galaxies are found. Magenta gives the interval 10\% - 20\%, yellow 5\% - 10\%, green 1\% - 5\%, and gray 0.1\% - 1\%. (\textbf{b}) The same as in panel a) for the E galaxies of the MANGA survey. (\textbf{c}) The same as in panel a) for S0 galaxies of WINGS. (\textbf{d}) The same as in panel a) for S0 in MANGA. On top of the figure, we indicate the size of the whole sample for  both data-sets. Within the box, we provide the total number of galaxies of that morphological type. The dashed line marks the ZoE.\label{fig16}}
\end{figure}

\begin{figure}[]
\begin{adjustwidth}{-\extralength}{-3cm}
\centering
\includegraphics[width=8.0cm]{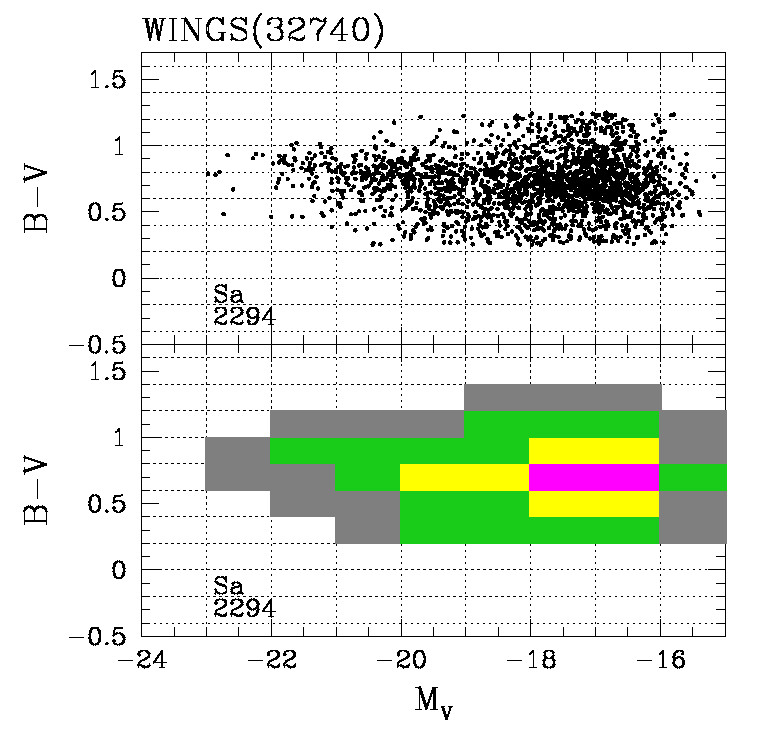}
\includegraphics[width=8.0cm]{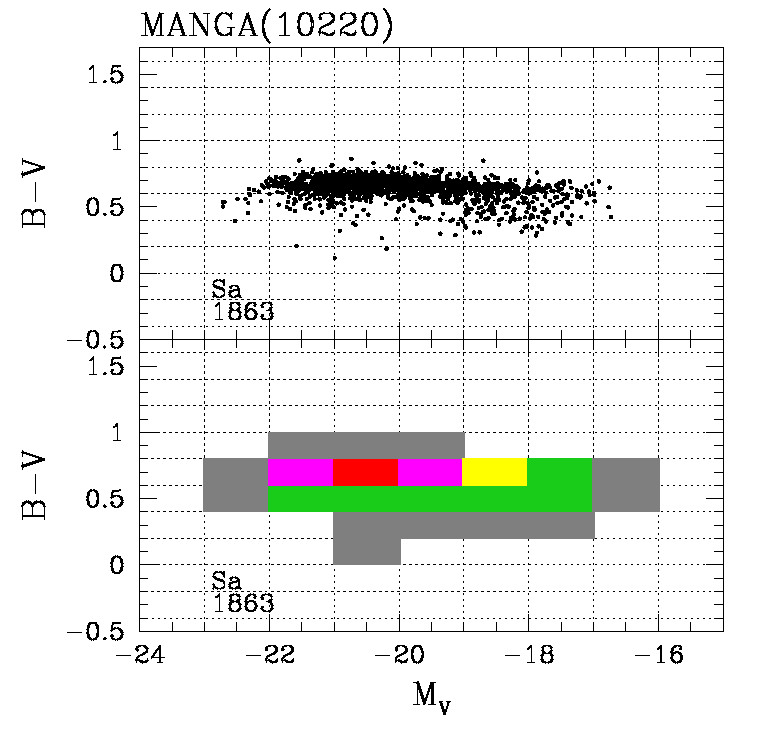}\\
\includegraphics[width=8.0cm]{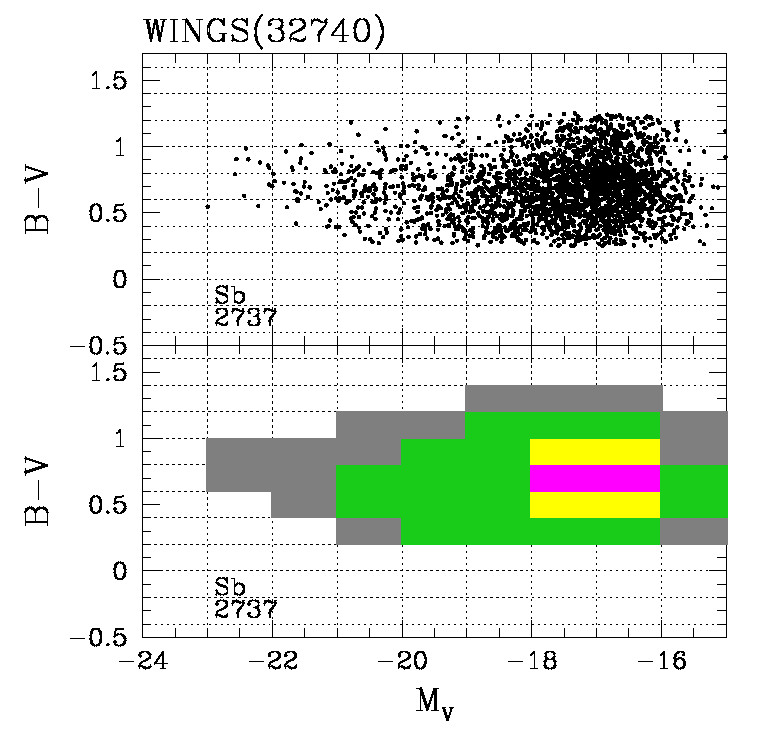}
\includegraphics[width=8.0cm]{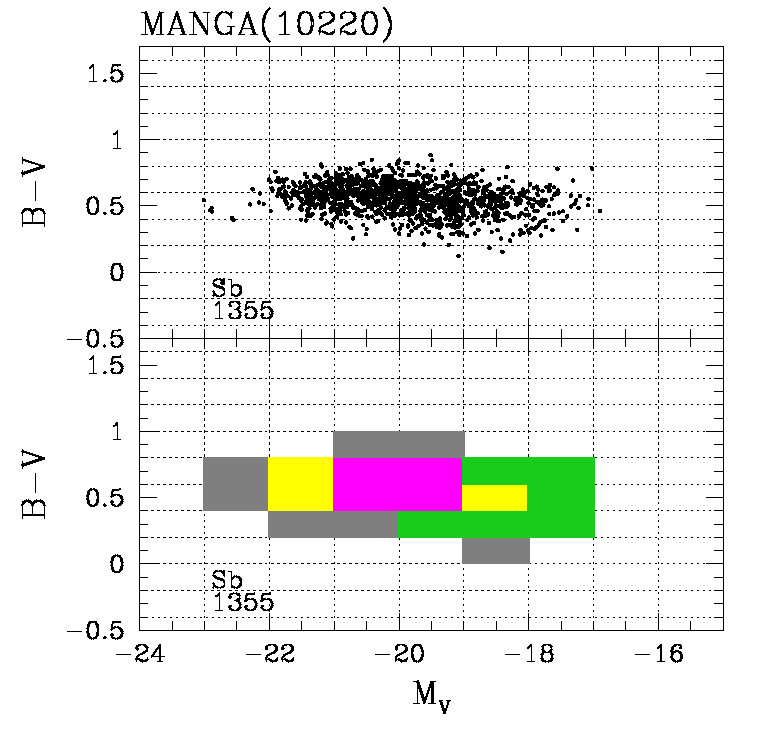}
\end{adjustwidth}
\caption{The \MVBmV\ plane for Sa (top panels) and Sb (bottom panels) galaxies: (\textbf{a}) The top panel shows with black dots the whole distribution of galaxies of the WINGS survey classified as Sa. The bottom panel shows with different colors the areas with the large number density of objects given in percent. Red corresponds to regions where more than 20\% of the galaxies are found. Magenta gives the interval 10\% - 20\%, yellow 5\% - 10\%, green 1\% - 5\%, gray 0.1\% - 1\%. (\textbf{b}) The same as in panel a) but for the Sa galaxies of the MANGA survey. (\textbf{c}) The same as in panel a) but for Sb galaxies of WINGS. (\textbf{d}) The same as in panel a) but for Sb in MANGA. On top of the figure, we indicate the whole sample size of both data-sets. Within the box, we provide the total number of galaxies of that morphological type. The dashed line marks the ZoE.\label{fig17}}
\end{figure}

\begin{figure}[]
\begin{adjustwidth}{-\extralength}{-3cm}
\centering
\includegraphics[width=8.0cm]{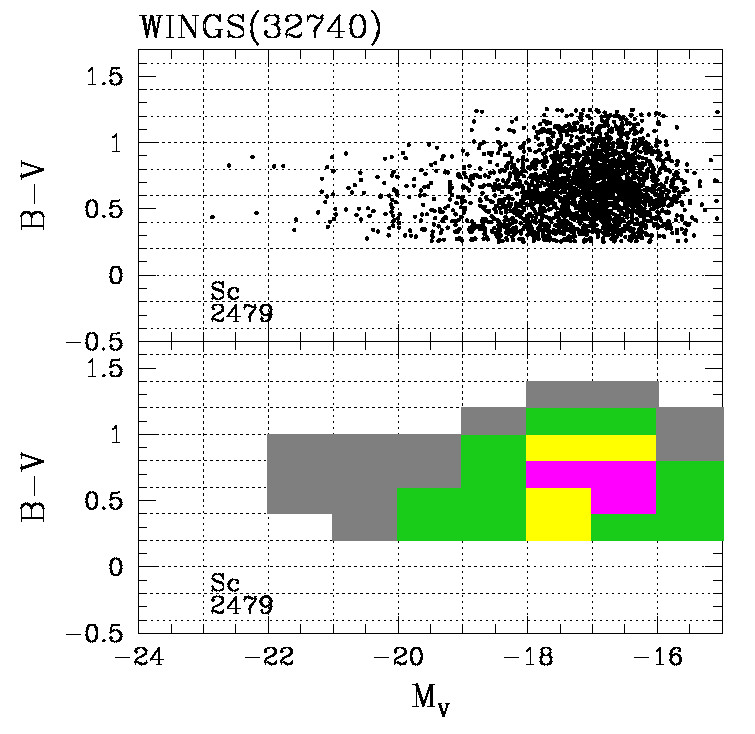}
\includegraphics[width=8.0cm]{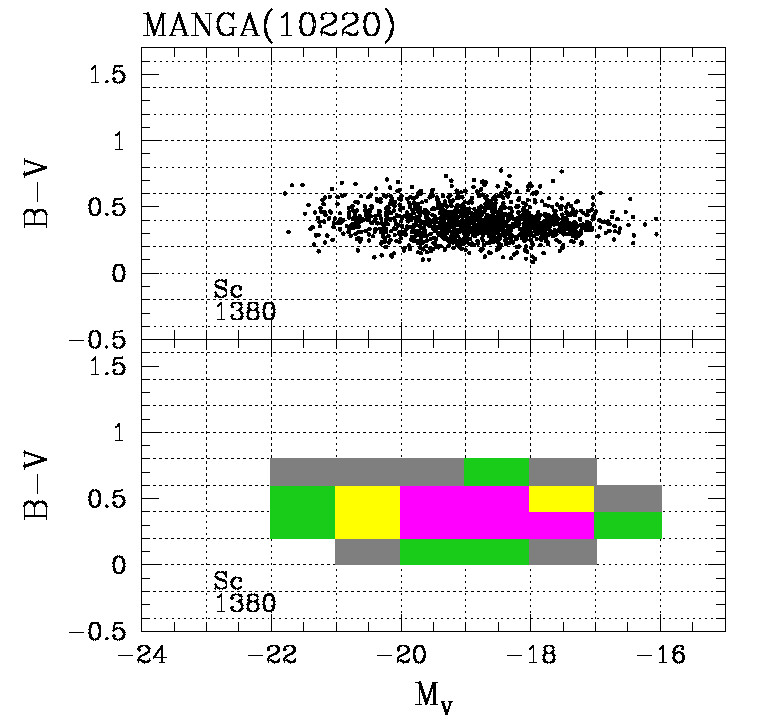}\\
\includegraphics[width=8.0cm]{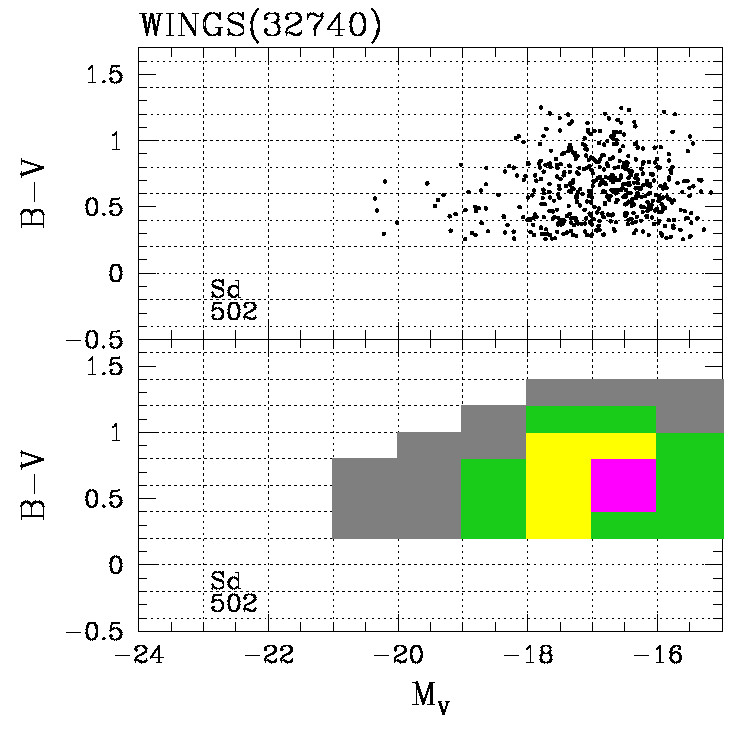}
\includegraphics[width=8.0cm]{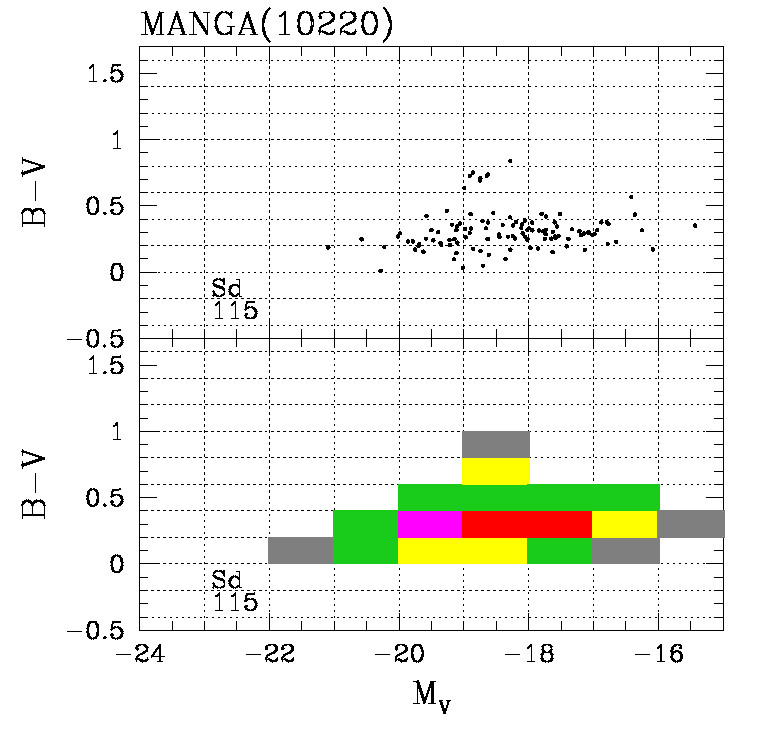}
\end{adjustwidth}
\caption{The \MVBmV\ plane for Sc (top panels) and Sd (bottom panels) galaxies: (\textbf{a}) The top panel shows with black dots the whole distribution of galaxies of the WINGS survey classified as Sc. The bottom panel shows with different colors the areas with the large number density of objects given in percent. Red corresponds to regions where more than 20\% of the galaxies are found. Magenta gives the interval 10\% - 20\%, yellow 5\% - 10\%, green 1\% - 5\%, and gray 0.1\% - 1\%. (\textbf{b}) The same as in panel a) but for the Sc galaxies of the MANGA survey. (\textbf{c}) The same as in panel a) but for the Sd galaxies of WINGS. (\textbf{d}) The same as in panel a) but for Sd in MANGA. On top of the figure, we indicate the size of the whole sample for both data-sets. Within the box, we provide the total number of galaxies of that morphological type. The dashed line marks the ZoE.\label{fig18}}
\end{figure}

\begin{figure}[]
\begin{adjustwidth}{-\extralength}{-3cm}
\centering
\includegraphics[width=8.0cm]{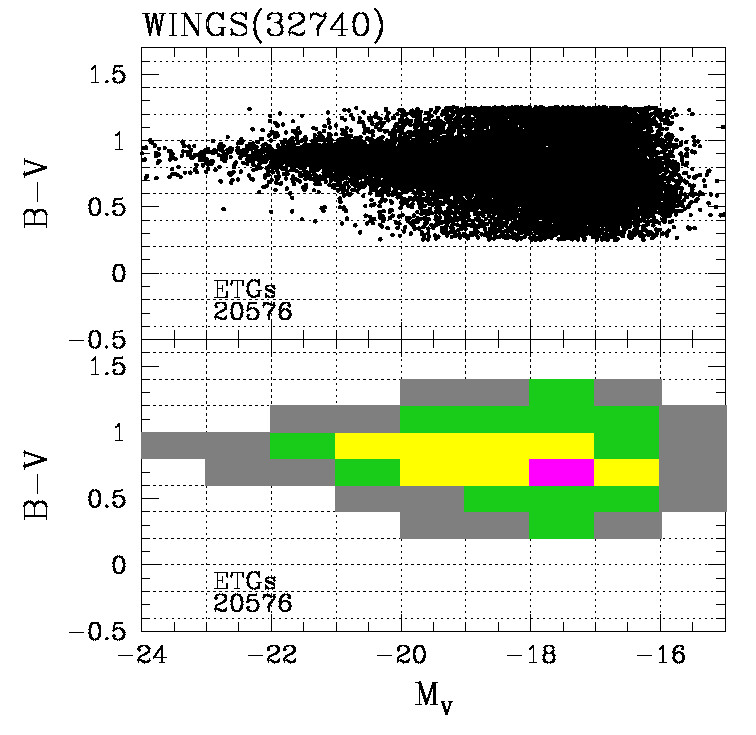}
\includegraphics[width=8.0cm]{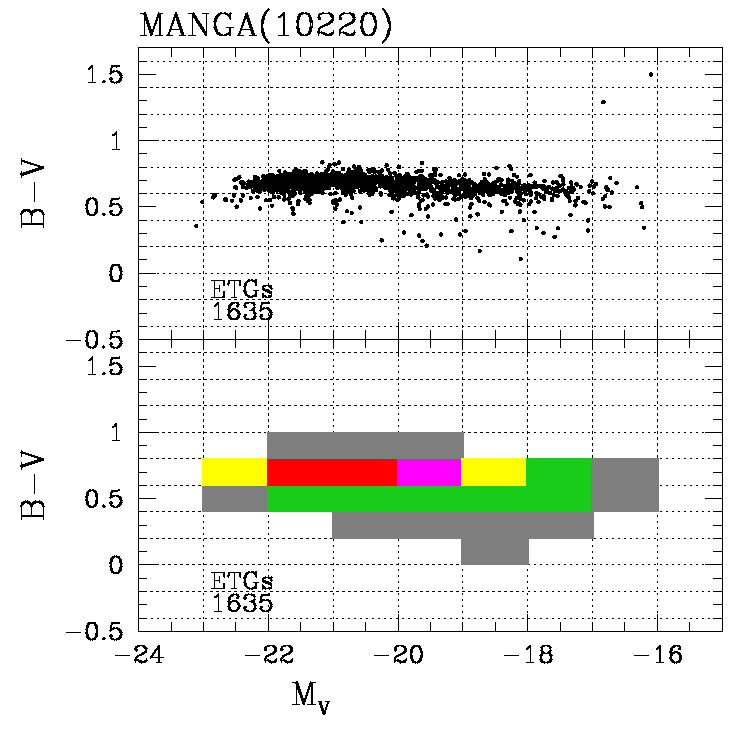}\\
\includegraphics[width=8.0cm]{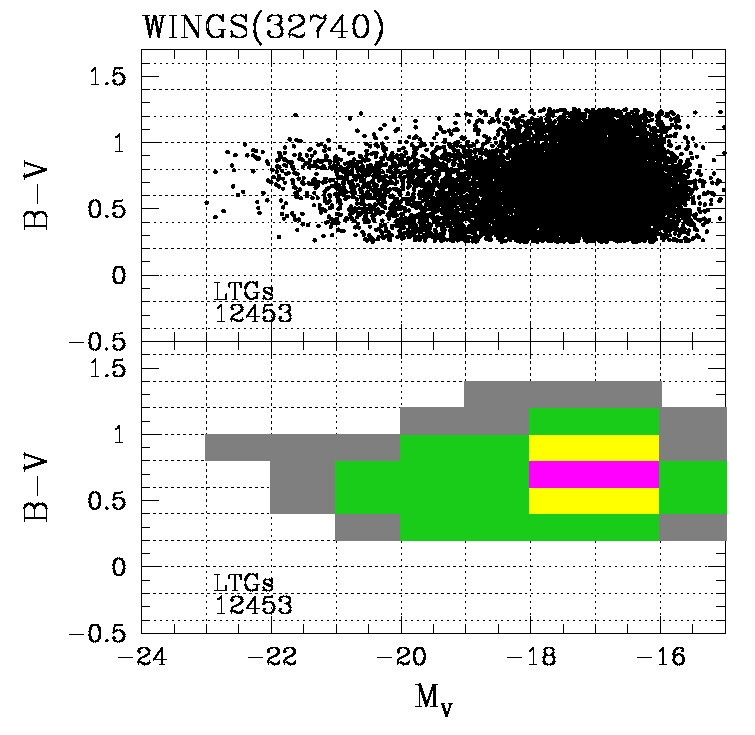}
\includegraphics[width=8.0cm]{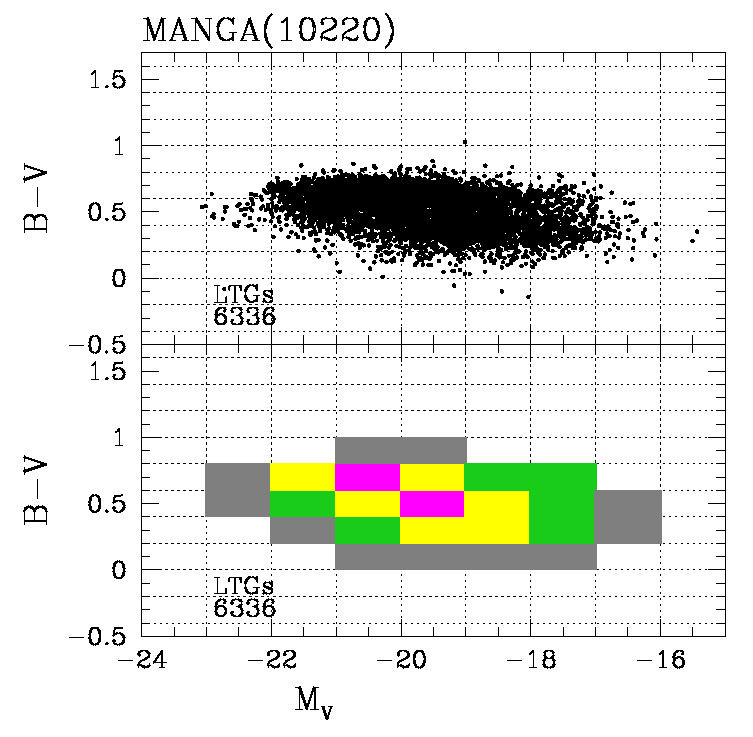}
\end{adjustwidth}
\caption{The \MVBmV\ plane for ETGs (top panels) and LTGs (bottom panels) galaxies: (\textbf{a}) The top panel shows with black dots the whole distribution of galaxies of the WINGS survey classified as early-types. The bottom panel shows with different colors the areas with the large number density of objects given in percent. Red corresponds to regions where more than 20\% of the galaxies are found. Magenta gives the interval 10\% - 20\%, yellow 5\% - 10\%, green 1\% - 5\%, and Gray 0.1\% - 1\%. (\textbf{b}) The same as in panel a) but for the ETGs galaxies of the MANGA survey. (\textbf{c}) The same as in panel a) but for LTGs galaxies of WINGS. (\textbf{d}) The same as in panel a) but for LTGs in MANGA. On top of the figure, we is indicated the size of whole sample for both data-sets. Within the box, we provide the total number of galaxies of that morphological type. The dashed line marks the ZoE.\label{fig19}}
\end{figure}

The whole distribution of ETGs and LTGs is visible in Fig. \ref{fig20}. The two samples suggest very different distributions.
It is clear from these plots that any {attempt at comparing} the galaxies properties should start from an accurate analysis of the characteristics of the samples and from standard procedures of data analysis. With these data the answer to the question: "Which is the percentage of galaxies within a given range of magnitudes and colors at the present epoch?" is not possible.

\section{The \nL\ plane}\label{sec:7}

{For ETGs, the S\'ersic index $n$ (\cite{Sersic1968}),  a proxy of shape of the luminosity profile, increases with the luminosity or the stellar mass, according to 
$\log(n)\propto\alpha Y + \beta$, where $Y \equiv L$  or  $M$. } In faint galaxies $n$ is close to an exponential ($n\sim1-2$), while in bright ellipticals $n\sim 4-10$. The S\'ersic index correlates also with the effective radius \re, with larger galaxies having higher $n$.
These trends imply that the galaxy structure is not scale-invariant,  {that is} they are not homologous. The S\'ersic index encodes information about { the formation history and the dynamical structure}. { Low values of $n$ are typical} of disk-like, rotationally supported systems, formed through dissipative processes, while high $n$ are typical of spheroidal systems, dominated by the velocity dispersion, likely formed via major mergers or violent relaxation. The relation reflects gradual changes in stellar density profile with mass, driven by feedback, merger history, and dissipation fraction.

{\cite{Caonetal1993} first provided a quantitative estimate of the correlation between $n$ and luminosity and radius using data of the Virgo and Fornax clusters. They showed that $n$ increases smoothly with the luminosity and radius and therefore that galaxies are non-homologous systems.}
This prompted several studies on the tilt of the Fundamental Plane (e.g., \cite{GrahamColless1997}). \cite{Trujilloetal2001} modeled the effects of non-homology on dynamical and photometric scaling laws, quantifying its contribution {to the tilt of the FP ($\sim$30–40\%). }

\begin{figure}[]
\begin{adjustwidth}{-\extralength}{-3cm}
\centering
\includegraphics[width=8.0cm]{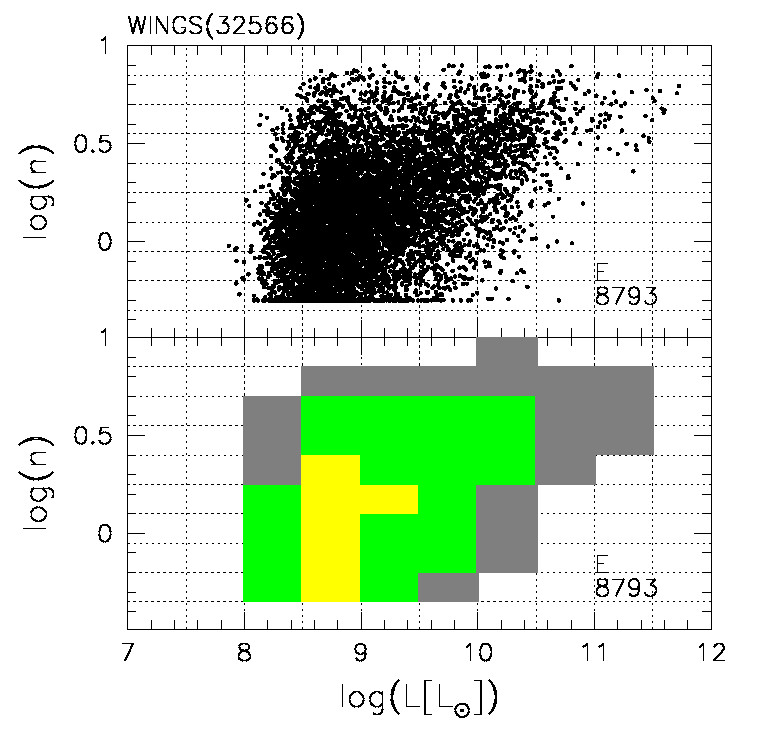}
\includegraphics[width=8.0cm]{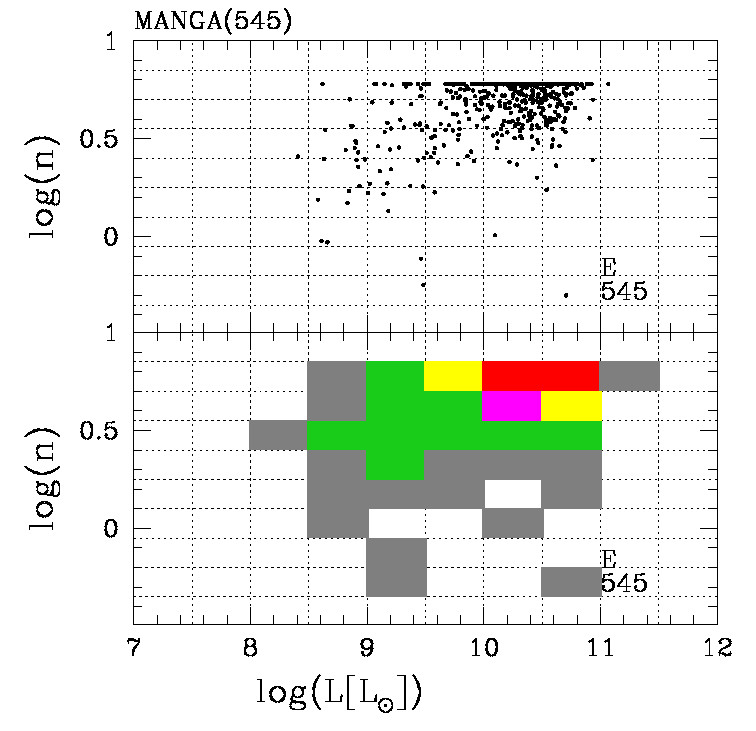}\\
\includegraphics[width=8.0cm]{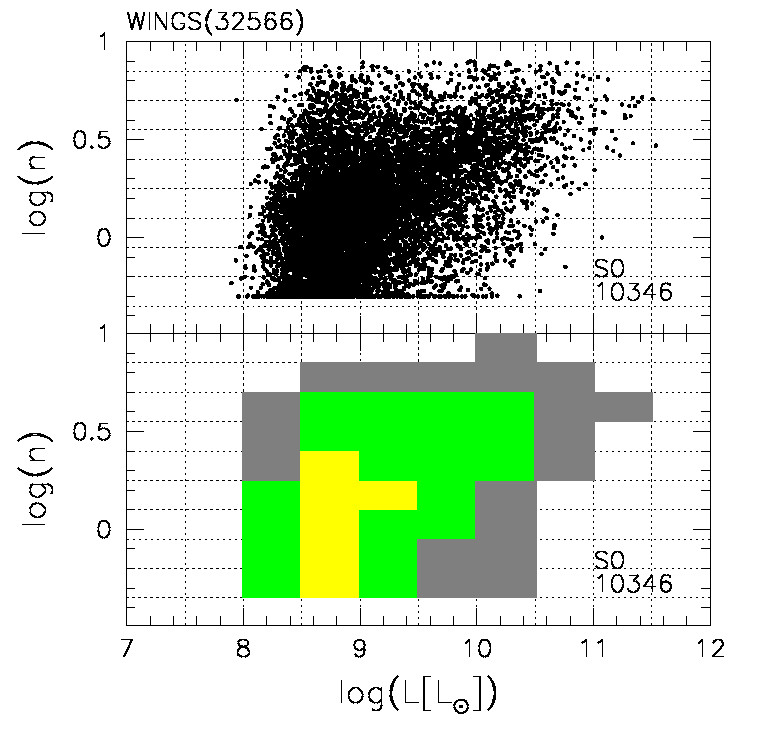}
\includegraphics[width=8.0cm]{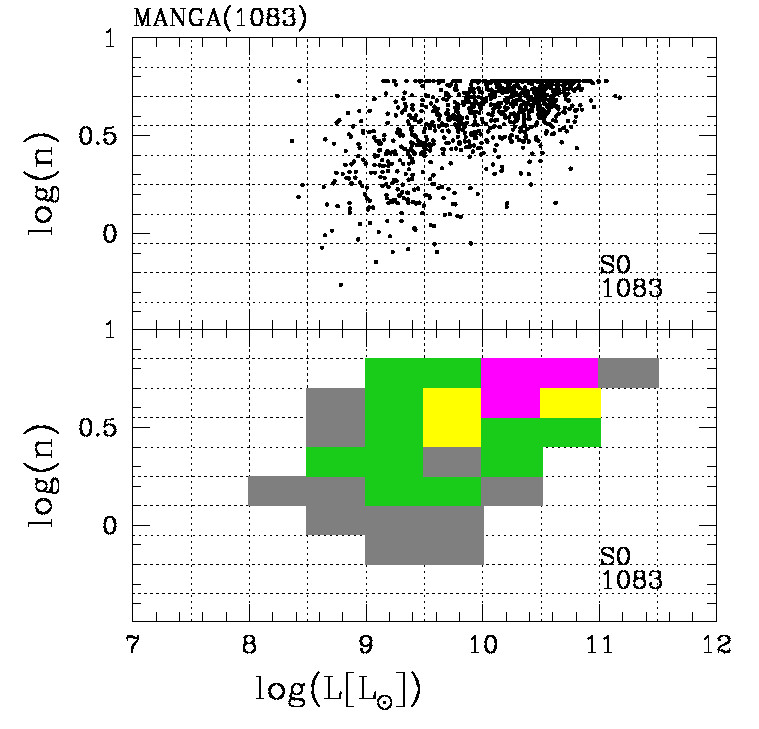}
\end{adjustwidth}
\caption{The \nL\ plane for E (top panels) and S0 (bottom panels) galaxies: (\textbf{a}) The top panel shows with black dots the whole distribution of galaxies of the WINGS survey classified as E. The bottom panel shows with different colors the areas with the large number density of objects given in percent. Red corresponds to regions where more than 20\% of the galaxies are found. Magenta gives the interval 10\% - 20\%, yellow 5\% - 10\%, green 1\% - 5\%, and gray 0.1\% - 1\%. (\textbf{b}) The same as in panel a) but for the E galaxies of the MANGA survey. (\textbf{c}) The same as in panel a) but for thre S0 galaxies of WINGS. (\textbf{d}) The same as in panel a) but for S0 in MANGA. On top of the figure, we indicate the size of the whole sample for both data-sets. Within the box, we provide the total number of galaxies of that morphological type. The dashed line marks the ZoE.\label{fig21}}
\end{figure}

The lack of Es and S0s of low luminosity in the MANGA sample is the reason of the large  difference observed in Fig. \ref{fig21}. {  The WINGS data clearly show a main trend  where  $n$ increases with increasing $L$. In any case there is a small group of galaxies of low luminosity that does not follow the main trend. This is better seen in Fig. \ref{fig24}, where the cumulative distribution of ETGs and LTGs is plotted. The trend is similar for both populations, but in MANGA the second population is missing. } The different technique used to get $n$ is likely at the origin of the observed difference. WINGS uses the fit of the growth curves along the major and minor axes with a S\'ersic law convolved with a multi-gaussian PSF. MANGA fits the light profiles along the major and minor axes, but it is not clear if the convolution is applied. The colored panels indicate  a different percentage of galaxies in the various boxes.

{The WINGS galaxies that do not follow the main trend are objects generally very small in size. Basing on it,  one might suspect that the convolution with a multi-gaussian PSF yield erroneous high values for $n$.}
If, on the other hand, the distribution is real, one {could} conclude that all ScRs contain two main populations depending on their mass and luminosity. These two populations follow different trends in the ScRs.

\begin{figure}[]
\begin{adjustwidth}{-\extralength}{-3cm}
\centering
\includegraphics[width=8.0cm]{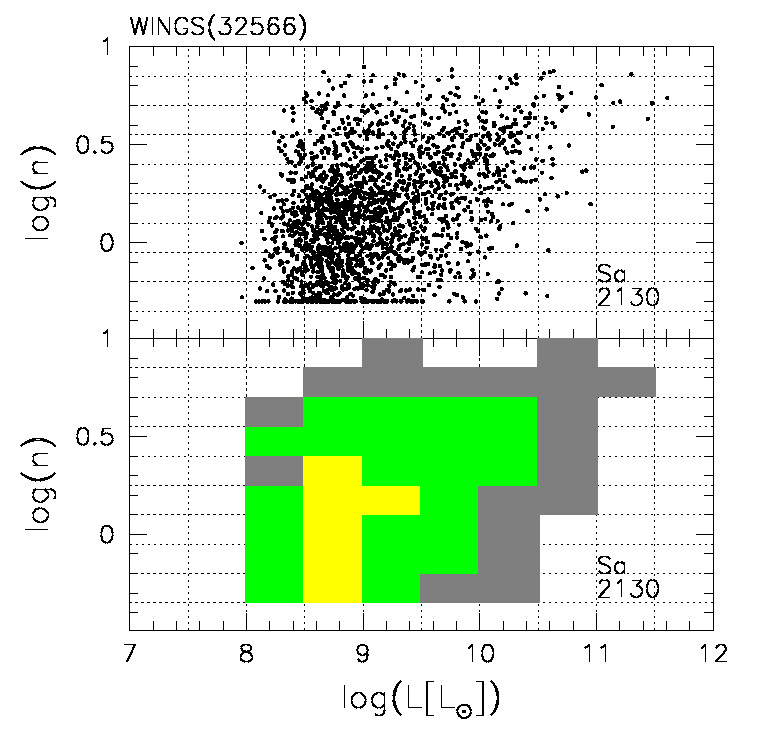}
\includegraphics[width=8.0cm]{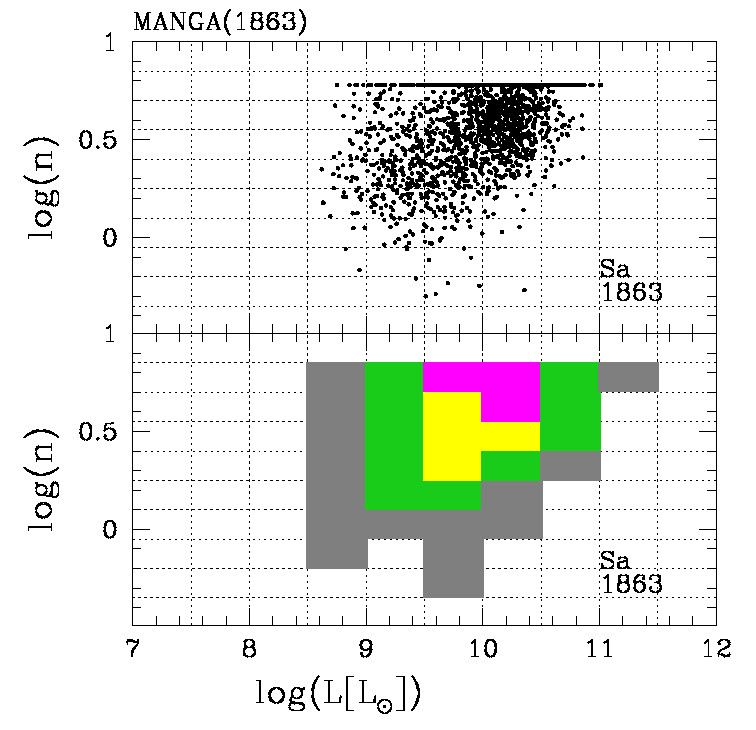}\\
\includegraphics[width=8.0cm]{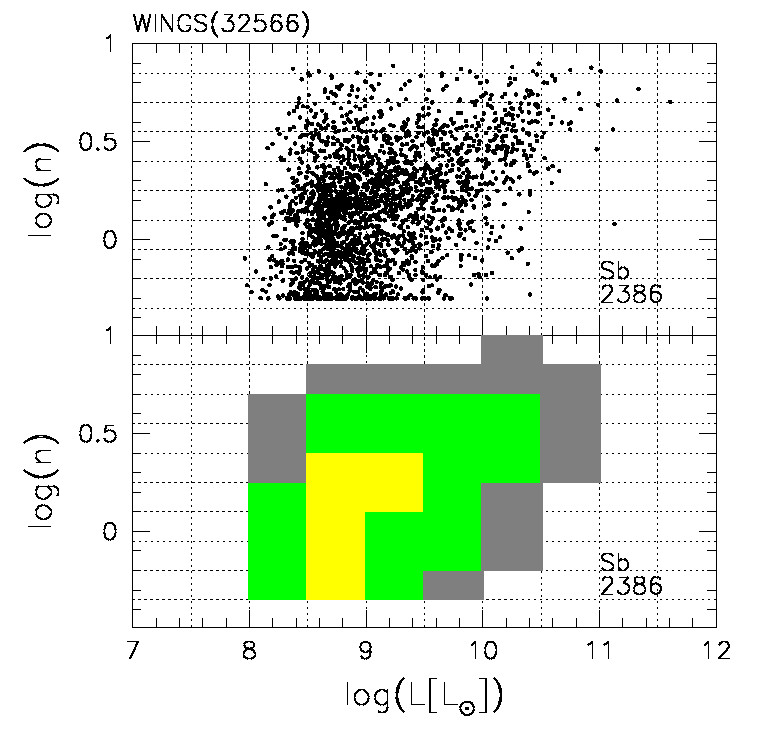}
\includegraphics[width=8.0cm]{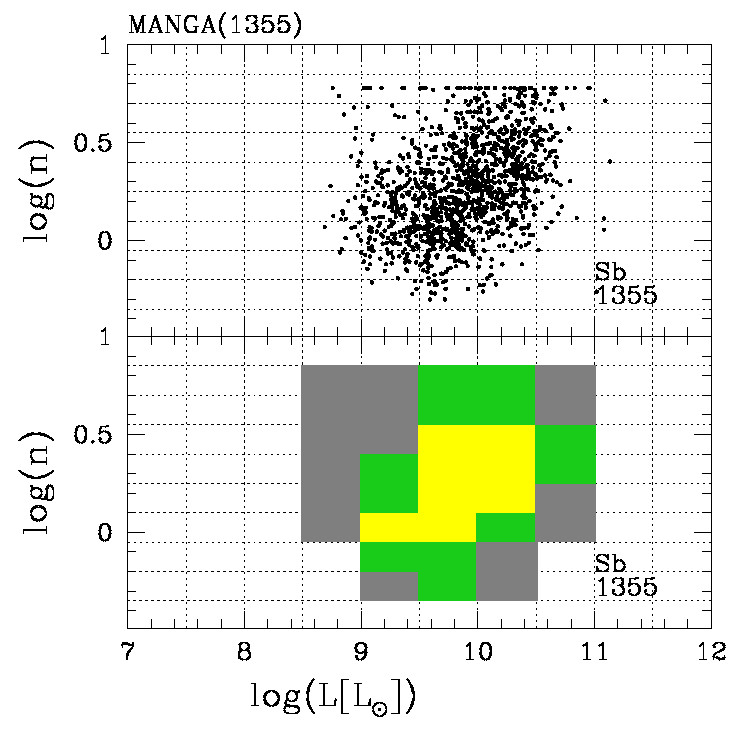}
\end{adjustwidth}
\caption{The \nL\ plane for Sa (top panels) and Sb (bottom panels) galaxies: (\textbf{a}) The top panel shows with black dots the whole distribution of galaxies of the WINGS survey classified as Sa. The bottom panel shows with different colors the areas with the large number density of objects given in percent. Red corresponds to regions where more than 20\% of the galaxies are found. Magenta gives the interval 10\% - 20\%, yellow 5\% - 10\%, green 1\% - 5\%, and gray 0.1\% - 1\%. (\textbf{b}) The same as in panel a) for the Sa galaxies of the MANGA survey. (\textbf{c}) The same as in panel a) but for Sb galaxies of WINGS. (\textbf{d}) The same as in panel a) but for Sb of MANGA. On top of the figure, we indicate the size of the whole sample for both data-sets. Within the box, we provide the total number of galaxies of that morphological type. The dashed line marks the ZoE.\label{fig22}}
\end{figure} 

\begin{figure}[]
\begin{adjustwidth}{-\extralength}{-3cm}
\centering
\includegraphics[width=8.0cm]{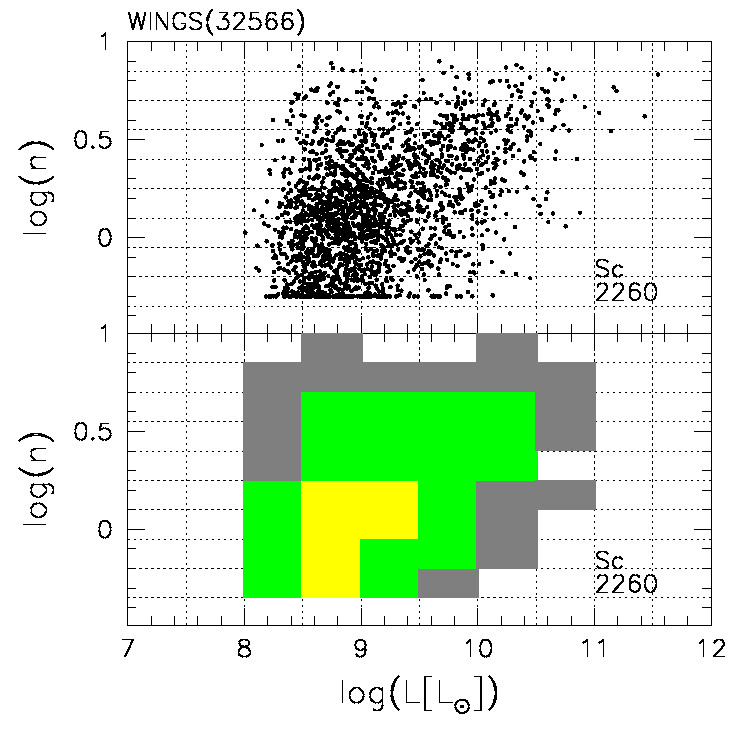}
\includegraphics[width=8.0cm]{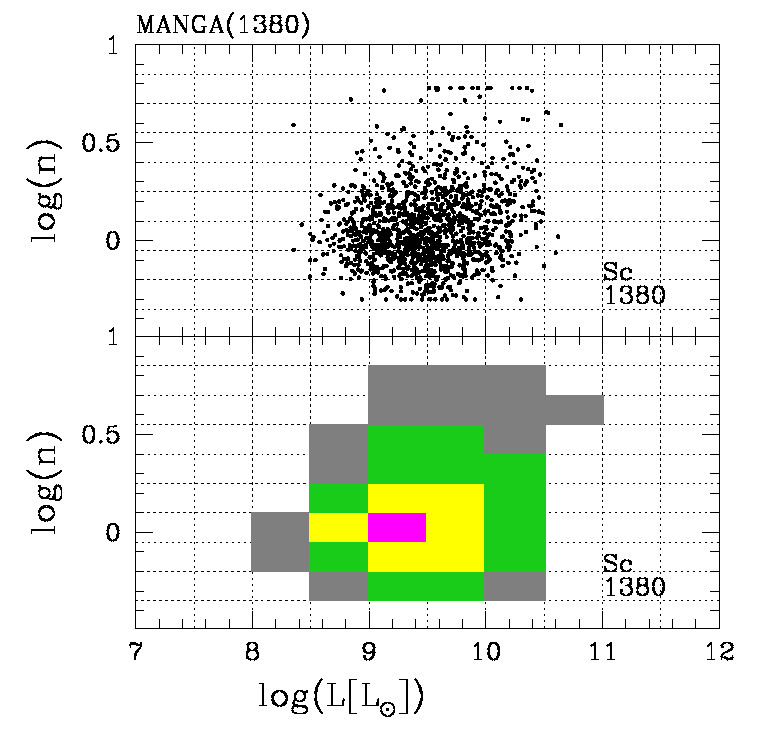}\\
\includegraphics[width=8.0cm]{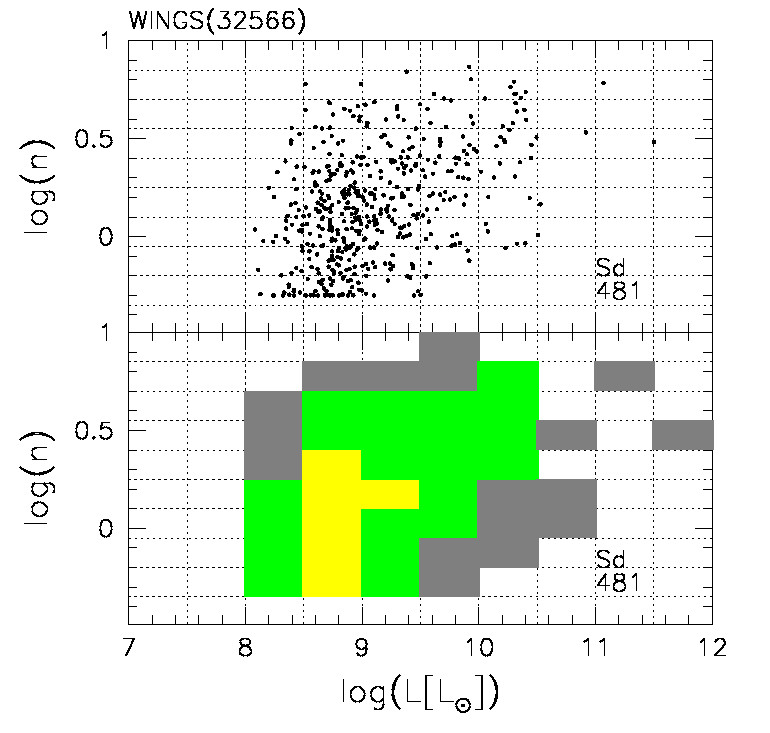}
\includegraphics[width=8.0cm]{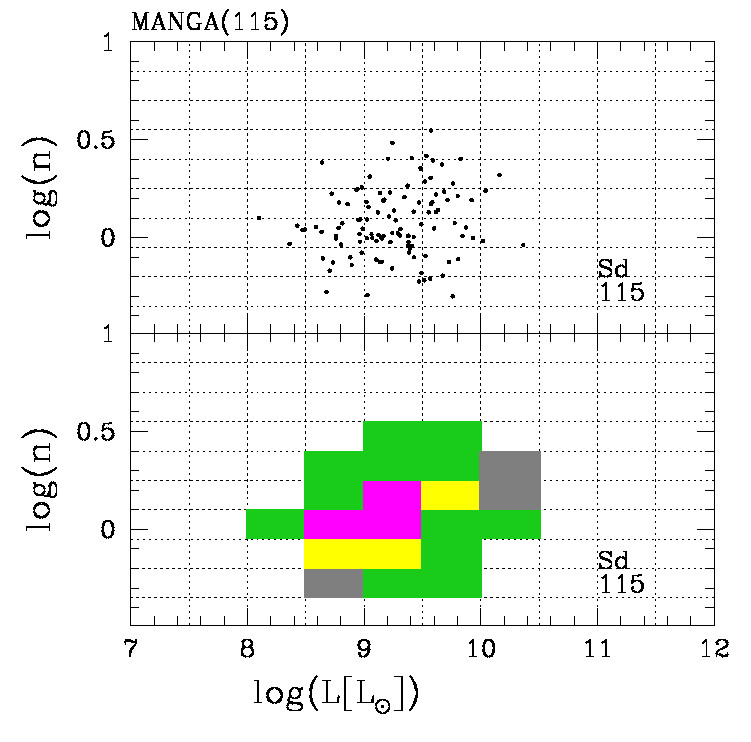}
\end{adjustwidth}
\caption{The \nL\ plane for Sc (top panels) and Sd (bottom panels) galaxies: (\textbf{a}) The top panel shows with black dots the whole distribution of galaxies of the WINGS survey classified as Sc. The bottom panel shows with different colors the areas with the large number density of objects given in percent. Red corresponds to regions where more than 20\% of the galaxies are found. Magenta gives the interval 10\% - 20\%, yellow 5\% - 10\%, green 1\% - 5\%, and gray 0.1\% - 1\%. (\textbf{b}) The same as in panel a) but for the Sc galaxies of the MANGA survey. (\textbf{c}) The same as in panel a) but for Sd galaxies of WINGS. (\textbf{d}) The same as in panel a) but for the Sd in MANGA. On top of the figure, we indicate the size of the whole sample  for both data-sets. Within the box, we provide the total number of galaxies of that morphological type. The dashed line marks the ZoE.\label{fig23}}
\end{figure} 

\begin{figure}[]
\begin{adjustwidth}{-\extralength}{-3cm}
\centering
\includegraphics[width=8.0cm]{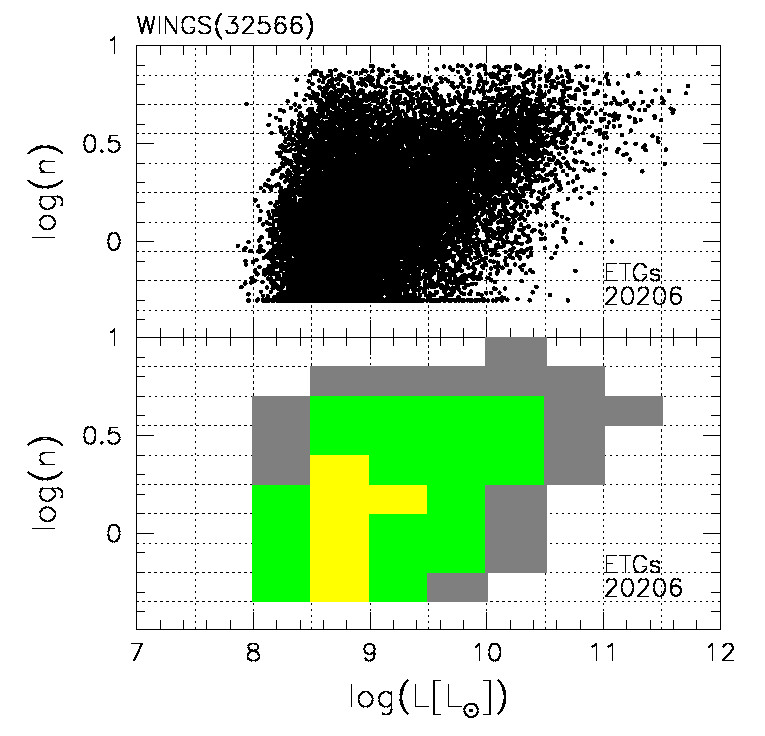}
\includegraphics[width=8.0cm]{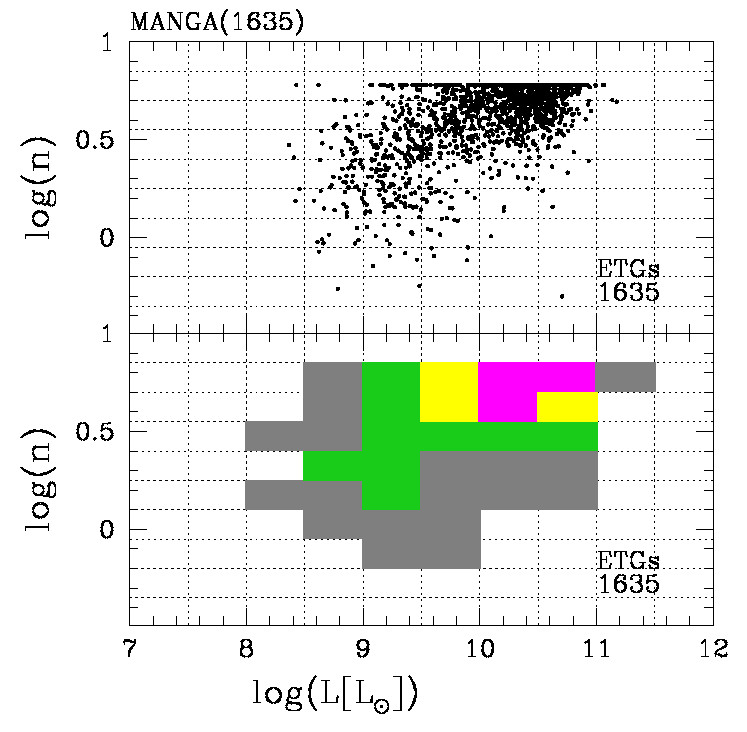}\\
\includegraphics[width=8.0cm]{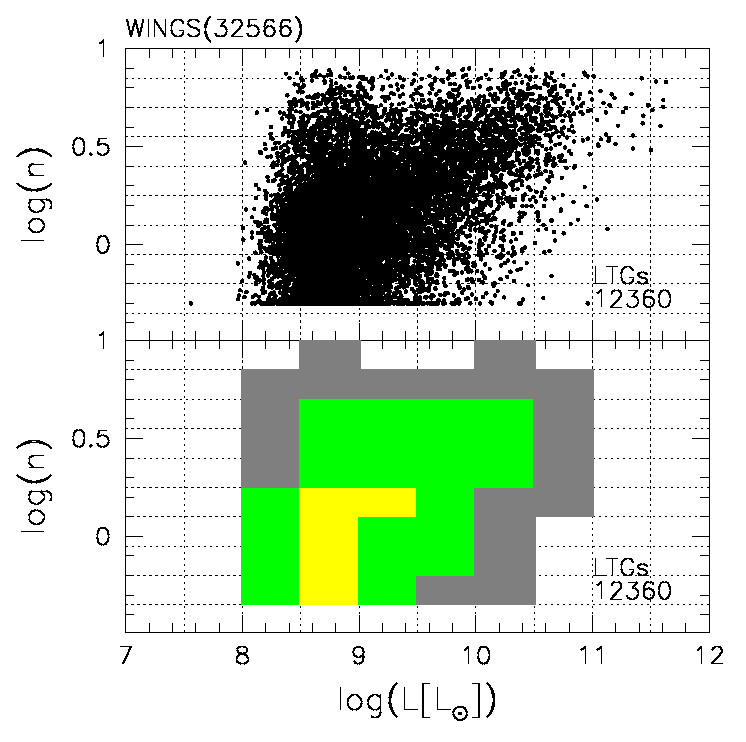}
\includegraphics[width=8.0cm]{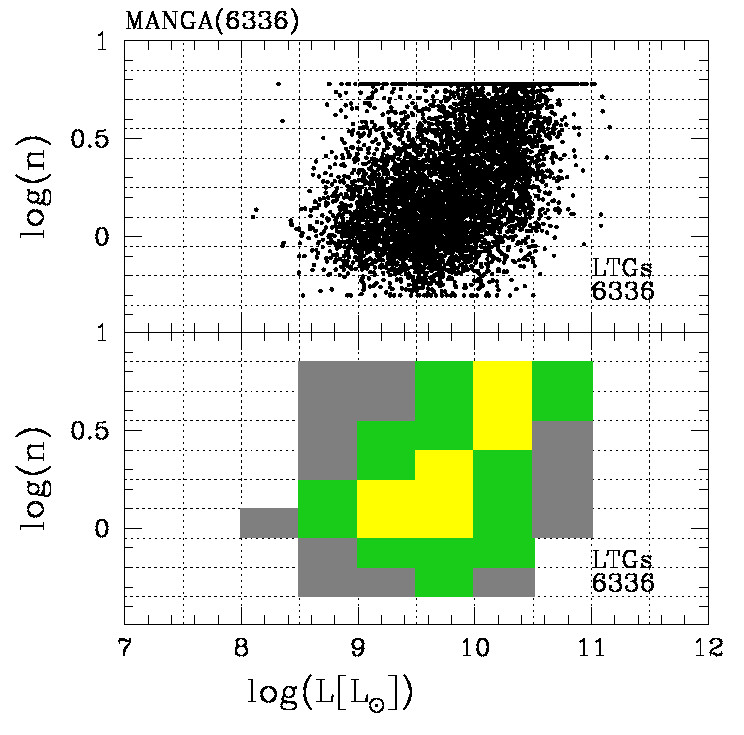}
\end{adjustwidth}
\caption{The \nL\ plane for ETGs (top panels) and LTGs (bottom panels) galaxies: (\textbf{a}) The top panel shows with black dots the whole distribution of galaxies of the WINGS survey classified as early-types. The bottom panel shows with different colors the areas with the large number density of objects given in percent. Red corresponds to regions where more than 20\% of the galaxies are found. Magenta gives the interval 10\% - 20\%,  yellow 5\% - 10\%, green 1\% - 5\%, and gray 0.1\% - 1\%. (\textbf{b}) The same as in panel a) but for the ETGs galaxies of the MANGA survey. (\textbf{c}) The same as in panel a) but for LTGs galaxies of WINGS. (\textbf{d}) The same as in panel a) but for LTGs in MANGA. On top of the figure, we indicate the size of the whole sample  for both data-sets. Within the box, we provide the total number of galaxies of that morphological type. The dashed line marks the ZoE.\label{fig24}}
\end{figure} 

\section{The \Lsig\ plane}\label{sec:8}

{The Faber–Jackson relation (FJ) \cite{FaberJackson1976} expresses the empirical correlation between the luminosity $L$ and the central stellar velocity dispersion $\sigma$ of a galaxy.} In logarithmic form we write $\log(L)=\alpha+\gamma\log(\sigma)$.
{For ETGs (ellipticals and bulges), the slope $\gamma$ is typically around 4 in the optical bands, though it varies with the sample, wavelength, and selection criteria }(values between 3 and 5 are commonly reported).
The \Lsig\ relation reflects the virial equilibrium of spheroidal systems and the connection between gravitational potential ($\sigma$) and stellar content ($L$).

Researchers used the \Lsig\ relation to probe the dynamical structure of galaxies, that is  to test if ellipticals are in virial equilibrium and how structural parameters (like $n$) affect it. {  Some studies addressed the evolution of galaxies with the redshift by looking at the shifts of the relation zero-points  that can trace the luminosity evolution and hence stellar population aging. Others  looked at  the evolution of the mass-to-light ratio by comparing of the \Lsig\ slope/offset at different $z$ (this gives the ratio $\Delta\log(M/L)$) over cosmic time). 
The \Lsig\ relation was also used to compare different environments, by investigating if cluster and field ellipticals obey the same scaling law (environmental dependence) and constraining the feedback and star formation histories (the slope variations across bands (e.g. B, V, K) indicate color–magnitude effects and metallicity trends). }

\cite{Schechter1980} first provided a theoretical explanation of the \Lsig\ relation based on virial equilibrium and explored the effect of $M/L$ variations. \cite{Guzmanetal1993} studied the validity of the relation for low-luminosity ellipticals, finding flatter slopes ($\gamma\sim2-3$) for dwarfs, i.e. structural non-homology. \cite{Jorgensenetal1996} analyzed the cluster ETGs at different redshifts ($z\sim0.02-0.8$) quantifying the evolution of $M/L$ from FJ offset. \cite{Treuetal2005} using field ETGs up to $z\sim1$ found faster $M/L$ evolution in low-mass galaxies (downsizing). \cite{Bernardietal2003} with massive SDSS galaxy samples quantified the slope dependence on luminosity, band, and environment.

\begin{figure}[]
\begin{adjustwidth}{-\extralength}{-3cm}
\centering
\includegraphics[width=8.0cm]{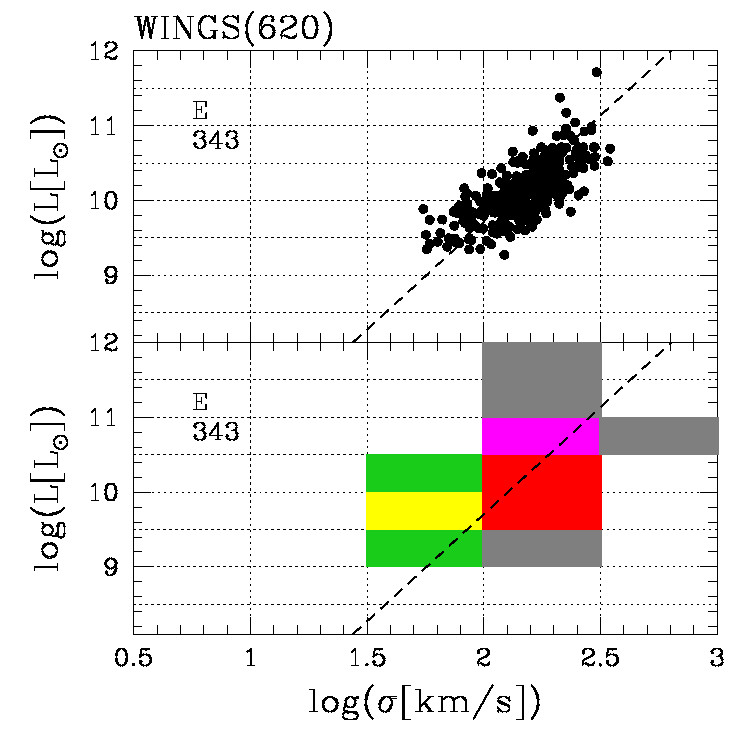}
\includegraphics[width=8.0cm]{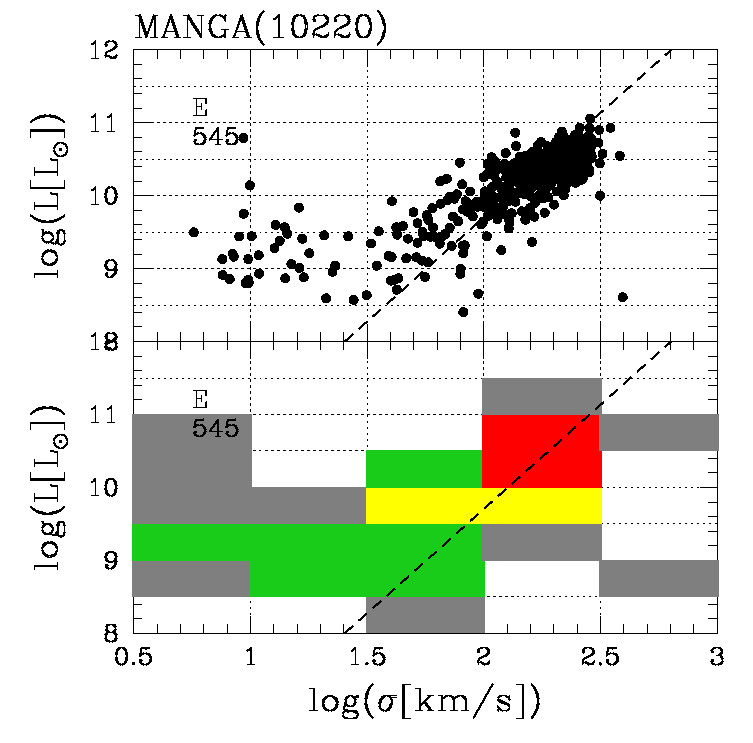}\\
\includegraphics[width=8.0cm]{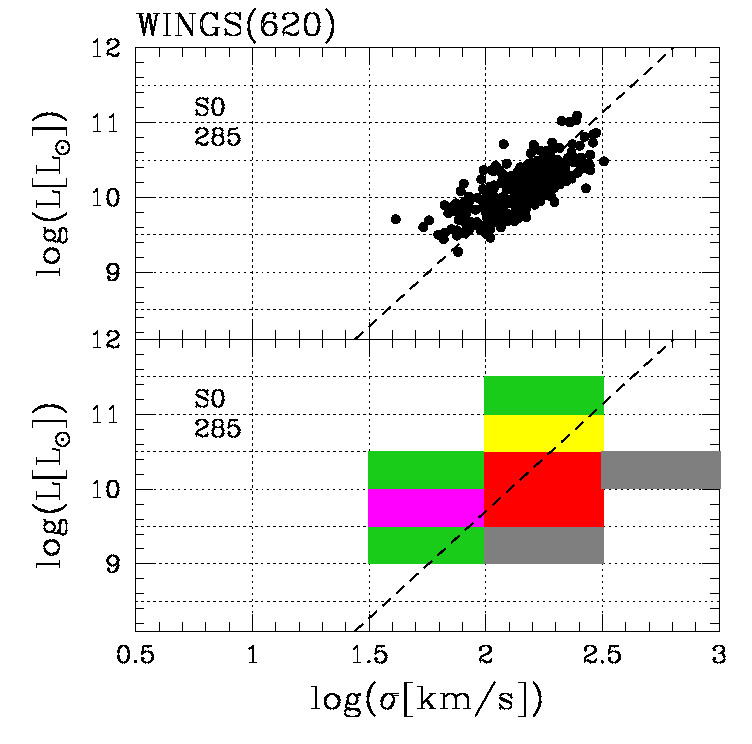}
\includegraphics[width=8.0cm]{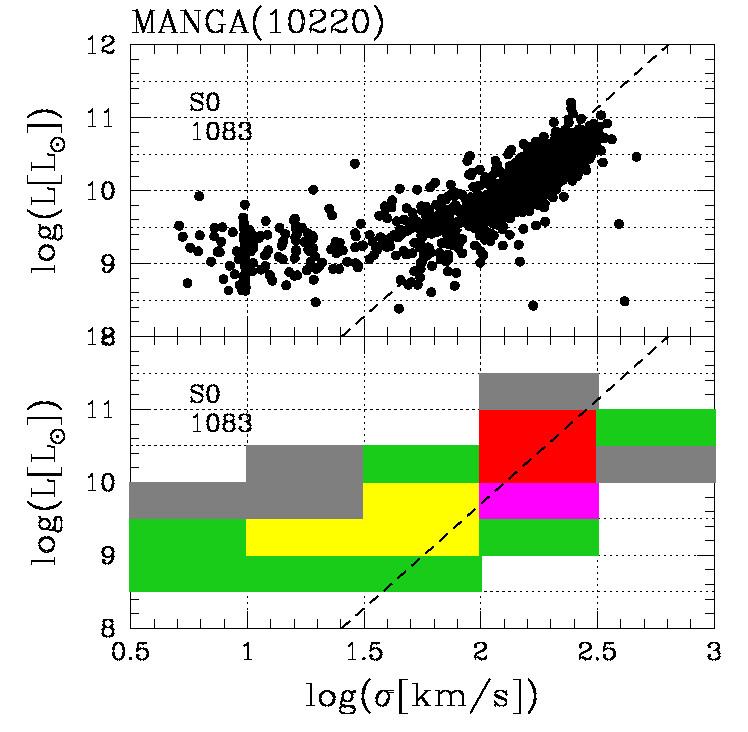}
\end{adjustwidth}
\caption{The \Lsig\ plane for E (top panels) and S0 (bottom panels) galaxies: (\textbf{a}) The top panel shows with black dots the whole distribution of galaxies of the WINGS survey classified as E. The bottom panel shows with different colors the areas with the large number density of objects given in percent. Red corresponds to regions where more than 20\% of the galaxies are found. Magenta gives the interval 10\% - 20\%, yellow 5\% - 10\%, green 1\% - 5\%, and gray 0.1\% - 1\%. (\textbf{b}) The same as in panel a) but for the E galaxies of the MANGA survey. (\textbf{c}) The same as in panel a) but for the S0 galaxies of WINGS. (\textbf{d}) The same as in panel a) but for the S0 in MANGA. On top of the figure, we indicate the size of the whole sample for both data-sets. Within the box, we provide the total number of galaxies of that morphological type. The dashed line marks the ZoE.\label{fig25}}
\end{figure}

WINGS {does not list data} for the central velocity dispersion of LTGs, so the comparison with MANGA {is limited to the ETGs only}. 
Figure \ref{fig25} shows the \Lsig\ plane for E and S0 galaxies. The two samples follow the same trend for objects with $\sigma>50$ \kms.
MANGA provides a long horizontal tail of galaxies of low luminosity that do not follow the main relation. This tail is present for all the morphological types. In particular Sc and Sd galaxies do not have objects that are distributed along the main trend of ETGs. The peaks of the distributions are somewhat  shifted toward high luminosity for the MANGA galaxies.

The existence of a curvature in the \Lsig\ relation has been debated several times (see e.g. \cite{Donofrioetal2024Univ}). MANGA suggests that {faint objects do not follow any \Lsig\ relationship.} These galaxies are largely supported by rotation and present very low velocity dispersions.

\begin{figure}[]
\begin{adjustwidth}{-\extralength}{-3cm}
\centering
\includegraphics[width=8.0cm]{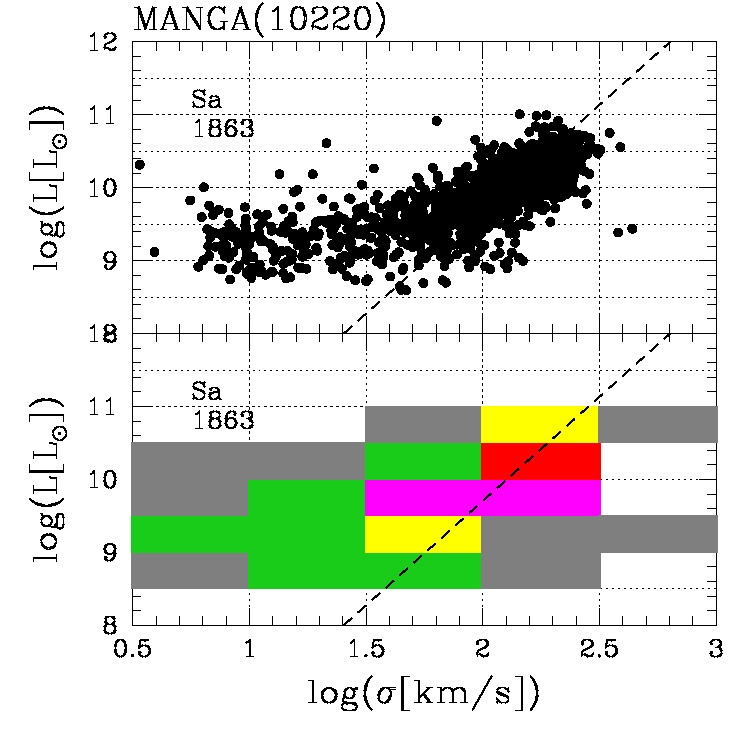}
\includegraphics[width=8.0cm]{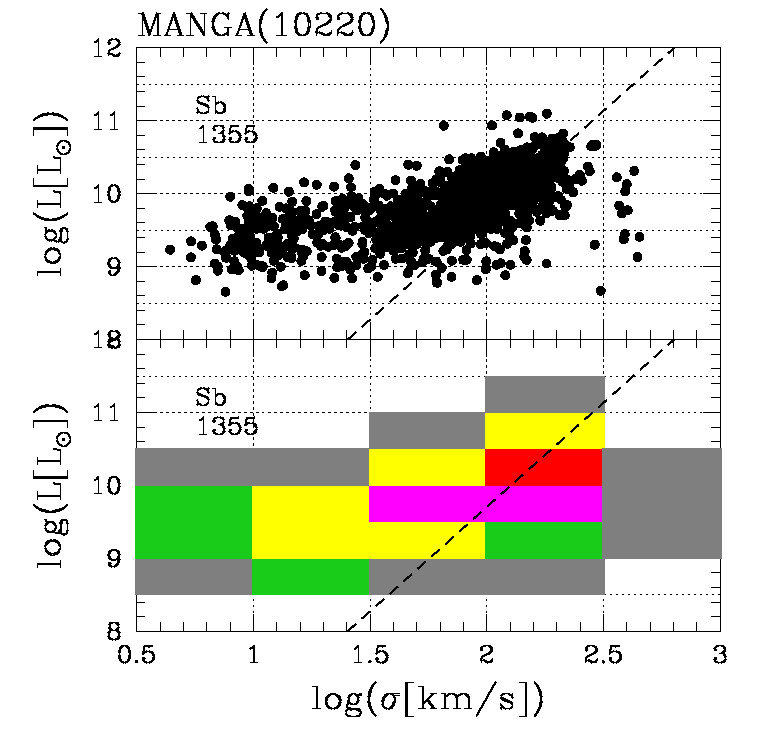}\\
\includegraphics[width=8.0cm]{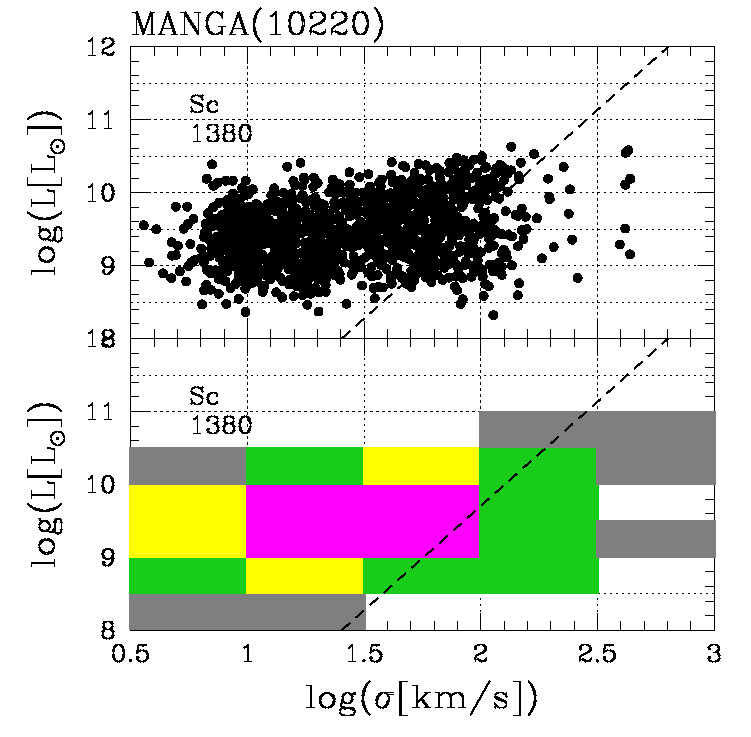}
\includegraphics[width=8.0cm]{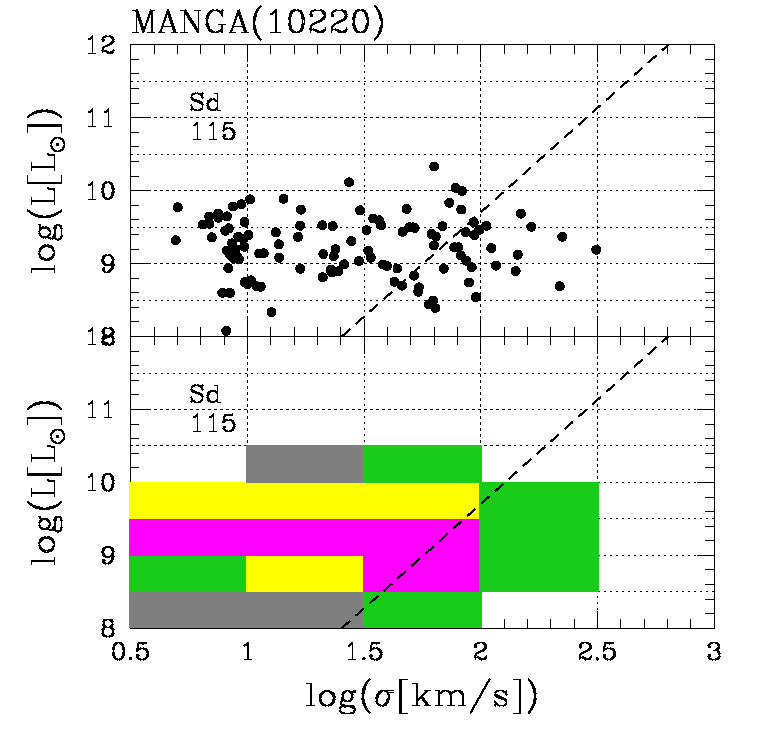}
\end{adjustwidth}
\caption{The \Lsig\ plane for Sa  and Sb galaxies (top panels) and Sc and Sd galaxies (bottom panels) of the MANGA survey: in all the panels, black dots mark the observed distribution. Colors mark the percentage of objects in each area, as in the above figures. \label{fig26}}
\end{figure} 

\section{Comparison with the Illustris simulation}\label{sec:9}

{In this section we address the comparison of WINGS and MANGA with two sets of data obtained by the Illustris simulation.
Ilustris simulation } is one of the landmark large-scale cosmological hydrodynamical simulations designed to study the formation and evolution of galaxies within a $\Lambda$CDM {  Universe}. Developed by the Illustris Collaboration (\cite{Vogelsberger_2014a, Vogelsberger_2014b, Genel_etal_2014, Nelsonetal2015}), it combines high-resolution gravity, hydrodynamics, and subgrid physics to follow the growth of cosmic structures {  from shortly after} the Big Bang to the present day.
Illustris evolves the coupled system of dark matter, gas, stars, supermassive black holes, and magnetic fields within a cosmological volume of (106.5 Mpc$^3$), using the moving-mesh hydrodynamics code AREPO. This numerical method allows for accurate and flexible treatment of gas flows, shocks, and mixing processes compared to traditional SPH or fixed-grid codes.

The simulation incorporates detailed subgrid models for key astrophysical processes, including star formation and stellar evolution,{ chemical enrichment and metal-line cooling, supernova feedback, black hole formation, accretion,  AGN feedback, and magneto-hydrodynamics.}
By capturing both large-scale cosmological structures and small-scale baryonic physics, Illustris qualitatively reproduces {  numerous observational properties of galaxies,} such as: the stellar mass–halo mass relation, the cosmic star formation history, galaxy morphologies and color distributions, gas fractions and metallicity trends.

Illustris was followed by the improved IllustrisTNG (The Next Generation) suite, which refined AGN feedback, chemical enrichment, and galaxy kinematics to better match observational constraints.

The first set of Illustris data analysed in this study is that of \cite{Rodriguez-Gomezetal2019} who provided the structural parameters of the TNG50 sample. The second one is that of \cite{Ferreiraetal2025}, a much larger sample of $\sim10100$ objects from the TNG100 data.
The galaxies of the first sample span the range of mass $\log(M^*/M_\odot)\sim 9.16\div12.4$ for objects with {redshift} $z<0.05$. The synthetic images were created with the SKIRT radiative transfer code. The simulation includes the effects of dust attenuation, scattering, and radiative transfer. The analysis, made with the aid of the  "Statmorph" code, provided a series of non-parametric morphological diagnostics parameters, such as the Gini–M20, the concentration–asymmetry–smoothness parameters, as well as the S\'ersic index, ellipticity, radii, and luminosity. In total, we got $\sim1500$ artificial galaxies that can be identified as ETGs or LTGs, on the basis of the morphological diagnostic parameters, and used them to build the ScRs analyzed here.

The second sample contains 10121 galaxies at redshifts $z<0.1$ from the TNG100-1 simulation. Most of them (more than 9000) are ETGs. For all of them radii, luminosities, velocity dispersions, and S\'ersic indeces were derived. Here the distinction between ETGs and LTGs is based only on the S\'ersic index ($n>2$ for ETGs). The range of mass is 10.17-11.96 dex in solar mass. The effective radius is circularized in both sets.

\begin{figure}[]
\begin{adjustwidth}{-\extralength}{-3cm}
\centering
\includegraphics[width=8.0cm]{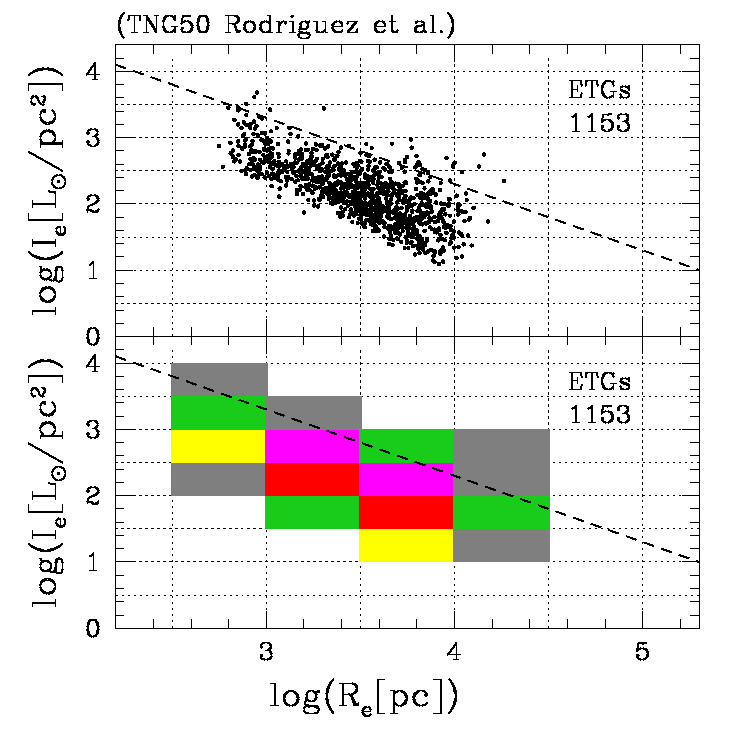}
\includegraphics[width=8.0cm]{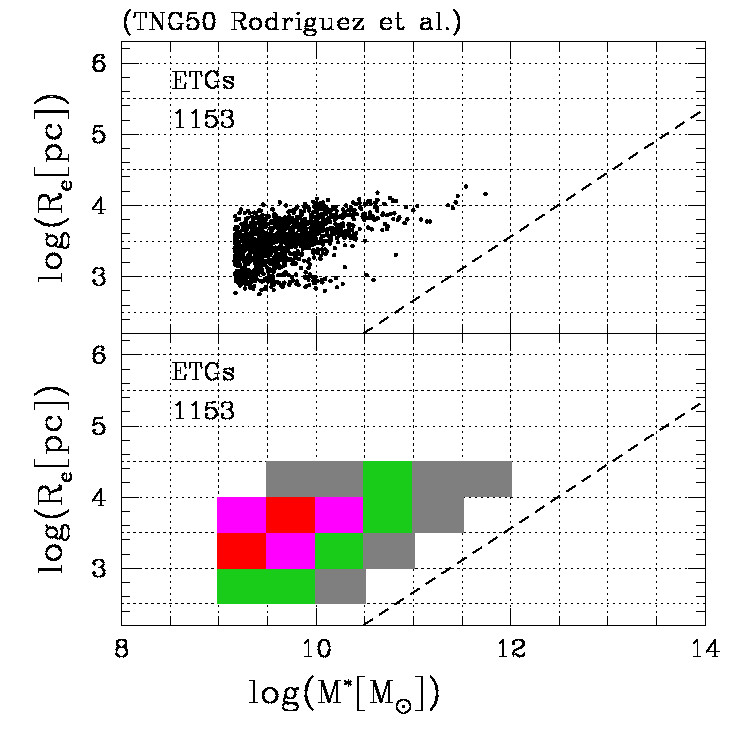}\\
\includegraphics[width=8.0cm]{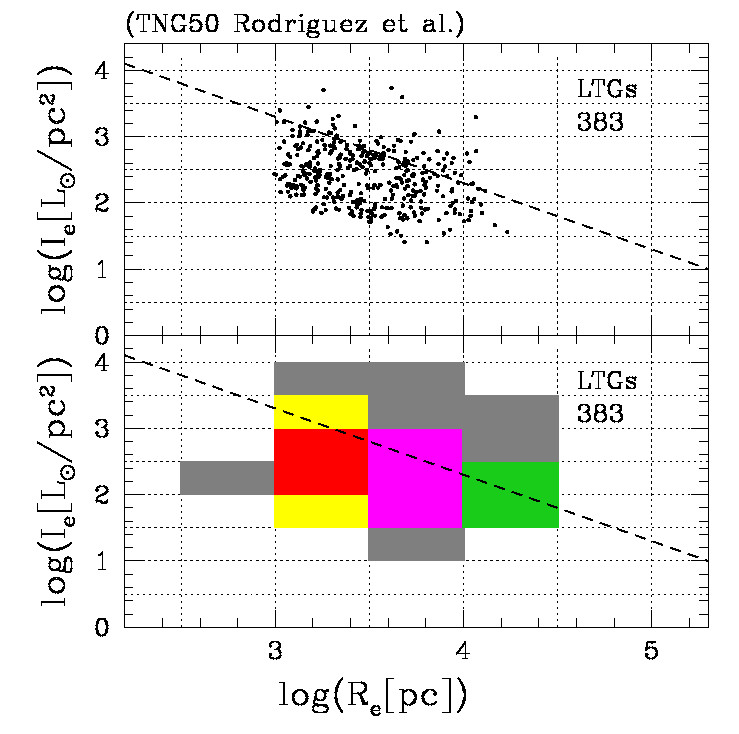}
\includegraphics[width=8.0cm]{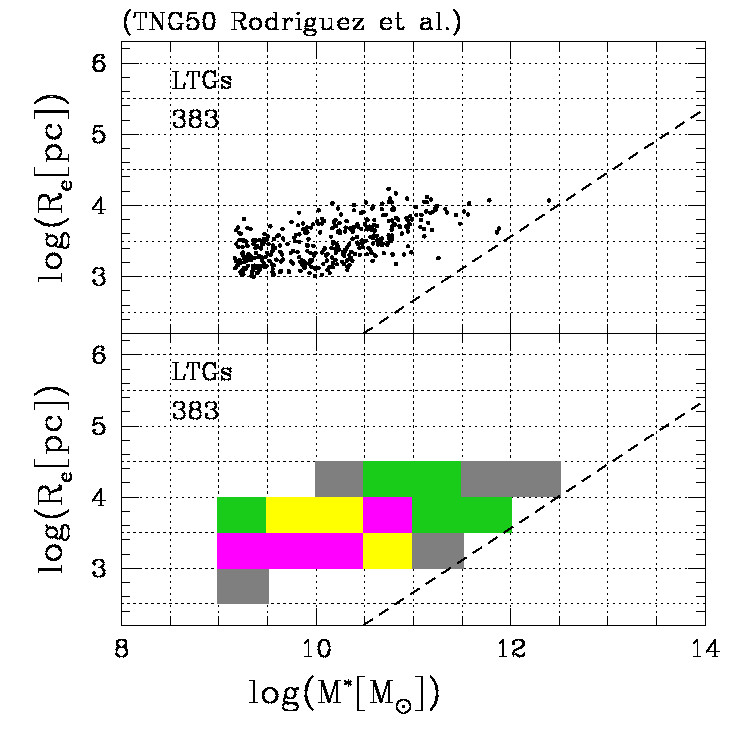}
\end{adjustwidth}
\caption{The \IeRe\ and \MRa\ planes for ETGs (top panels) and LTGs (bottom panels) for the simulated galaxies of \cite{Rodriguez-Gomezetal2019}: in all the panels, the black dots mark the observed distribution. Colors mark the percentage of objects in each area, as in the above figures. \label{fig27}}
\end{figure} 

\begin{figure}[]
\begin{adjustwidth}{-\extralength}{-3cm}
\centering
\includegraphics[width=8.0cm]{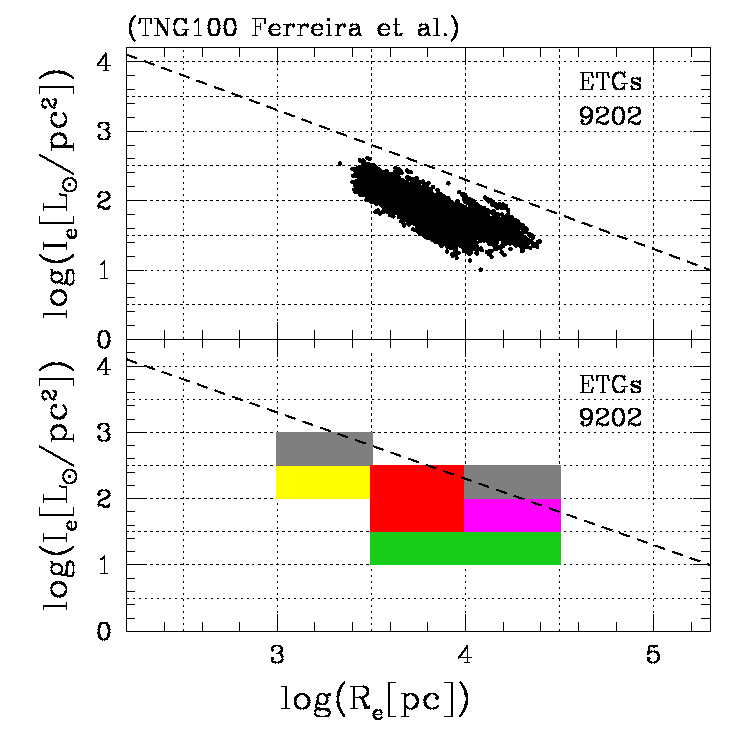}
\includegraphics[width=8.0cm]{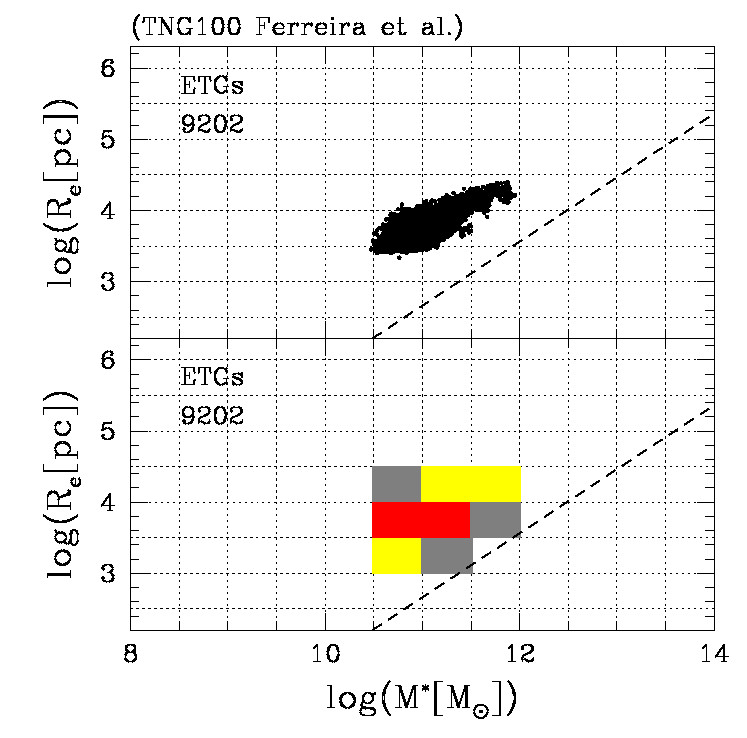}\\
\includegraphics[width=8.0cm]{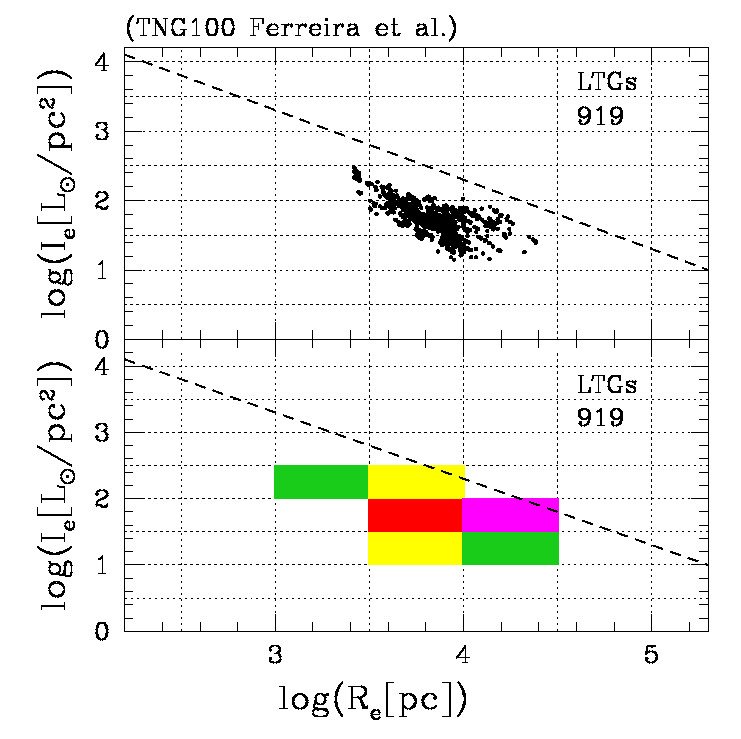}
\includegraphics[width=8.0cm]{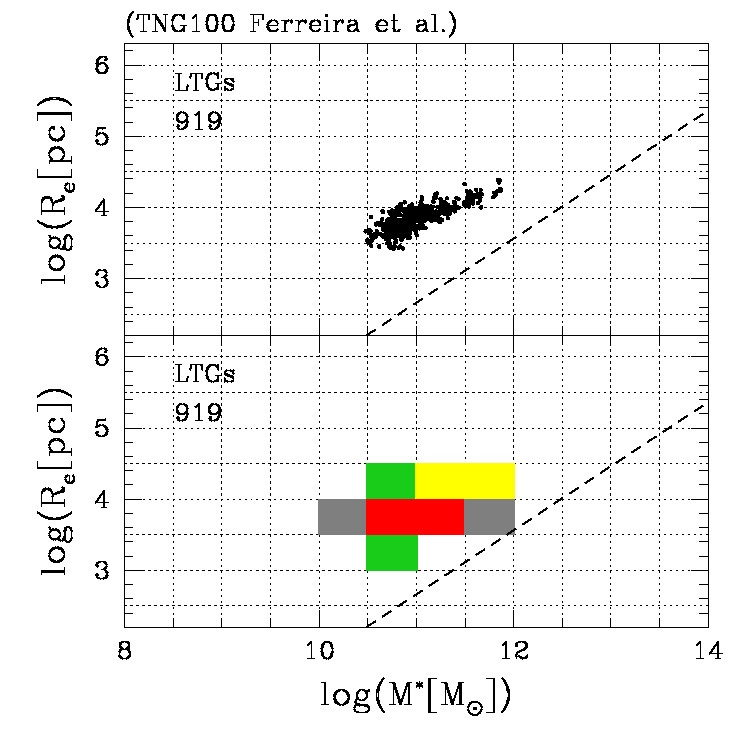}
\end{adjustwidth}
\caption{The \IeRe\ and \MRa\ planes for ETGs (top panels) and LTGs (bottom panels) for the simulated galaxies of \cite{Ferreiraetal2025}: in all the panels, black dots mark the observed distribution. Colors mark the percentage of objects in each area, as in the above figures. \label{fig28}}
\end{figure} 

Figure \ref{fig27} shows the \IeRe\ (left panel) and \MRa\ plane (right panel) for the first set of artificial galaxies. As before, the lower panels provide the percentage of objects in each box interval. The upper panels give the trend for ETGs, the lower panels that of LTGs. 

Qualitatively one can say that simulations cover approximately the same area of real galaxies in both diagrams.

Figure \ref{fig28} is similar to Fig. \ref{fig27}, but for the second set of artificial galaxies. Now the range covered by simulations appears a bit smaller. {  Both ETGs and LTGs } are confined in a short range of radii, brightnesses,  and masses.

\begin{figure}[]
\begin{adjustwidth}{-\extralength}{-3cm}
\centering
\includegraphics[width=8.0cm]{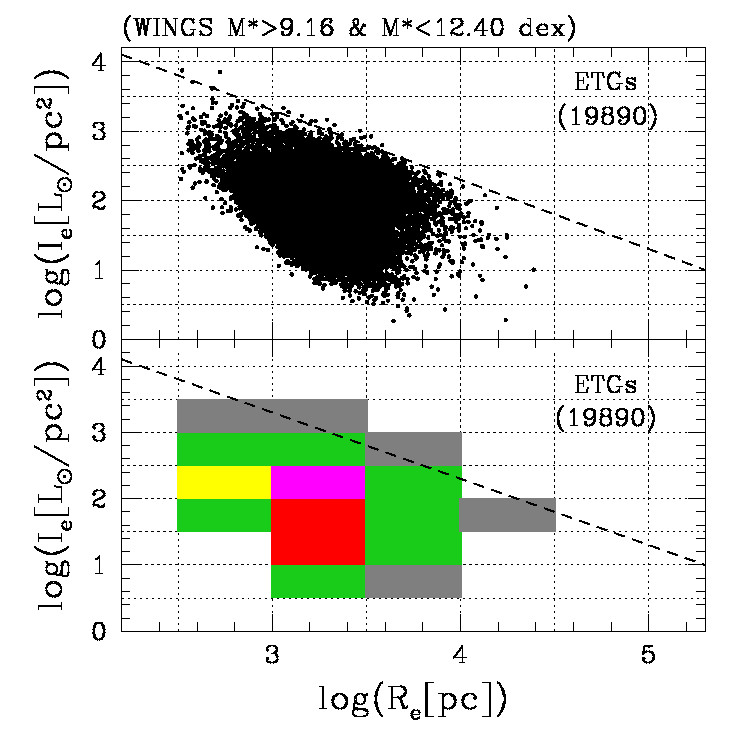}
\includegraphics[width=8.0cm]{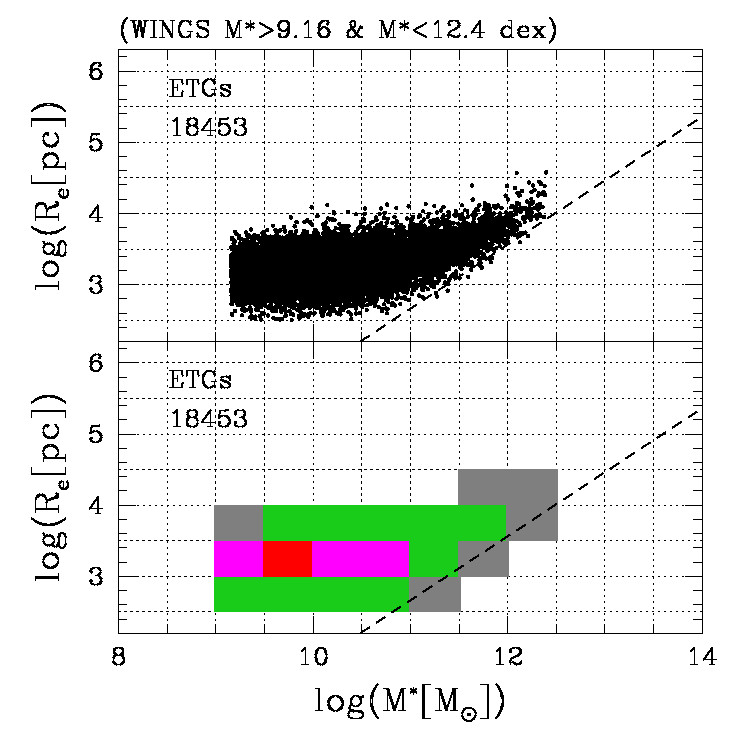}\\
\includegraphics[width=8.0cm]{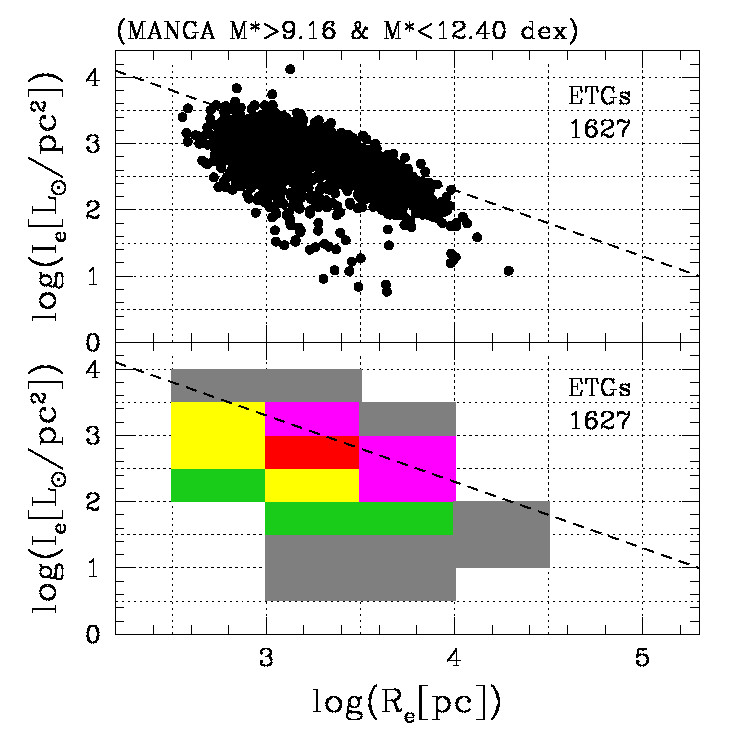}
\includegraphics[width=8.0cm]{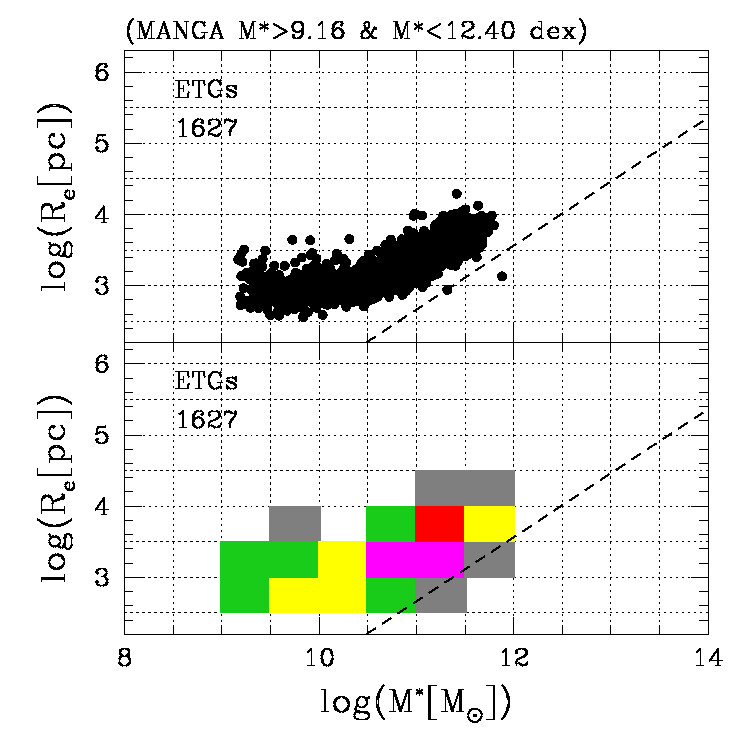}
\end{adjustwidth}
\caption{The \IeRe\ (left) and \MRa\ (right) planes for ETGs of the WINGS survey (top panels) and MANGA survey (bottom panels). The limits in stellar mass are those of the simulated galaxies of \cite{Rodriguez-Gomezetal2019}: in all the panels black dots mark the observed distribution. Colors mark the percentage of objects in each area, as in the above figures. \label{fig29}}
\end{figure}

{To compare observations with simulations} we need to limit the mass interval of WINGS and MANGA to that of artificial galaxies. Figure \ref{fig29} presents the \IeRe\ and \MRa\ plane when the mass range of the first dataset is adopted. The WINGS data for ETGs and LTGs are in the upper panels. MANGA is in the lower panels.

Apart from the larger number of objects, if we look at the lower panels with the colored boxes giving the percentage of objects in them, we note that  WINGS and MANGA  distributions are very different.

\begin{figure}[]
\begin{adjustwidth}{-\extralength}{-3cm}
\centering
\includegraphics[width=8.0cm]{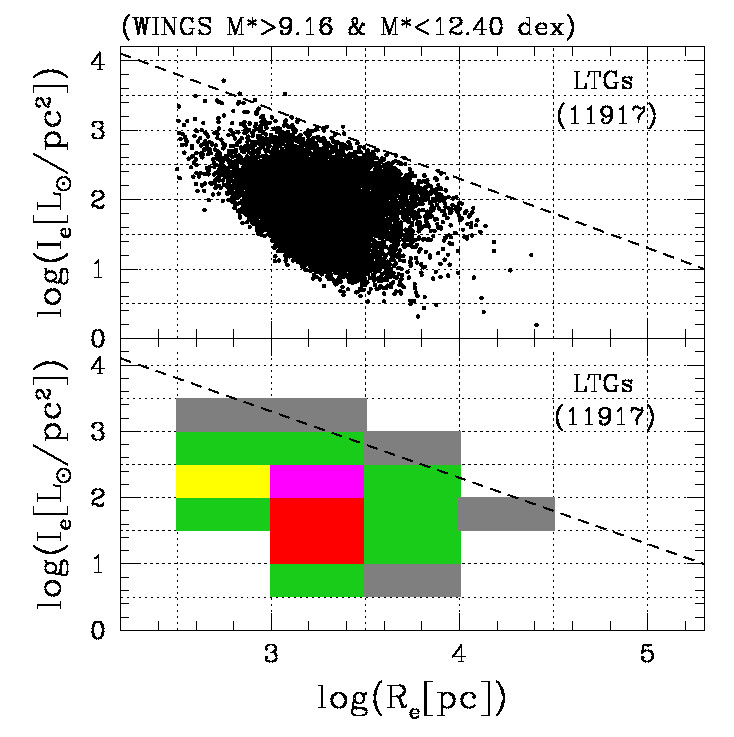}
\includegraphics[width=8.0cm]{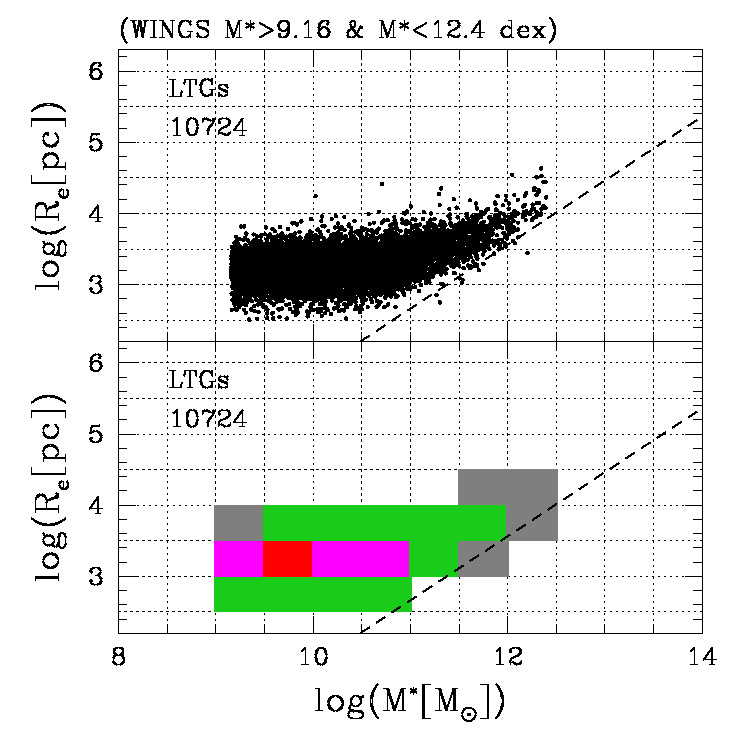}\\
\includegraphics[width=8.0cm]{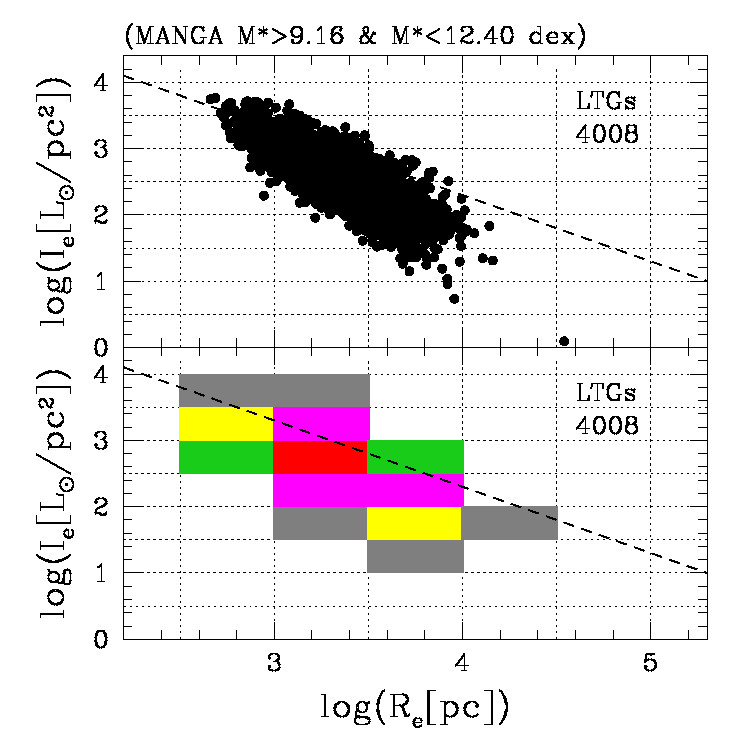}
\includegraphics[width=8.0cm]{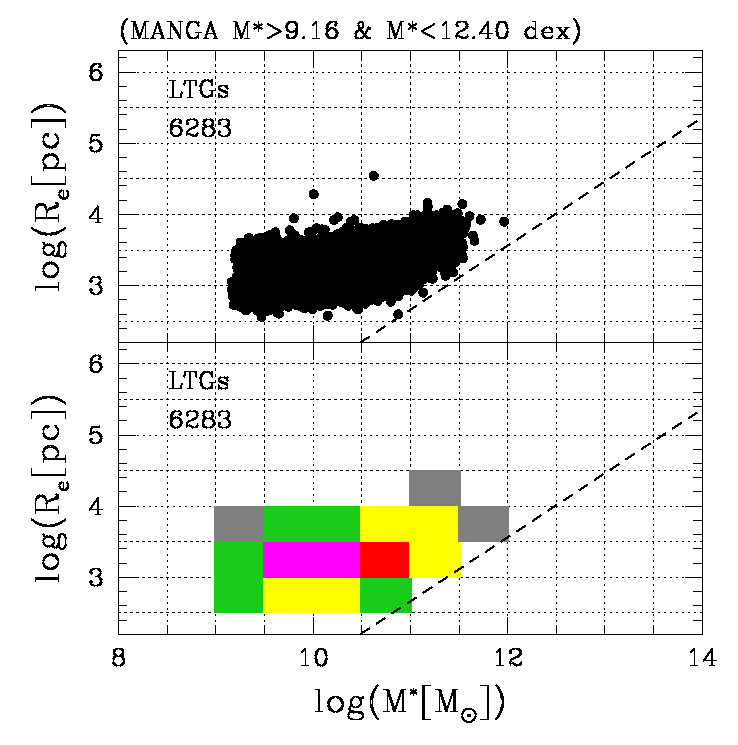}
\end{adjustwidth}
\caption{The \IeRe\ (left) and \MRa\ (right) planes for LTGs of the WINGS survey (top panels) and MANGA survey (bottom panels). The limits in stellar mass are those of the simulated galaxies of \cite{Rodriguez-Gomezetal2019}: in all the panels black dots mark the observed distribution. Colors mark the percentage of objects in each area, as in the above figures. \label{fig30}}
\end{figure}

Figure \ref{fig30} {shows the same, but limiting to the mass interval to that of the second set of simulations. Also in this case, } the percentage 2D distribution is different. Simulations reproduce the ScRs only qualitatively. The density of galaxies in each region of the diagrams is not reproduced.

\begin{figure}[]
\begin{adjustwidth}{-\extralength}{-3cm}
\centering
\includegraphics[width=8.0cm]{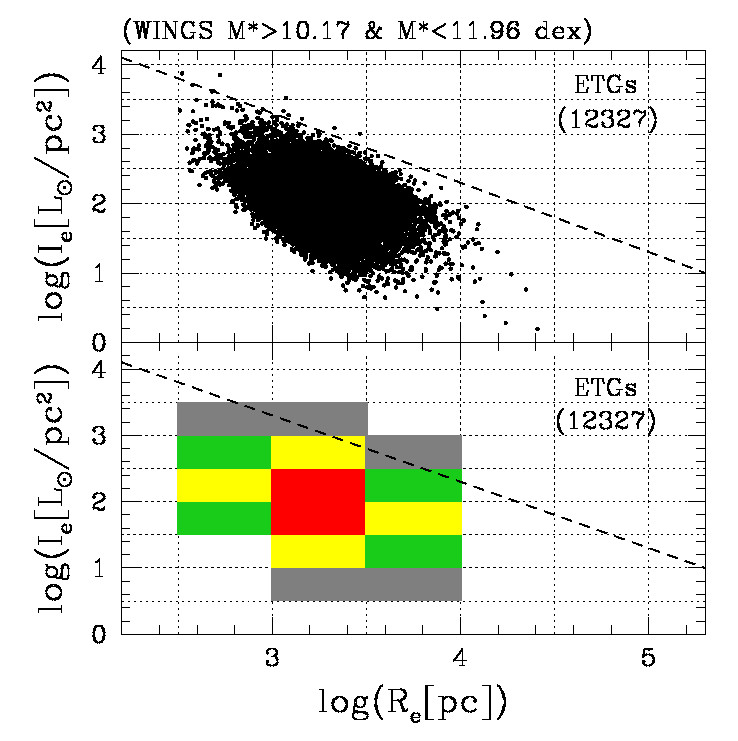}
\includegraphics[width=8.0cm]{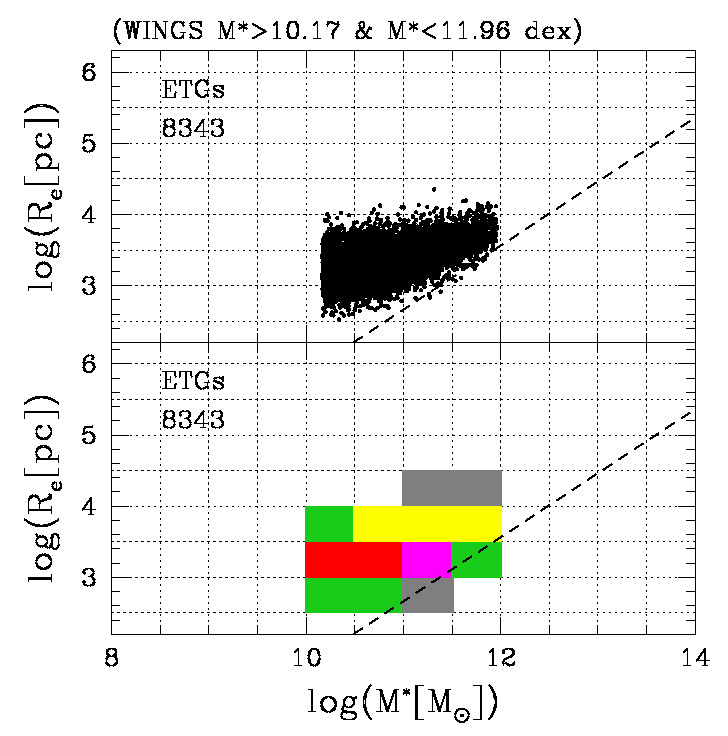}\\
\includegraphics[width=8.0cm]{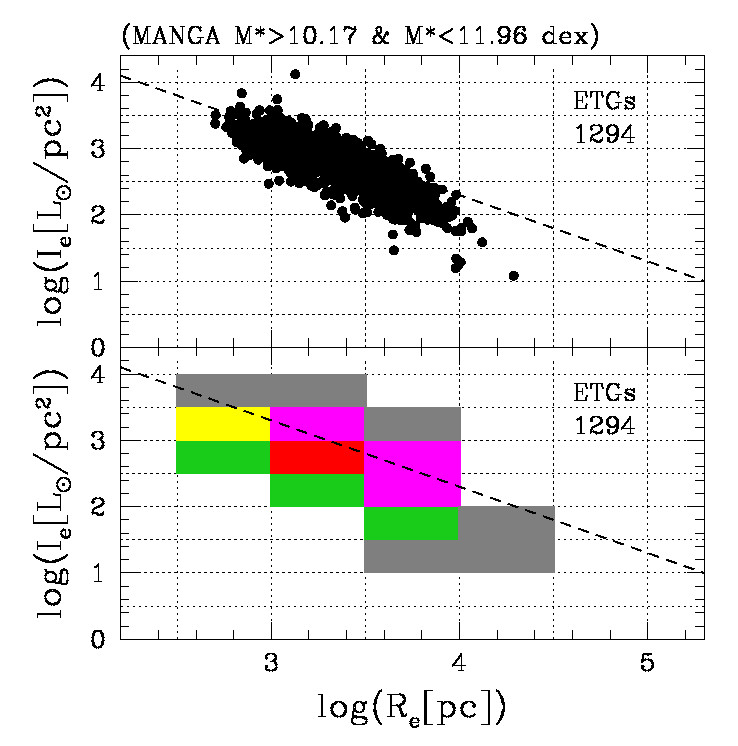}
\includegraphics[width=8.0cm]{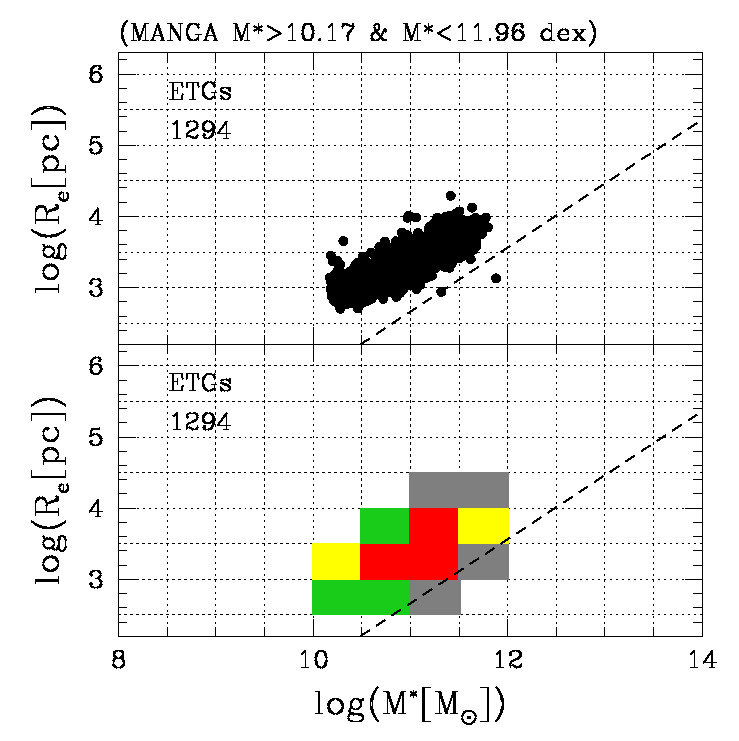}
\end{adjustwidth}
\caption{The \IeRe\ (left) and \MRa\ (right) planes for ETGs of the WINGS survey (top panels) and MANGA survey (bottom panels). The limits in stellar mass are those of the simulated galaxies of \cite{Ferreiraetal2025}: in all the panels, black dots mark the observed distribution. Colors mark the percentage of objects in each area, as in the above figures. \label{fig31}}
\end{figure} 

\begin{figure}[]
\begin{adjustwidth}{-\extralength}{-3cm}
\centering
\includegraphics[width=8.0cm]{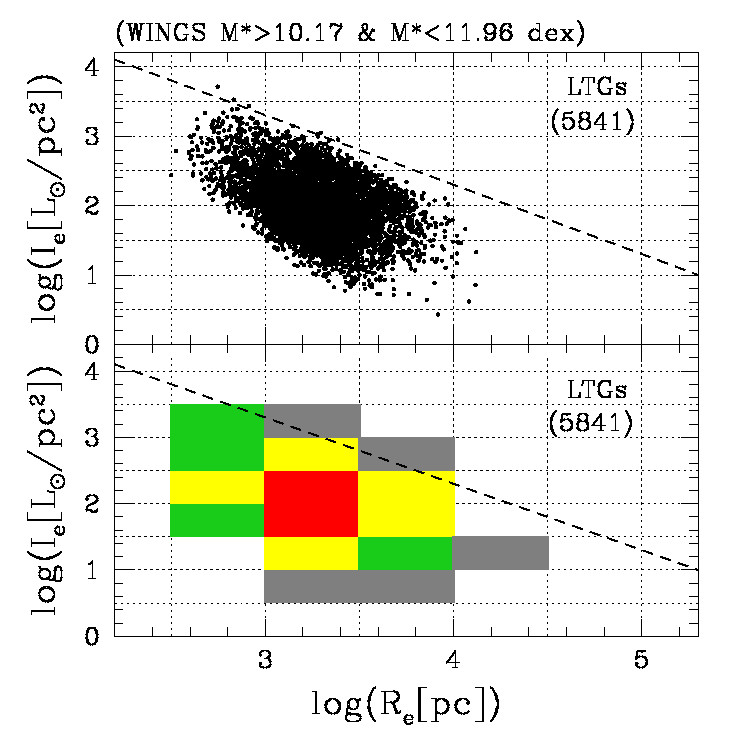}
\includegraphics[width=8.0cm]{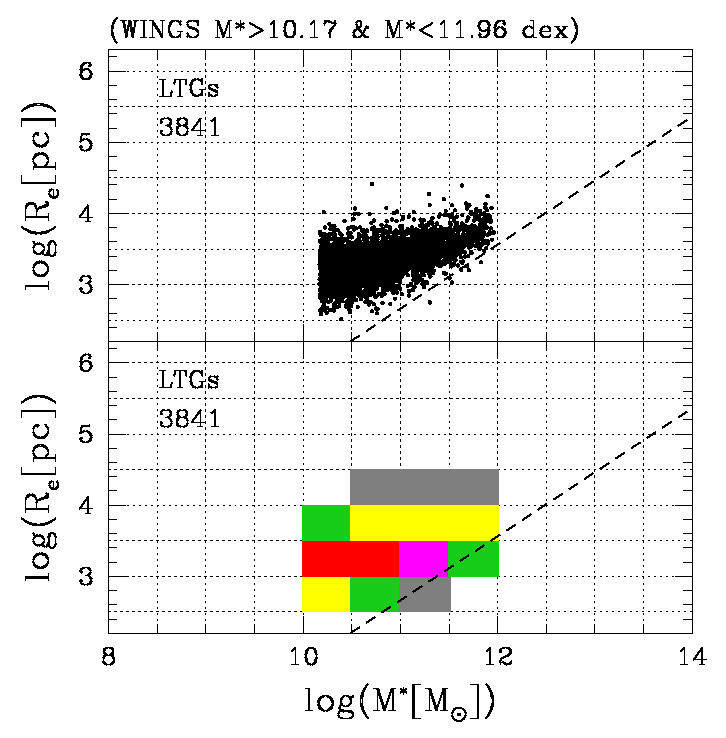}\\
\includegraphics[width=8.0cm]{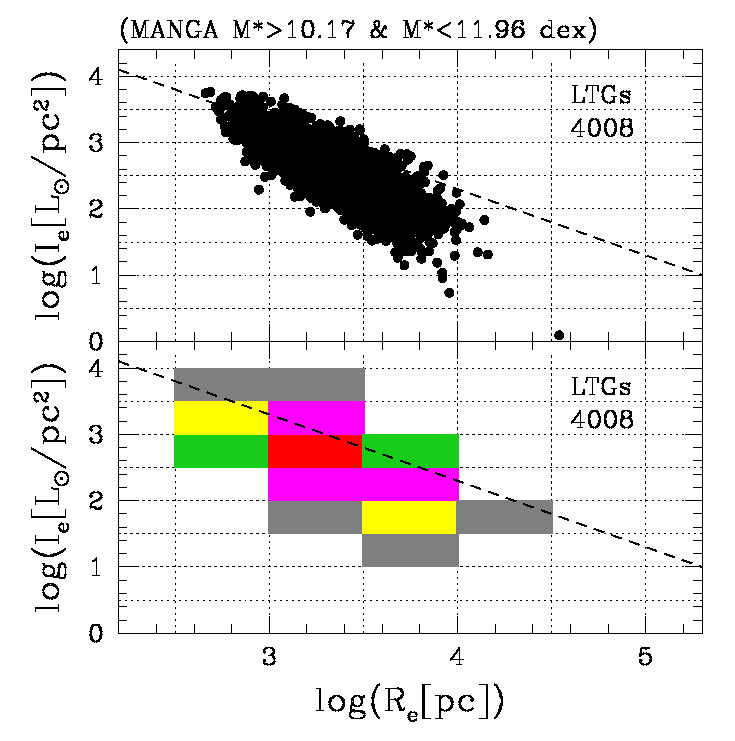}
\includegraphics[width=8.0cm]{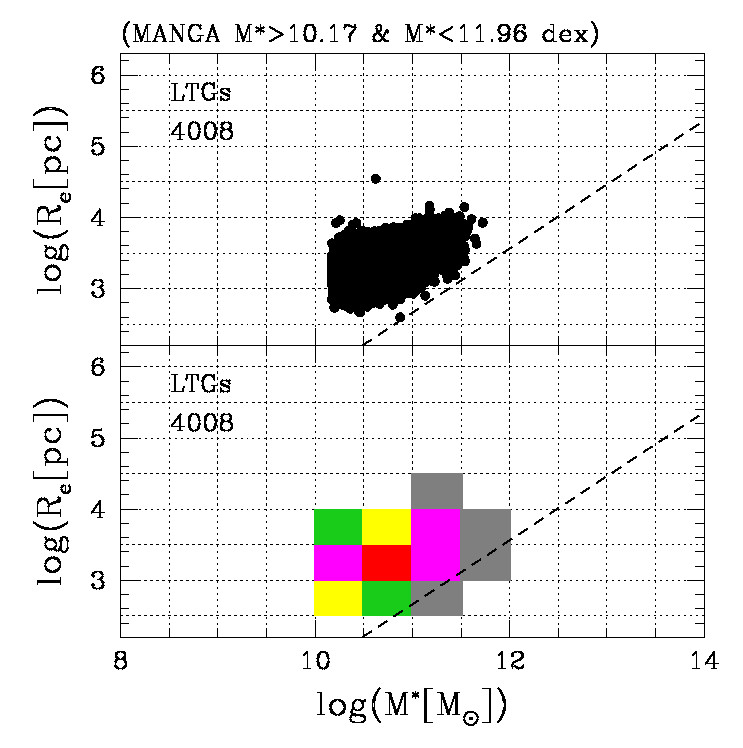}
\end{adjustwidth}
\caption{The \IeRe\ (left) and \MRa\ (right) planes for LTGs of the WINGS survey (top panels) and MANGA survey (bottom panels). The limits in stellar mass are those of the simulated galaxies of \cite{Ferreiraetal2025}: in all the panels, black dots mark the observed distribution. Colors mark the percentage of objects in each area, as in the above figures. \label{fig32}}
\end{figure} 

\begin{figure}[]
\begin{adjustwidth}{-\extralength}{-3cm}
\centering
\includegraphics[width=8.0cm]{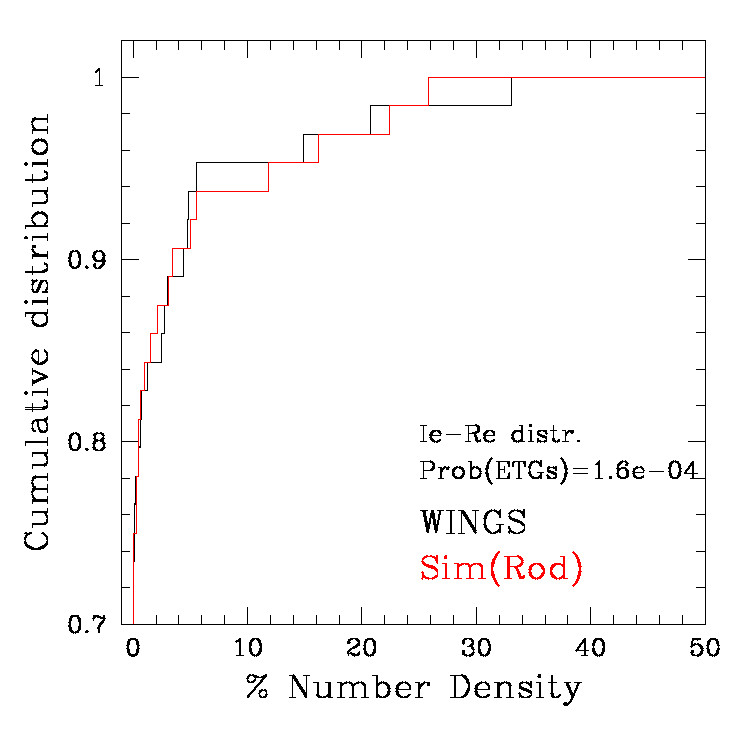}
\includegraphics[width=8.0cm]{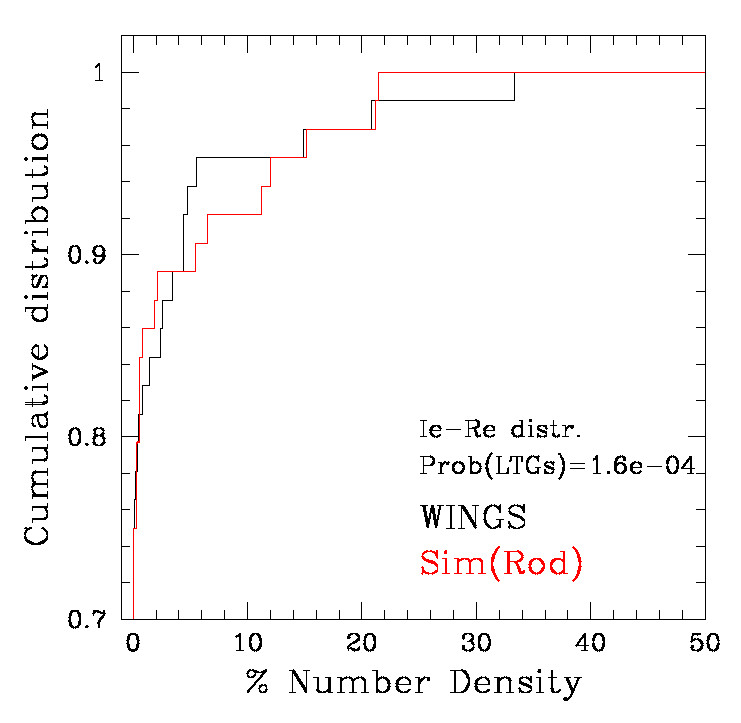}\\
\includegraphics[width=8.0cm]{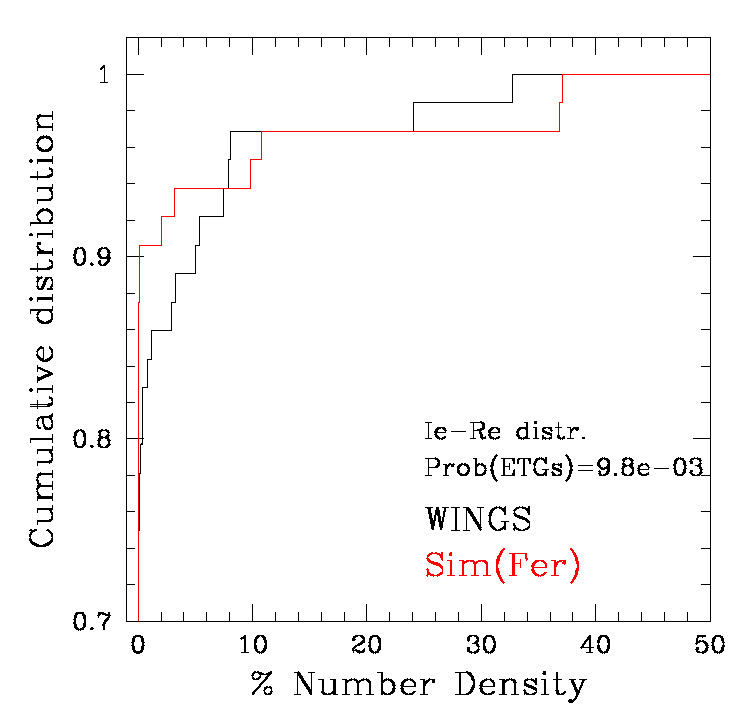}
\includegraphics[width=8.0cm]{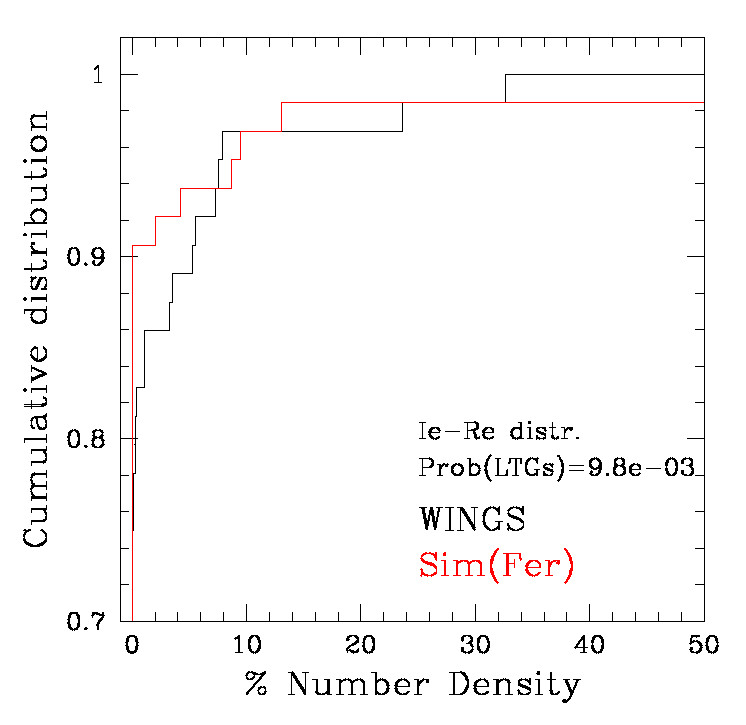}
\end{adjustwidth}
\caption{Kolmogorov-Smirnov test of the percentage distribution of galaxies in the \IeRe\ plane. The upper panels show the cumulative distribution vs the percentage for ETGs (left) and LTGs (right) of the WINGS galaxies (black line) and the simulated galaxies of \cite{Rodriguez-Gomezetal2019} (red lines). The lower panels do the same for the comparison between WINGS and \cite{Ferreiraetal2025}. \label{fig33}}
\end{figure}

\begin{figure}[]
\begin{adjustwidth}{-\extralength}{-3cm}
\centering
\includegraphics[width=8.0cm]{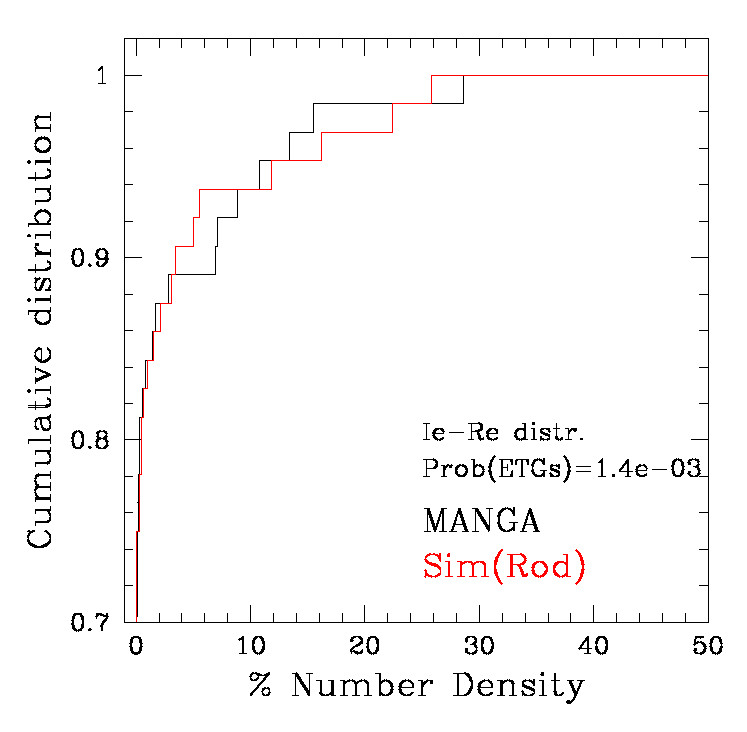}
\includegraphics[width=8.0cm]{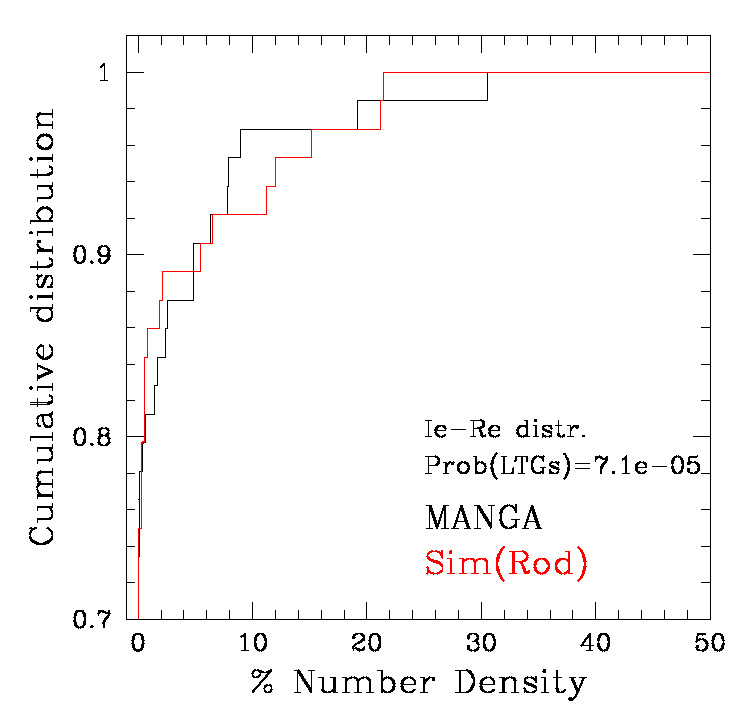}\\
\includegraphics[width=8.0cm]{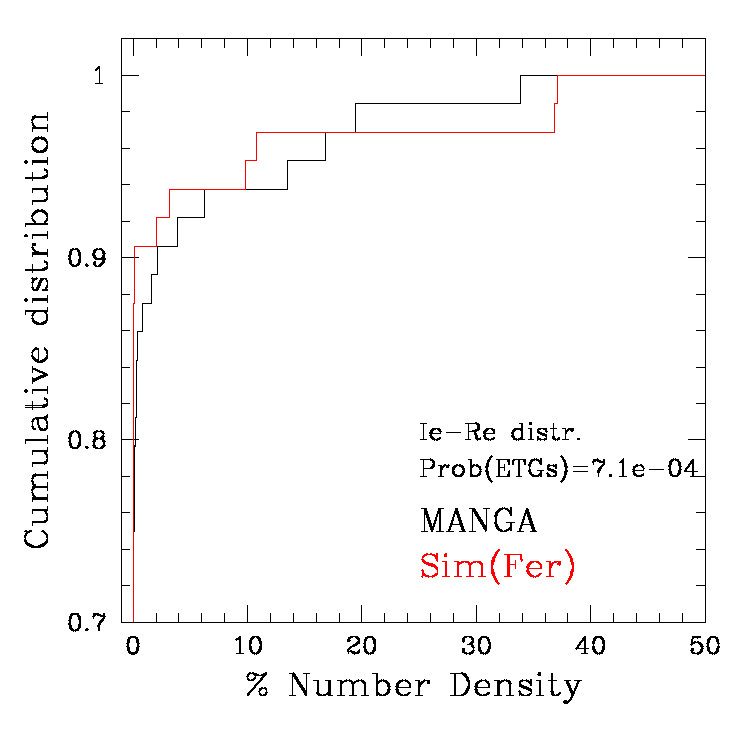}
\includegraphics[width=8.0cm]{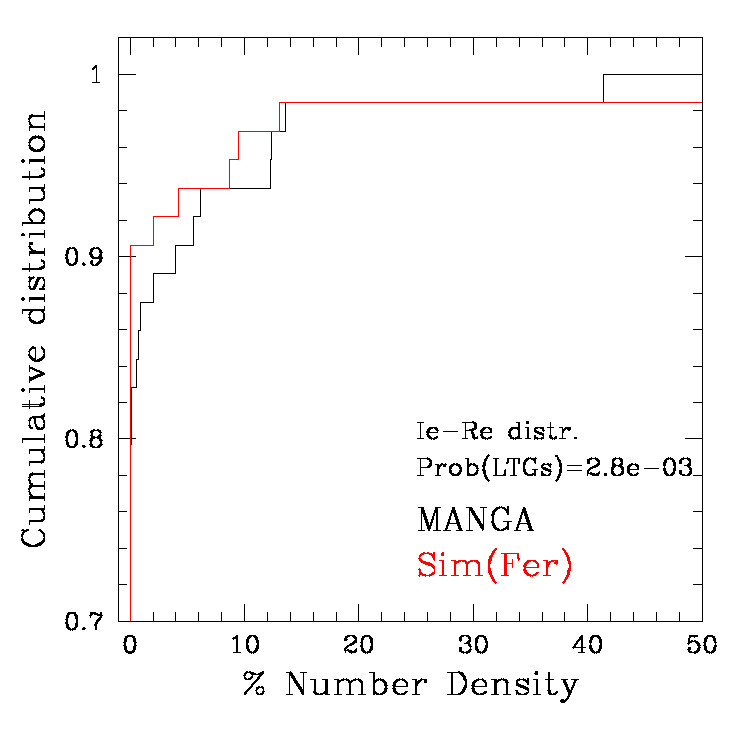}
\end{adjustwidth}
\caption{Kolmogorov-Smirnov test of the percentage distribution of galaxies in the \IeRe\ plane. The upper panels show the cumulative distribution vs the percentage for ETGs (left) and LTGs (right) of the MANGA galaxies (black line) and the simulated galaxies of \cite{Rodriguez-Gomezetal2019} (red lines). The lower panels do the same for the comparison between MANGA and \cite{Ferreiraetal2025}. \label{fig34}}
\end{figure}

A more quantitative comparison {  of the number density distribution given by data of real galaxies with the ones predicted by the simulations  is presented in  Fig. \ref{fig33}. }The figure shows the results of a Kolmogorov-Smirnov test between the two vectors containing the percentage distribution of galaxies in each area of the \IeRe\ plane, obtained for the WINGS galaxies and the simulated galaxies. The upper and lower panels give the test with respect to the first and second sets of simulations respectively. The comparison is made by selecting only the WINGS galaxies that are in the same mass range covered by the two simulations.{   It is soon evident } that in all cases the probability that the two samples (real and simulated galaxies) are extracted from the same population is very low.
A similar result is obtained in the comparison of the MANGA data with the simulated data (Fig. \ref{fig34}). 

The same conclusion  is true when we compare the distribution of real and simulated galaxies in  other ScRs. {  In other words, the lack of a reference sample does not allow us to statistically significantly verify the ability of the simulations to reproduce the experimental data.
The comparison is limited to a qualitative evaluation. 
Simulations should not only reproduce the observed trends in the ScRs, but also fit the numbers of objects detected (expected) in each region of the whole 2D distribution. Ideally, the percentage of galaxies in each region of the planes should be accounted for.
This requires comparisons of similar volumes of the Universe, with galaxies in similar range of masses and luminosities. }

\begin{table}[]
    \centering
    \begin{tabular}{|c|c|c|c|}
    \hline
    \multicolumn{4}{|c|}{ETGs: WINGS vs TNG50} \\
    \hline
    Lum. interval &  Weight. Energ. Dist. &  Nr. of Gal. &   Permutation p-val. \\
    \hline\hline
     8-12  & 0.179  & 20568-1153 & 0.545 \\
     9-11  & $7.7\times10^{-2}$  & 11660-416 & 1.00 \\
10.0-10.5  & 0.128  & 1954-70 & 1.00 \\
    \hline
    
    \hline
    \multicolumn{4}{|c|}{LTGs: WINGS vs TNG50} \\
    \hline
    Lum. interval &  Weight. Energ. Dist. &  Nr. of Gal. &   Permutation p-val. \\
    \hline\hline
     8-12  & 0.274  & 12436-383 & 0.182 \\
     9-11  & 0.316  & 4395-162 & $9.0\times10^{-2}$ \\
10.0-10.5  & 0.275  & 481-16 & 0.363 \\
    \hline   

    \hline
    \multicolumn{4}{|c|}{ETGs: WINGS vs TNG100} \\
    \hline
    Lum. interval &  Weight. Energ. Dist. &  Nr. of Gal. &   Permutation p-val. \\
    \hline\hline
     8-12  & 0.482  & 20568-9198 & $9.0\times10^{-2}$ \\
     9-11  & 0.421  & 11160-4177 & 0.454 \\
10.0-10.5  & 0.321  & 1954-623 & 0.636 \\
    \hline   

    \hline
    \multicolumn{4}{|c|}{LTGs: WINGS vs TNG100} \\
    \hline
    Lum. interval &  Weight. Energ. Dist. &  Nr. of Gal. &   Permutation p-val. \\
    \hline\hline
     8-12  & 0.594  & 12436-919 & 0.182 \\
     9-11  & 0.589  & 4395-399 & 0.182 \\
10.0-10.5  & 0.593  & 481-60 & $9.0\times10^{-2}$ \\
    \hline   

    \multicolumn{4}{|c|}{ETGs: MANGA vs TNG50} \\
    \hline
    Lum. interval &  Weight. Energ. Dist. &  Nr. of Gal. &   Permutation p-val. \\
    \hline\hline
     8-12  & 0.309  & 1635-1153 & 0.272 \\
     9-11  & 0.336  & 1156-1099 & 0.272 \\
10.0-10.5  & 0.382  & 586-422 & 0.454 \\
    \hline
    
    \hline
    \multicolumn{4}{|c|}{LTGs: MANGA vs TNG50} \\
    \hline
    Lum. interval &  Weight. Energ. Dist. &  Nr. of Gal. &   Permutation p-val. \\
    \hline\hline
     8-12  & $8.9\times10^{-2}$  & 6336-383 & 0.818 \\
     9-11  & $9.3\times10^{-2}$  & 5970-368 & 0.818 \\
10.0-10.5  & 0.142  & 2008-136 & 0.909 \\
    \hline   

    \hline
    \multicolumn{4}{|c|}{ETGs: MANGA vs TNG100} \\
    \hline
    Lum. interval &  Weight. Energ. Dist. &  Nr. of Gal. &   Permutation p-val. \\
    \hline\hline
     8-12  & 0.813  & 1635-9202 & $9.0\times10^{-2}$ \\
     9-11  & 0.846  & 1156-9116 & $9.0\times10^{-2}$  \\
10.0-10.5  & 0.823  & 586-6672 & 0.182 \\
    \hline   

    \hline
    \multicolumn{4}{|c|}{LTGs: MANGA vs TNG100} \\
    \hline
    Lum. interval &  Weight. Energ. Dist. &  Nr. of Gal. &   Permutation p-val. \\
    \hline\hline
     8-12  & 1.089  & 6336-919 & 0.182 \\
     9-11  & 1.140  & 5970-917 & 0.182 \\
10.0-10.5  & 1.257  & 2008-625 & 0.182 \\
    \hline   

    \end{tabular}
    \caption{Results of the Energy test. WINGS and MANGA vs TNG50 and TNG100. The table provides in each column: the luminosity interval considered in the comparison, the value of the weighted energy distance, the number of WINGS and MANGA objects available for the calculation of the percentage of galaxies in each box of the \IeRe\ plane and the values of the probability of similarity of the two distributions when 10 random permutations are applied to the data.}
    \label{tab:2}
\end{table}

In order to improve the analysis, the Energy Distance test was also applied for the comparison between WINGS, MANGA and the Illustris datasets (TNG50 and TNG100) (see \ref{tab:2}).
The last column of the table gives an idea of the similarity of the 2D distributions in the \IeRe\ plane.

The following remarks can be made: 1. the ETGs of the TNG50 match well the WINGS data but much less the MANGA data. Viceversa the WINGS LTGs do not match the TNG50 simulations, while MANGA do it much better; 2. WINGS ETGs are statistically consistent with the TNG100 data, while MANGA ETGs do not. The LTGs of TNG100 are inconsistent both with WINGS and MANGA.

{  As before, the test shows that when the magnitude interval is reduced the agreement  among samples increases. This again would require  samples with similar distributions of objects in intervals of interest.}


\section{Conclusions}\label{sec:10}

In this paper, we presented a panoramic overview of the most commonly used scaling relations (ScRs), derived from galaxies of different morphological types selected from two large surveys: WINGS and MANGA. Our goal was to highlight the importance of adopting consistent and well-defined procedures when constructing and analysing galaxy scaling relations.

Our detailed comparisons show that, although the two samples yield broadly similar ScRs, the intrinsic differences in the 2D-distributions of structural parameters do not allow us to conclude that the galaxies under examinations belong to the same parent population. The Kolmogorov–Smirnov test applied to the vectors representing the percentage density of galaxies in each region of the 2D parameter space consistently returns very low probabilities for the hypothesis that the two samples are drawn from the same distribution. This seems confirmed by the Energy Distance test.

This lack of similarity indicates that {preliminar studies of the selection criteria and  the methods adopted in the data analysis are necessary before attempting any comparison. Consequently,} we do not yet possess a standard, homogeneous sample of galaxies with robustly measured structural parameters suitable for direct comparison with simulations. Even more relevant, this indicates that the comparison between real and simulated galaxies cannot be purely qualitative: the simulations should reproduce not only the general morphological appearance of galaxies, but also their statistical distribution in the local Universe. Our results show that neither real–real nor real–simulated comparisons of the two-dimensional ScRs distributions yield a positive match. The relations may look qualitatively similar, but their full 2D distributions are significantly different.

These results underline the need to set up  standardized scaling relations, based on statistically well-defined samples and consistent definitions of the structural parameters. Such parameters must be measured in ways that are compatible between observations and simulations if we wish to perform meaningful, quantitative comparisons.

The discrepancies found among observational and theoretical ScRs imply that the current data-sets are not yet robust enough to serve as direct tests for cosmological simulations. If simulations should reproduce not only the qualitative characteristics but also the detailed statistical distributions of galaxies, then observational benchmarks should be standardized and tightly controlled. Our results suggest that meaningful progress in comparing real and simulated galaxies will require coordinated developments on both sides: improved homogeneity among observational samples and more realistic theoretical simulations.

\authorcontributions{Conceptualization, M.D. and C.C.; methodology, M.D. and F. B.; software, M.D.; validation, M.D., C.C. and F.B.; formal analysis, M.D.; investigation, M.D.; resources, M.D.; data curation, F.B.; writing---original draft preparation, M.D. and C.C.; writing---review and editing, M.D.; visualization, M.D.; supervision, C.C.; project administration, M.D.; funding acquisition, M.D. All authors have read and agreed to the published version of the manuscript.}

\acknowledgments{C.C. thanks the Department of Physics and Astronomy for the hospitality.}

\conflictsofinterest{The authors declare no conflicts of interest.} 

\noindent 



\begin{adjustwidth}{-\extralength}{0cm}

\reftitle{References}

\bibliography{Paperbib.bib}

\PublishersNote{}
\end{adjustwidth}
\end{document}